\documentclass{article}

\usepackage{algorithm} \usepackage[noend]{algorithmic}
\usepackage{amsmath,amssymb,amsthm,bm,epic,geometry,graphicx,subfigure}

\newtheorem{thm}{Theorem}[section]

\newtheorem{lem}[thm]{Lemma}
\newtheorem{defn}[thm]{Definition}

\newcommand{\accumulatedirectlocalmoment}{\textsc{AccumulateDirectLocalMoment}}
\newcommand{\accumulatefarfieldmoment}{\textsc{AccumulateFarFieldMoment}}

\newcommand{\kdtree}{\textsc{BuildKdTree}}
\newcommand{\choosebestmethod}{\textsc{ChooseBestMethod}}
\newcommand{\computehermitefunction}{\textsc{ComputeHermiteFunction}}
\newcommand{\computepartialderivatives}{\textsc{ComputePartialDerivatives}}
\newcommand{\computemultiindexexpansion}{\textsc{MultiIndexExpansion}}

\newcommand{\dtkde}{\textsc{DFGT}}
\newcommand{\dtkdebase}{\textsc{DFGTBase}}
\newcommand{\dtkdeinitquerytree}{\textsc{DFGTInitQ}}
\newcommand{\dtkdeinitreferencetree}{\textsc{DFGTInitR}}
\newcommand{\dtkdemain}{\textsc{DFGTMain}}
\newcommand{\dtkdepost}{\textsc{DFGTPost}}

\newcommand{\cansummarize}{\textsc{CanSummarize}}

\newcommand{\summarize}{\textsc{Summarize}}

\newcommand{\dualtree}{\textsc{DualTree}}
\newcommand{\dualtreebase}{\textsc{DualTreeBase}}
\newcommand{\evaluatefarfieldexpansion}{\textsc{EvalFarFieldExpansion}}
\newcommand{\evaluatelocalexpansion}{\textsc{EvalLocalExpansion}}

\newcommand{\multiindextoposition}{\textsc{MultiIndexToPosition}}
\newcommand{\naivekde}{\textsc{NaiveKDE}}

\newcommand{\positiontomultiindex}{\textsc{PositionToMultiindex}}

\newcommand{\translatefartofarfield}{\textsc{TransFarToFar}}
\newcommand{\translatefartolocal}{\textsc{TransFarToLocal}}
\newcommand{\translatelocaltolocal}{\textsc{TransLocalToLocal}}
\newcommand{\farfieldorder}{\textsc{FarFieldOrder}}
\newcommand{\localaccumulationorder}{\textsc{LocalAccumulationOrder}}
\newcommand{\convertfarfieldtolocalorder}{\textsc{ConvertFarFieldToLocalOrder}}

\begin{document}

\title{Dual-Tree Fast Gauss Transforms}

 \author{Dongryeol Lee\\ School of Computational Science and
   Engineering\\Georgia Institute of Technology, Atlanta,
   GA. USA.\\ dongryel@cc.gatech.edu\\ \ \\Alexander G. Gray\\School of
   Computational Science and Engineering\\Georgia Institute of
   Technology, Atlanta, GA. USA.\\agray@cc.gatech.edu\\ \ \\Andrew
   W. Moore\\ Robotics Institute\\Carnegie Mellon University\\
   Pittsburgh, PA. USA.\\awm@cs.cmu.edu }

\maketitle

\begin{abstract}
Kernel density estimation (KDE) is a popular statistical technique for
estimating the underlying density distribution with minimal
assumptions. Although they can be shown to achieve asymptotic
estimation optimality for any input distribution, cross-validating for
an optimal parameter requires significant computation dominated by
kernel summations. In this paper we present an improvement to the
dual-tree algorithm, the first practical kernel summation algorithm
for general dimension. Our extension is based on the series-expansion
for the Gaussian kernel used by fast Gauss transform. First, we derive
two additional analytical machinery for extending the original
algorithm to utilize a hierarchical data structure, demonstrating the
first truly hierarchical fast Gauss transform. Second, we show how to
integrate the series-expansion approximation within the dual-tree
approach to compute kernel summations with a user-controllable
relative error bound. We evaluate our algorithm on real-world datasets
in the context of optimal bandwidth selection in kernel density
estimation. Our results demonstrate that our new algorithm is the only
one that guarantees a hard relative error bound and offers fast
performance across a wide range of bandwidths evaluated in cross
validation procedures.
\end{abstract}

\section{Introduction}
Kernel density estimation (KDE) is the most widely used and studied
nonparametric density estimation method. The model is the reference
dataset $\mathcal{R}$ itself, containing the reference points indexed
by natural numbers. Assume a local kernel function $K_h(\cdot)$
centered upon each reference point, and its scale parameter $h$ (the
'bandwidth'). The common choices for $K_h(\cdot)$ include the
spherical, Gaussian and Epanechnikov kernels. We are given the query
dataset $\mathcal{Q}$ containing query points whose densities we want
to predict. The density estimate at the $i$-th query point $q_i \in
\mathcal{Q}$ is: {\small
\begin{equation}
\hat{p}_h(q_i) = \frac{1}{|\mathcal{R}|} \sum\limits_{r_j \in
\mathcal{R}} \frac{1}{V_{Dh}} K_h \left( ||q_i - r_j|| \right)
\end{equation}
}
\noindent where $||q_i - r_j||$ denotes the Euclidean distance between the
$i$-th query point $q_i$ and the $j$-th reference point $r_j$, $D$ the
dimensionality of the data, $|\mathcal{R}|$ the size of the reference
dataset, and $V_{Dh} = \int\limits_{-\infty}^{\infty} K_h(z) dz$, a
normalizing constant depending on $D$ and $h$. With no assumptions on
the true underlying distribution, if $h \rightarrow 0$ and $|
\mathcal{R} | h \rightarrow \infty$ and $K(\cdot)$ satisfy some mild
conditions: {\small
\begin{equation}
\int |\hat{p}_h(x) - p(x)|dx \rightarrow 0
\end{equation}}
\noindent as $| \mathcal{R} | \rightarrow \infty$ with probability
1. As more data are observed, the estimate converges to the true
density.
\begin{algorithm}[t]

\caption{$\mbox{\naivekde}(\mathcal{Q}, \mathcal{R})$: A brute-force
  computation of KDE.}

\begin{algorithmic}

\FOR{each $q_i \in \mathcal{Q}$}

\STATE{$G(q_i, \mathcal{R}) \leftarrow 0$}

\FOR{each $r_j \in \mathcal{R}$}

\STATE{$G(q_i, \mathcal{R}) \leftarrow G(q_i, \mathcal{R}) + K_h(||q_i
- r_j||)$}

\ENDFOR

\STATE{Normalize each $G(q_i, \mathcal{R})$}

\ENDFOR
\end{algorithmic}
\label{alg:bruteforce_alg}
\end{algorithm}
In order to build our model for evaluating the densities at each $q_i
\in \mathcal{Q}$, we need to find the initially unknown asymptotically
optimal bandwidth $h^*$ for the given reference dataset
$\mathcal{R}$. There are two main types of cross-validation methods
for selecting the asymptotically optimal bandwidth. Cross-validation
methods use the reference dataset $\mathcal{R}$ as the query dataset
$\mathcal{Q}$ (i.e. $\mathcal{Q} = \mathcal{R}$). {\it Likelihood
  cross-validation} is derived by minimizing the Kullback-Leibler
divergence $\int p(x) \log \frac{p(x)}{\hat{p}_h(x)} dx$, which yields
the score:
{\small
\begin{equation}
CV_{LK}(h) = \frac{1}{|\mathcal{R}|} \sum\limits_{r_j \in \mathcal{R}
  } \log \hat{p}_{h,-j}(r_j)
\end{equation}
}
\noindent where the $-j$ subscript denotes an estimate using all $|\mathcal{R}|$
points except the $j$-th reference point. The bandwidth
$h^*_{CV_{LK}}$ that maximizes $CV_{LK}(h)$ is an asymptotically
optimal bandwidth in likelihood cross validation sense. {\it
Least-squares cross-validation} minimizes the integrated squared
error\\ $\int \left(\hat{p}_h(x) - p(x)\right)^2 dx$, yielding the
score:
{\small
\begin{equation}
CV_{LS}(h) = \frac{1}{|\mathcal{R}|} \sum\limits_{r_j \in \mathcal{R}}
\left( \hat{p}^{*}_{-j}(r_j) - 2 \hat{p}_{-j}(r_j) \right)
\end{equation}
}
\noindent where $\hat{p}^{*}_{-j}(\cdot)$ is evaluated using the convolution
kernel $K_h(\cdot) * K_h(\cdot)$. For the Gaussian kernel with
bandwidth of $h$, the convolution kernel $K_h(\cdot) * K_h(\cdot)$ is
the Gaussian kernel with bandwidth of $2h$. Both cross validation
scores require $|\mathcal{R}|$ density estimate based on
$|\mathcal{R}| - 1$ points, yielding a brute-force computational cost
scaling {\bf quadratically} (that is $O(|\mathcal{R}|^2)$) (see
Algorithm~\ref{alg:bruteforce_alg}). To make matters worse,
nonparametric methods require a large number of reference points for
convergence to the true underlying distribution and this has prevented
many practitioners from applying nonparametric methods for function
estimation.

\subsection{Efficient Computation of Gaussian Kernel Sums}
One of the most commonly used kernel function is the Gaussian kernel,
$K_h(||q_i - r_j||) = e^{\frac{-||q_i - r_j||^2}{2h^2}}$, although it
is not the asymptotically optimal kernel. In this paper we focus on
evaluating the Gaussian sums efficiently for each $q_i \in
\mathcal{Q}$: {\small
\begin{equation}
G(q_i, \mathcal{R}) = \sum\limits_{r_j \in \mathcal{R}}
e^{\frac{-||q_i - r_j||^2}{2h^2}}
\label{eq:gaussian_kernel_sums}
\end{equation}}which is proportional to $\hat{p}(q_i)$ using the Gaussian
kernel. This computationally expensive sum is evaluated many times
when cross-validating for an asymptotically optimal bandwidth for the
Gaussian kernel. Algorithms have been developed to approximate the
Gaussian kernel sums at the expense of reduced precision. We consider
the following two error bound criteria that measure the quality of the
approximation with respect to the true value.
\begin{defn}{\bf (Bounding the absolute error) }
An approximation algorithm guarantees $\epsilon$ absolute error bound,
if for each exact value $\Phi(q_i, \mathcal{R} ) $, it computes an
approximation $\widetilde{\Phi}(q_i, \mathcal{R})$ such that $|
\widetilde{\Phi}(q_i, \mathcal{R}) - \Phi(q_i, \mathcal{R} ) | \leq
\epsilon$.
\end{defn}
\begin{defn}{\bf (Bounding the relative error) }
An approximation algorithm guarantees $\epsilon$ relative error bound,
if for each exact value $\Phi(q_i, \mathcal{R} )$, it computes an
approximation $\widetilde{\Phi}(x_q, \mathcal{R} )$ such that $|
\widetilde{\Phi}(q_i, \mathcal{R} ) - \Phi(q_i, \mathcal{R} ) | \leq
\epsilon | \Phi(q_i, \mathcal{R} ) |$.
\label{defn:bound_relative_error}
\end{defn}
Bounding the relative error is much harder because the error bound is
in terms of the initially unknown exact quantity. Many previous
methods~\cite{greengard1991fgt, yang2003improved} have focused on
bounding the absolute error. Nevertheless, the relative error bound
criterion is preferred to the absolute error bound criterion in
statistical applications. Therefore, our experiment will evaluate the
performance of the algorithms for achieving the user-specified
relative error tolerance. Our new algorithm which builds
upon~\cite{gray2003nde,gray2003rem,gray2003vfm} is the only one to
guarantee both the absolute error and the relative error bound
criterion for all density estimates.

\subsection{Previous Approaches}
\begin{figure}[!t]
\subfigure[]{\noindent\scalebox{0.5}{\includegraphics{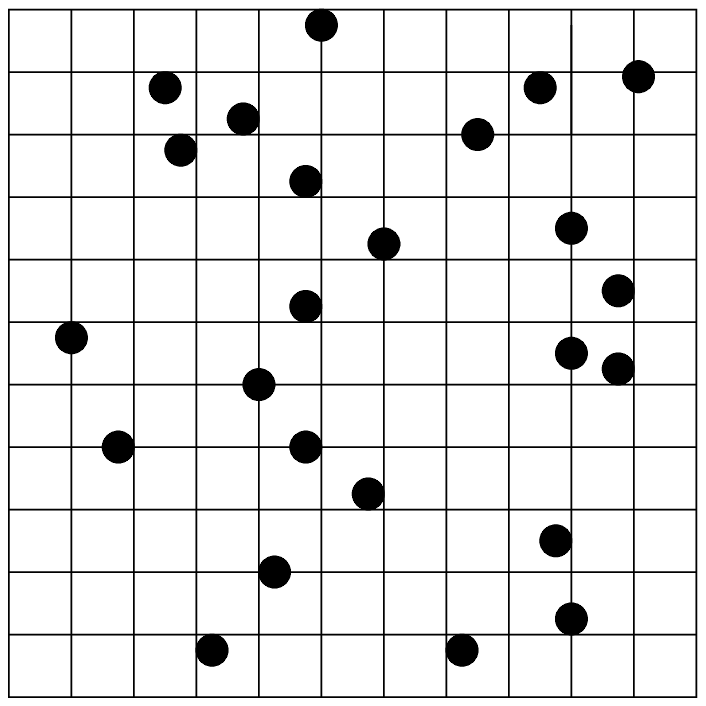}}
\label{fig:fgt_grid}}
\hspace{0.1in}
\subfigure[]{\noindent\scalebox{0.5}{\includegraphics{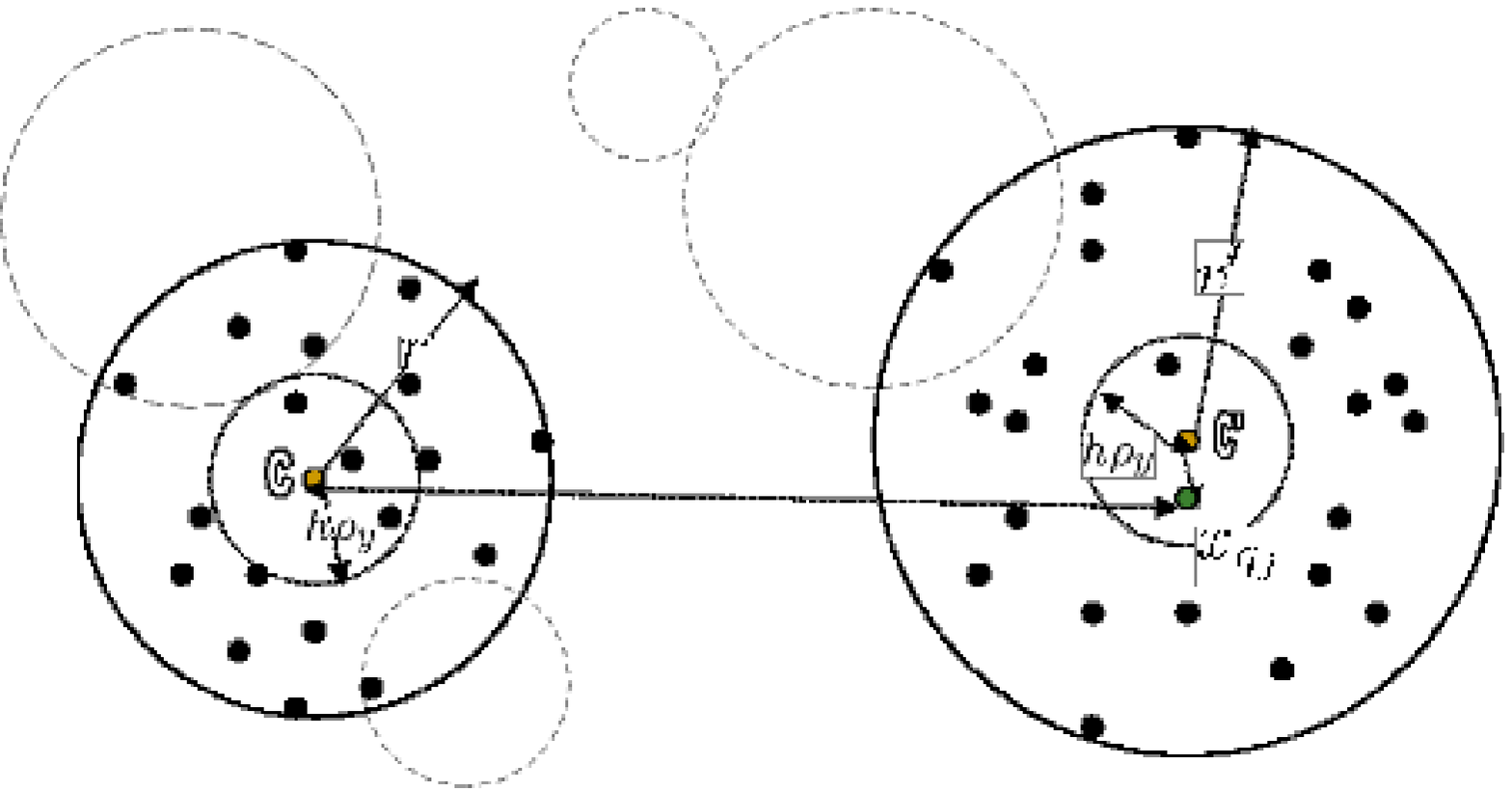}}
\label{fig:ifgt_cluster}}
\caption{(a) Grid structure used in fast Gauss transform and
multidimensional fast Fourier transform. (b) Single-level Clustering
structure used in improved fast Gauss transform.}
\end{figure}
There are three main approaches proposed for overcoming the
computational barrier in evaluating the Gaussian kernel sums:
\begin{enumerate}
\item{to expand the kernel sum as a power
  series~\cite{greengard1991fgt,yang2003improved,raykar2005fast}
  using a grid or a flat-clustering.}
\item{to express the kernel sum as a convolution sum by using the grid
of field charges created from the dataset~\cite{wand94}.}
\item{to utilize an adaptive hierarchical structure to group data
  points based on
  proximity~\cite{gray2003nde,gray2003rem,gray2003vfm}.}
\end{enumerate}
Now we briefly describe the strengths and the weaknesses of these
methods.

{\ \\ \noindent \bf The Fast Gauss Transform (FGT). }
FGT~\cite{greengard1991fgt} belongs to a family of methods called the
Fast Multipole Methods (FMM). These family of methods come with
rigorous error bound on the kernel sums. Unlike other FMM algorithms,
FGT uses a grid structure (see Figure~\ref{fig:fgt_grid}) whose
maximum side length is restricted to be at most the bandwidth $h$ used
in cross-validation due to the error bound criterion.
FGT has not been widely used in higher dimensional statistical
contexts. First, the number of the terms in the power series expansion
for the kernel sums grows exponentially with dimensionality $D$; this
causes computational bottleneck in evaluating the series expansion or
translating a series expansion from one center to another. Second, the
grid structure is extremely inefficient in higher dimensions since the
storage cost is exponential in $D$ and many of the boxes will be
empty. 

{\noindent \bf The Improved Fast Gauss Transform (IFGT). }IFGT is
similar to FMM but utilizes a flat clustering to group data points
(see Figure~\ref{fig:ifgt_cluster}), which is more efficient than a
grid structure used in FGT. The number of clusters $k$ is chosen in
advance. A partition of the data points into $C_1$, $C_2$, $\cdots$,
$C_k$ is formed so that each reference point $r_j \in \mathcal{R}$ is
grouped according to its proximity to the set of representative points
$c_1$, $c_2$, $\cdots$, $c_k$. That is, $r_j \in C_m$ (whose
representative point is $c_m$) if and only if $||r_j - c_m|| \leq
||r_j - c_l||$ for $1 \leq l \leq k$.

Furthermore, IFGT proposes using a different series expansion that
does not require translation of expansion centers as done in FGT. The
original algorithm~\cite{yang2003improved} required tweaking of
multiple parameters which did not offer for a user to control the
accuracy of the approximation.  The latest
version~\cite{raykar2005fast} is now fully automatic in choosing the
approximation parameter for the absolute error bound, but is still
inefficient except on large bandwidth parameters. We will discuss this
further in Section~\ref{sec:experimental_results}.

\begin{figure}[t]
\centering
\subfigure[Nearest Neighbor Binning Rule $(A = 1, B = C = D = 0)$]{
  \begin{picture}(150,150)
  \put(15,2){\makebox(10,10){A(0,0)}}
  \put(15,132){\makebox(10,10){B(0,150)}}
  \put(120,132){\makebox(10,10){C(150,150)}}
  \put(120,2){\makebox(10,10){D(150,0)}}
  \put(50,50){\circle{5}}
  \put(63,55){\makebox(10,10){(50,50)}}
  \drawline(0,0)(0,150)(150,150)(150,0)(0,0)
  \dashline{3}(50,0)(50,150)
  \dashline{3}(0,50)(150,50)
  \end{picture}
  \label{NNBinRuleFigure}
}
\hspace{0.25in} \subfigure[Linear Binning Rule $(A = \frac{4}{9}, B =
\frac{2}{9}, C = \frac{1}{9}, D = \frac{2}{9})$]{
  \begin{picture}(150,150)
  \put(15,2){\makebox(10,10){A(0,0)}}
  \put(15,132){\makebox(10,10){B(0,150)}}
  \put(120,132){\makebox(10,10){C(150,150)}}
  \put(120,2){\makebox(10,10){D(150,0)}}
  \put(63,55){\makebox(10,10){(50,50)}}
  \drawline(0,0)(0,150)(150,150)(150,0)(0,0)
  \put(50,50){\circle{5}}
  \dashline{3}(50,0)(50,150)
  \dashline{3}(0,50)(150,50)
  \end{picture}
  \label{LBinRuleFigure}
}
\caption[Binning Rules for Multidimensional Fast Fourier
Transform]{Two possible binning rules for KDE using multidimensional
fast Fourier transform. Consider a data point falling in a
two-dimensional rectangle. In \ref{NNBinRuleFigure}, the entire weight
is assigned to the nearest grid point. In \ref{LBinRuleFigure}, the
weight is distributed to all neighboring grid points by linear
interpolation.}
\end{figure}
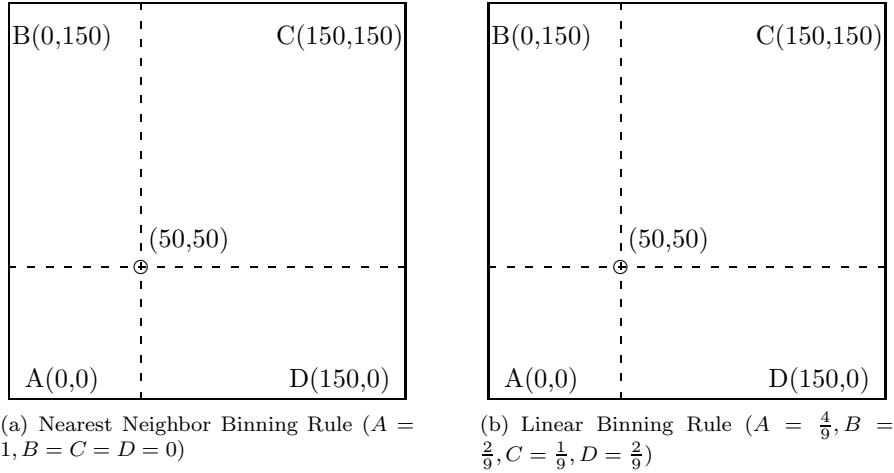
{\ \\ \noindent \bf Fast Fourier Transform (FFT). } FFT is often
quoted as the solution to the computational problem in evaluating the
Gaussian kernel sums. Gaussian kernel summation using FFT is described
in~\cite{silverman1982kernel} and ~\cite{wand94}. ~\cite{silverman1982kernel} discusses the
implementation of KDE only in a univariate case, while~\cite{wand94}
extends~\cite{silverman1982kernel} to handle more than one dimension. It uses a
grid structure shown in Figure~\ref{fig:fgt_grid} by specifying the
number of grid points along each dimension.

The algorithm first computes the $M_1 \times \cdot\cdot\cdot \times
M_D$ matrix by binning the data assigning the raw data to neighboring
grid points using one of the binning rules. This involves computing
the minimum and maximum coordinate values ($g_{i, M_i}, g_{i, 1}$),
and the grid width $\delta_i = \frac{g_{i, M_i} - g_{i, 1}}{M_i - 1}$
for each $i$-th dimension. This essentially divides each $i$-th
dimension into $M_i - 1$ intervals of equal length. In particular,
\cite{wand94} discusses two different types of binning rules - linear
binning, which is recommended by Silverman, and nearest-neighbor
binning. \cite{wand94} states that nearest-neighbor binning rule
performs poorly, so we will test the implementation using the linear
binning rule, as recommended by both authors.
In addition, we compute the $L_1 \times \cdot\cdot\cdot \times L_D$
kernel weight matrix, where $L_i = \min \left (\left\lfloor \frac{\tau
  h}{\delta_i} \right \rfloor, M_i - 1 \right )$, with $\tau \approx
4$ and $K_l = \prod\limits_{k = 1}^d e^{\frac{-0.5l_k\delta_k}{h^2}}$,
$-L_k \leq l_k \leq L_k$, for $l = [l_1, ..., l_D]^T \in
\mathbb{Z}^D$.

To reduce the wrap-around effects of fast Fourier transform near the
dataset boundary, we appropriately zero-pad the grid count and the
kernel weight matrices to two matrices of the dimensionality $P_1
\times \cdot\cdot\cdot P_D$, where $P_i = 2^{\log_2 \lceil M_i + L_i
  \rceil}$. The key ingredient in this method is the use of
Convolution Theorem for Fourier transforms. The structure of the
computed grid count matrix and the kernel weight matrix is crafted to
take advantage of the fast Fourier transform. For every grid point $g
= (g_{1j_1}, ..., g_{dj_D})$, $\tilde{s_k}(g_j) = \sum\limits_{l_1 =
  -L_1}^{L_1} \cdots \sum\limits_{l_D = -L_D}^{L_D} c_{j - l} K_{k,
  l}$ can be computed using the Convolution Theorem for Fourier
Transform. After taking the convolution of the grid count matrix and
the kernel weight matrix, the $M_1 \times \cdots \times M_D$
sub-matrix in the upper left corner of the resultant matrix contains
the kernel density estimate of the grid points. The density estimate
of each query point is then linearly interpolated using the density
estimates of neighboring grid points inside the cell it falls into.
\begin{figure}[!t]
\centering The grid count matrix: $c^Z = \begin{pmatrix} c_{1, 1}
&\cdots & c_{1, M_2}\\ \vdots & \ddots & \vdots & \textbf{0}\\ c_{M_1,
1} & \cdots & c_{M_1, M_2}\\ & \textbf{0} & & \textbf{0}\end{pmatrix}$

The kernel weight matrix: $\textbf{K}^Z = \begin{pmatrix} K_{00} &
\cdots & K_{0L_2} & & K_{0L_2} & \cdots & K_{01} \\ \vdots & \ddots &
\vdots & \textbf{0} & \vdots & \ddots & \vdots \\ K_{L_10} & \cdots &
K_{L_1L_2} & & K_{L_1L_2} & \cdots & K_{L_11} \\ & \textbf{0} & &
\textbf{0} & & \textbf{0} \\ K_{L_10} & \cdots & K_{L_1L_2} & &
K_{L_1L_2} & \cdots & K_{L_11} \\ \vdots & \ddots & \vdots &
\textbf{0} & \vdots & \ddots & \vdots \\ K_{10} & \cdots & K_{1L_2} &
& K_{1L_2} & \cdots & K_{11}
\end{pmatrix}$

where $K_{l_1, l_2} = e^{\frac{-0.5((l_1\delta_1)^2 +
(l_2\delta_2)^2)}{h^2}}$.
\caption[Grid Count and Kernel Weight Matrix in KDE using FFT]{The
grid count and the kernel weight matrix formed for a two-dimensional
dataset. They are formed by appropriately zero-padding for taking the
boundary-effects of fast Fourier transform based algorithms into
account.}
\end{figure}
However, performing a calculation on equally-spaced grid points
introduces artifacts at the boundaries of the data. The linear
interpolation of the data points by assigning to neighboring grid
points introduce further errors. Increasing the number of grid points
to use along each dimension can provide more accuracy but also require
more space to store the grid. Moreover, it is impossible to directly
quantify incurred error on each estimate in terms of the number of
grid points.

{\noindent \bf Dual-tree KDE. }In terms of discrete algorithmic
structure, the dual-tree framework of \cite{gray2001nbp} generalizes
all of the well-known kernel summation algorithms.  These include the
Barnes-Hut algorithm \cite{barnes1986hof}, the Fast Multipole Method
\cite{greengard1987fap}, Appel's algorithm
\cite{appel1985efficient}, and the WSPD \cite{callahan1995dwh}: the
dual-tree method is a node-node algorithm (considers query regions
rather than points), is fully recursive, can use
distribution-sensitive data structures such as {\it kd}-trees, and is
bichromatic (can specialize for differing query set $\mathcal{Q}$ and
reference set $\mathcal{R}$).  It was applied to the problem of kernel
density estimation in \cite{gray2003nde} using a simple variant of a
centroid approximation used in~\cite{appel1985efficient}.

This algorithm is currently the fastest Gaussian kernel summation
algorithm for general dimensions. Unfortunately, when performing
cross-validation to determine the (initially unknown) optimal
bandwidth, both sub-optimally small and large bandwidths must be
evaluated. Section~\ref{sec:experimental_results} demonstrates that
the dual-tree method tends to be efficient at the optimal bandwidth
and at bandwidths below the optimal bandwidth and at very large
bandwidths. However, its performance degrades for intermediately large
bandwidths.

\subsection{Our Contribution}
In this paper we present an improvement to the dual-tree
algorithm~\cite{gray2003nde,gray2003rem,gray2003vfm}, the first
practical kernel summation algorithm for general dimension. Our
extension is based on the series-expansion for the Gaussian kernel
used by fast Gauss transform~\cite{greengard1991fgt}. First, we
derive two additional analytical machinery for extending the original
algorithm to utilize a adaptive hierarchical data structure called
$kd$-trees~\cite{bentley1975multidimensional}, demonstrating the first truly hierarchical
fast Gauss transform, which we call the Dual-tree Fast Gauss Transform
(DFGT). Second, we show how to integrate the series-expansion
approximation within the dual-tree approach to compute kernel
summations with a user-controllable relative error bound. We evaluate
our algorithm on real-world datasets in the context of optimal
bandwidth selection in kernel density estimation. Our results
demonstrate that our new algorithm is the only one that guarantees a
relative error bound and offers fast performance across a wide range
of bandwidths evaluated in cross validation procedures.

\subsection{Structure of This Paper}
This paper builds on~\cite{lee2006dtf} where the Dual-Tree Fast Gauss
Transform was presented briefly. It adds details on the approximation
mechanisms used in the algorithm and provides a more thorough
comparison with the other algorithms. In
Section~\ref{sec:computational_strategy}, we introduce a general
computational strategy for efficiently computing the Gaussian kernel
sums. In Section~\ref{sec:higher_order_dfgt}, we describe our
extensions to the dual-tree algorithm to handle higher-order series
expansion approximations. In Section~\ref{sec:experimental_results},
we provide performance comparison with some of the existing methods
for evaluating the Gaussian kernel sums.

\subsection{Notations}
The general notation conventions used throughout this paper are as
follows. $\mathcal{Q}$ denotes the set of {\it query points} for which
we want to make the density computations. $\mathcal{R}$ denotes the
set of {\it reference points} which are used to construct the kernel
density estimation model. {\it Query points} and {\it reference
points} are indexed by natural numbers $i, j \in \mathbb{N}$ and
denoted $q_i$ and $r_j$ respectively. For any set $S$, $|S|$ denotes
the number of elements in $S$. For any vector $v \in \mathbb{R}^D$ and
$1 \leq i \leq D$, let $v[i]$ denote the $i$-th component of $v$.

\section{Computational Technique}
\label{sec:computational_strategy}
We first introduce a hierarchical method for for organizing the data
points for computation, and describe the {\it generalized $N$-body
  approach}~\cite{gray2003nde,gray2003rem,gray2003vfm} that enables
the efficient computation of kernel sums using a tree.

\subsection{Spatial Trees} 
\label{sec:spatial_trees}
A {\it spatial tree} is a {\it hierarchical} data structure that
allows summarization and access of the dataset at different
resolutions. The recursive nature of hierarchical data structures
enables efficient computations that are not possible with single-level
data structures such as grids and flat clusterings. A hierarchical
data structure satisfies the following properties:
\begin{enumerate}
\item{There is one root node representing the entire dataset.}
\item{Each leaf node is a terminal node.}
\item{Each internal node $N$ points to two child nodes $N^L$
and $N^R$ such that $N^L \cap N^R = \emptyset$ and $N^L \cup N^R =
N$.}
\end{enumerate}
Since a {\it node} can be viewed as a collection of points, each term
will be used interchangeably with the other. A {\it reference node} is
a collection of reference points and a {\it query node} is a
collection of query points. We use a variant of $kd$-trees
\cite{bentley1975multidimensional} to form hierarchical groupings of points based on their
locations using the {\it recursive} procedure shown in
Algorithm~\ref{alg:kdtree}.
\begin{algorithm}[t]
\caption{$\mbox{\kdtree}(\mathcal{P})$: Builds a mid-point {\it
    kd}-tree from $\mathcal{P}$.}

\begin{algorithmic}

\STATE{$N \leftarrow \mbox{empty\ node}$, \ \ \ \ $N.\mathcal{P}
  \leftarrow \mathcal{P}$, \ \ \ \ $N^L \leftarrow \emptyset$,
  \ \ \ \ $N^R \leftarrow \emptyset$}

\FOR{each $d \in [1, D]$}

\STATE{$N.b[d].l \leftarrow \min\limits_{x \in P} x[d]$,
  \ \ \ \ \ \ \ \ \ \ \ \ \ \ \ \ \ \ \ $N.b[d].u \leftarrow \max\limits_{x \in P} x[d]$}

\ENDFOR

\IF{$|\mathcal{P}|$ is above leaf threshold}

\STATE{$N.\mathit{sd} \leftarrow \arg\max\limits_{1 \leq d \leq D}
N.b[d].u - N.b[d].l$}

\STATE{$N.\mathit{sc} \leftarrow \frac{N.b[N.\mathit{sd}].l +
N.b[N.\mathit{sd}].u}{2}$}

\STATE{$\mathcal{P}^L \leftarrow \{x \in \mathcal{P} |
  x[N.\mathit{sd}] \leq N.\mathit{sc} \}$, \ \ \ \ $\mathcal{P}^R
  \leftarrow \{x \in \mathcal{P} | x[N.\mathit{sd}] > N.\mathit{sc}
  \}$}

\STATE{$N^L \leftarrow \mbox{\kdtree}(\mathcal{P}^L)$, \ \ \ \ \ \ \ \ $N^R
  \leftarrow \mbox{\kdtree}(\mathcal{P}^R)$}

\ENDIF

\RETURN $N$

\end{algorithmic}
\label{alg:kdtree}
\end{algorithm}
In this procedure, the set of points in each node $N$ defines a
bounding hyper-rectangle $[N.b[1].l, N.b[1].u] \times [N.b[2].l,
N.b[2].u] \times \cdots \times [N.b[D].l, N.b[D].u]$ whose $i$-th
coordinates for $1 \leq i \leq D$ are defined by: $N.b[i].l =
\min\limits_{x \in N.\mathcal{P}} x[i]$ and $N.b[i].u = \min\limits_{x
\in N.P} x[i]$ where $N.\mathcal{P}$ is the set of points owned by the
node $N$. We also define the geometric center of each node, which is
{\small
$$N.c = \left[ \frac{N.b[1].l + N.b[1].u}{2}, \frac{N.b[2].l +
      N.b[2].u}{2}, \cdots, \frac{N.b[D].l + N.b[D].u}{2} \right]^T
  \in \mathbb{R}^D$$}The node $N$
is split along the widest dimension of the bounding hyper-rectangle
$N.\mathit{sd}$ into two equal halves at the splitting coordinate
$N.\mathit{sc}$. The algorithm continues splitting until the number of
points is below the {\it leaf threshold}. Computing a bounding
hyper-rectangle requires $O(|\mathcal{P}|)$ cost.

\subsection{Generalized $N$-body Approach}
\label{sec:generalized_nbody_approach}
\begin{figure}[t]
\scalebox{1.4}{
\includegraphics[trim=0em 15em 0em 0em, clip=true]{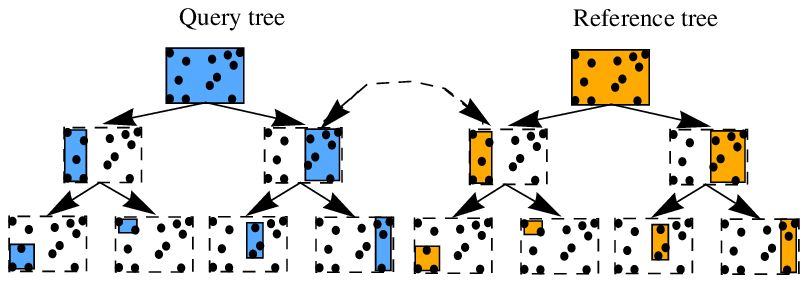}}
\scalebox{0.42}{
\includegraphics{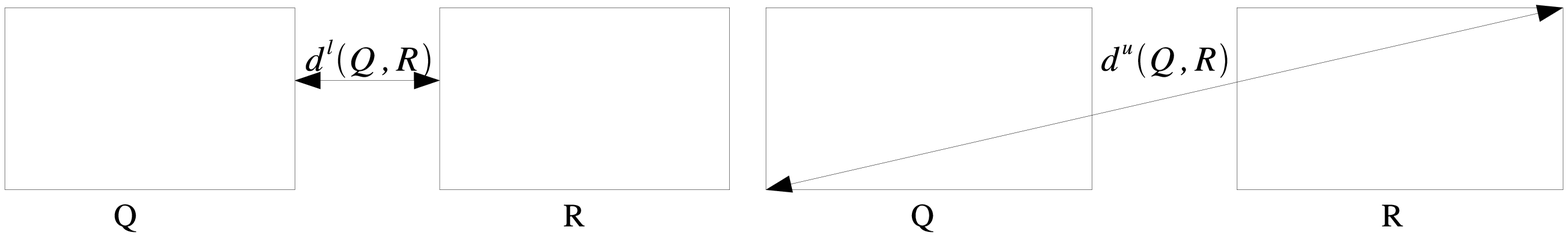}
}
\caption{{\bf Top:} A $kd$-tree partitions 2-dimensional points. Each
  node in the $kd$-tree records the bounding box for the subset of the
  dataset it contains (highlighted in color). In dual-tree recursion,
  a pair of nodes chosen from the query tree and the reference tree is
  considered at a time. {\bf Bottom:} the lower and upper bound on
  pairwise distances between the points contained in each of the
  query/reference node pair.}
\label{fig:dualtree}
\end{figure}

Recall that the computational task involved in KDE is defined as:
$\forall q_i \in \mathcal{Q}$, compute $G(q_i, \mathcal{R}) =
\sum\limits_{r_j \in \mathcal{R}} e^{\frac{-||q_i -
    r_j||^2}{2h^2}}$. The general framework for computing a summation
of this form is formalized
in~\cite{gray2003nde,gray2003rem,gray2003vfm}. This approach forms
{\it kd}-trees for both the query and reference data and then perform
a {\it dual-tree traversal} over pairs of nodes, demonstrated in
Figure~\ref{fig:dualtree} and Algorithm~\ref{alg:dualtree}. This
procedure is called with $Q$ and $R$ as the root nodes of the query
and the reference tree respectively. This allows us to compare chunks
of the query and reference data, using the bounding boxes and
additional information stored by the {\it kd}-tree to compute bounds on
distances as shown in Figure~\ref{fig:dualtree}.
\begin{algorithm}[t]
\caption{$\mbox{\dualtree}(Q, R)$: The dual-tree main routine.}

\begin{algorithmic}

\IF{\cansummarize $(Q, R, \epsilon)$}

\STATE{\summarize $(Q, R)$}

\ELSE

\IF{$Q$ is a leaf node}

\IF{$R$ is a leaf node}

\STATE $\mbox{\dualtreebase}(Q,R)$

\ELSE

\STATE{$\mbox{\dualtree}(Q, R^L)$, \ \ \ \ \ \ \ \ \ $\mbox{\dualtree}(Q, R^R)$}

\ENDIF

\ELSE

\IF{$R$ is a leaf node}

\STATE{$\mbox{\dualtree}(Q^L, R)$, \ \ \ \ \ \ \ \ \ $\mbox{\dualtree}(Q^R, R)$}

\ELSE

\STATE{$\mbox{\dualtree}(Q^L, R^L)$, \ \ \ \ \ \ \ $\mbox{\dualtree}(Q^L, R^R)$}

\STATE{$\mbox{\dualtree}(Q^R, R^L)$, \ \ \ \ \ \ \ $\mbox{\dualtree}(Q^R, R^R)$}

\ENDIF

\ENDIF

\ENDIF

\end{algorithmic}
\label{alg:dualtree}
\end{algorithm}
These distance bounds can be computed in $O(D)$ time using:
{ \small
\begin{align}
d^l(Q, R) &= \frac{1}{2} \sqrt{ \sum\limits_{k=1}^D
  \left( d^{l, u}_{j,i}[k] + \left | d^{l, u}_{j,i}[k] \right | +
d^{l,u}_{i,j}[k] + \left | d^{l,u}_{i,j}[k] \right |
\right )^2 }\\ d^u(Q, R) &= \sqrt{\sum\limits_{k=1}^D
  \left( \max \left \{ d^{u,l}_{j,i}[k], d_{i,j}^{u,l}[k] \right \} \right ) ^2 }
\end{align}
}where $d^{l,u}_{j,i}[k] = R^l[k] - Q^u[k]$, $d^{l, u}_{i, j}[k] =
Q^l[k] - R^u[k]$, $d^{u, l}_{j, i}[k] = R^u[k] - Q^l[k]$,\\ $d^{u,
  l}_{i, j}[k] = Q^u[k] - R^l[k]$.  The $\mbox{\cansummarize}$
function tests whether it is possible to summarize the sum
contribution of the given reference node for each query point in the
given query node. If possible, the $\mbox{\summarize}$ function
approximates the sum contribution of the given reference node; we then
say the given pair of the query node and the reference node has been
{\bf pruned}. The idea is to {\it prune} unneeded portions of the
dual-tree traversal, thereby minimizing the number of exhaustive
leaf-leaf computations.

\section{Dual-Tree Fast Gauss Transform}
\label{sec:higher_order_dfgt}

\subsection{Mathematical Preliminaries}
\label{sec:mathematical_preliminaries}
{\noindent \bf Univariate Taylor's Theorem. }The univariate Taylor's
theorem is crucial for the approximation mechanism in Fast Gauss
transform and our new algorithm:
\begin{thm}
If $n \geq 0$ is an integer and $f$ is a function which is $n$ times
continuously differentiable on the closed interval $[c, x]$ and $n +
1$ times differentiable on $(c, x)$ then
{\small
\begin{equation}
f(x) = \sum\limits_{i=0}^{n} f^{(i)}(c)\frac{(x-c)^i}{i!} +R_{n}
\end{equation}
}
where the Lagrange form of the remainder term is given by\\ $R_{n} =
f^{(n+1)}(\xi)\frac{(x-c)^{n+1}}{(n+1)!}$ for some $\xi \in (c, x)$.
\end{thm}

\ \\{\noindent \bf Multi-index Notation. }Throughout this paper, we
will be using the multi-index notation. A $D$-dimensional multi-index
$\alpha$ is a $D$-tuple of non-negative integers. For any
$D$-dimensional multi-indices $\alpha$, $\beta$ and any $x\in
\mathbb{R}^D$, {\small
\begin{itemize}
\item{$|\alpha| = \alpha[1] + \alpha[2] + \cdots + \alpha[D]$}
\item{$\alpha! = (\alpha[1])! (\alpha[2])! \cdots (\alpha[D])!$}
\item{$x^{\alpha} = (x[1])^{\alpha[1]} (x[2])^{\alpha[2]} \cdots
(x[D])^{\alpha[D]}$}
\item{$D^{\alpha} = \partial_1^{\alpha[1]} \partial_2^{\alpha[2]}
\cdots \partial_D^{\alpha[D]}$}
\item{$\alpha + \beta = (\alpha[1] + \beta[1], \cdots, \alpha[D] +
\beta[D])$}
\item{$\alpha - \beta = (\alpha[1] - \beta[1], \cdots, \alpha[D] -
\beta[D])$ for $\alpha \geq \beta$.}
\end{itemize}
} where $\partial_i$ is a $i$-th directional partial
derivative. Define $\alpha > \beta$ if $\alpha[d] > \beta[d]$, and
$\alpha \geq p$ for $p \in \mathbb{Z^+} \cup \{0 \}$ if $\alpha[d]
\geq p$ for $1 \leq d \leq D$ (and similarly for $\alpha \leq p$).

\ \\ {\noindent \bf Properties of the Gaussian Kernel.}
\label{PropertiesOfTheGaussianKernel}
Based on the univariate Taylor's Theorem stated
above,~\cite{greengard1991fgt} develops the series expansion
mechanism for the Gaussian kernel sum. Our development begins with
one-dimensional setting and generalizes to multi-dimensional
setting. We first define the Hermite polynomials by the Rodrigues'
formula: {\small
\begin{equation}
\begin{split}
H_n(t) = (-1)^n e^{t^2} D^n e^{-t^2}, t \in \mathbb{R}^1
\end{split}
\end{equation}
}
The first few polynomials include: $H_0(t) = 1$, $H_1(t) = 2t$,
$H_2(t) = 4t^2 - 2$.
The generating function for Hermite polynomials is defined by:
{\small
\begin{equation}
\begin{split}
e^{2ts - s^2} = \sum\limits_{n = 0}^{\infty} \frac{s^n}{n!} H_n(t)
\end{split}
\end{equation}
}
Let us define the Hermite functions $h_n(t)$ by
{\small
\begin{equation}
\begin{split}
h_n(t) = e^{-t^2}H_n(t)
\end{split}
\end{equation}
}
Multiplying both sides by $e^{-t^2}$ yields:
\begin{equation}
\begin{split}
e^{-(t - s)^2} = \sum\limits_{n = 0}^{\infty} \frac{s^n}{n!} h_n(t)
\end{split}
\end{equation}
We would like to use a ``scaled and shifted'' version of this
derivation for taking the bandwidth $h$ into account.
{\small
\begin{align}
e^{\frac{-(t - s)^2}{2h^2}} &= e^{\frac{-((t - s_0) - (s - s_0))^2}{2h^2}}
= \sum\limits_{n = 0}^{\infty} \frac{1}{n!} \left( \frac{s -
s_0}{\sqrt{2h^2}} \right)^n h_n \left( \frac{t - s_0}{\sqrt{2h^2}} \right)
\end{align}
}
Note that our $D$-dimensional multivariate Gaussian kernel can be
expressed as a product of $D$ one-dimensional Gaussian
kernel. Similarly, the multidimensional Hermite functions can be
written as a product of one-dimensional Hermite functions using the
following identity for any $t \in \mathbb{R^D}$.
{\small
\begin{equation}
\begin{split}
H_{\alpha}(t) &= H_{\alpha[1]}(t[1]) \cdots H_{\alpha[D]}(t[D])\\
h_{\alpha}(t) &= e^{-||t||^2}H_{\alpha}(t) = h_{\alpha[1]}(t[1])
\cdots h_{\alpha[D]}(t[D])\\
\end{split}
\end{equation}
}
where $||t||^2 = (t[1])^2 + \cdots + (t[D])^2$.
{\small
\begin{equation}
\begin{split}
e^{\frac{-||t - s||^2}{2h^2}} &= e^{\frac{-(t[1] - s[1])^2 - (t[2] -
s[2])^2 - \cdots - (t[D] - s[D])^2}{2h^2}}\\ &= e^{\frac{-(t[1] -
s[1])^2}{2h^2}}e^{\frac{-(t[2] - s[2])^2}{2h^2}} \cdots
e^{\frac{-(t[D] - s[D])^2}{2h^2}}\\
\end{split}
\end{equation}
}
We can also express the multivariate Gaussian about
another point $s_0 \in \mathbb{R}^D$ as:
{\small
\begin{equation}
\begin{split}
e^{\frac{-||t - s||^2}{2h^2}} &= \prod\limits_{d=1}^D \left(
\sum\limits_{n_d = 0}^{\infty} \frac{1}{n_d!} \left( \frac{s[d] -
s_0[d]}{\sqrt{2h^2}} \right)^{n_d} h_{n_d} \left( \frac{t[d] -
s_0[d]}{\sqrt{2h^2}} \right) \right)\\ &= \sum\limits_{\alpha \geq 0}
\frac{1}{\alpha!} \left( \frac{s - s_0}{\sqrt{2h^2}} \right)^{\alpha}
h_{\alpha} \left( \frac{t - s_0}{\sqrt{2h^2}} \right)\\
\end{split}
\label{eq:primal}
\end{equation}
}
The representation which is {\it dual} to Equation~\eqref{eq:primal}
is given by:
{\small
\begin{equation}
\begin{split}
e^{\frac{-||t - s||^2}{2h^2}} &= \prod\limits_{d=1}^D \left(
\sum\limits_{n_d = 0}^{\infty} \frac{(-1)^{n_d}}{n_d!} h_{n_d} \left(
\frac{t_0[d] - s(d)}{\sqrt{2h^2}} \right) \left( \frac{t[d] -
t_0}{\sqrt{2h^2}}\right)^{\beta} \right)\\ &= \sum\limits_{\beta \geq
0} \frac{(-1)^{\beta}}{\beta!}  h_{\beta} \left( \frac{t_0 -
s}{\sqrt{2h^2}} \right) \left( \frac{t -
t_0}{\sqrt{2h^2}}\right)^{\beta}
\end{split}
\label{eq:dual}
\end{equation}
}
The final property is the recurrence relation of the one-dimensional
Hermite function:
{\small
\begin{equation}
\begin{split}
h_{n + 1}(t) = 2t \cdot h_n(t) - 2n \cdot h_{n - 1}(t), t \in
\mathbb{R}^1
\end{split}
\end{equation}
}
\noindent and the Taylor expansion of the Hermite function $h_{\alpha}(t)$ about
$t_0 \in \mathbb{R^D}$.
{\small
\begin{equation}
\begin{split}
h_{\alpha}(t) = \sum\limits_{\beta \geq 0} \frac{(t -
t_0)^{\beta}}{\beta!}(-1)^{|\beta|} h_{\alpha + \beta}(t_0)\\
\end{split}
\end{equation}
}
\subsection{Notations in Algorithm Descriptions}
\label{sec:algorithmic_notations}
Here we summarize notations used throughout the descriptions and the
pseudocodes for our algorithms. The followings are notations that are
relevant to a query point $q_i \in \mathcal{Q}$ or a query node $Q$ in
the query tree.
\begin{itemize}
\item{$\mathcal{R_E}(\cdot)$: The set of reference points $r_{j_n} \in
  \mathcal{R}$ whose pairwise interaction is computed exhaustively for
  a query point $q_i \in \mathcal{Q}$ or a query node $Q$.}
\item{$\mathcal{R_T}(\cdot)$: The set of reference points $r_{j_n} \in
  \mathcal{R}$ whose contribution is pruned via centroid-based
  approximation for a given query point $q_i \in \mathcal{Q}$.}
\end{itemize}
The followings are notations relevant to a query point $q_i \in
\mathcal{Q}$.
\begin{itemize}
\item{$G(q_i, R)$: The true initially unknown kernel sum for a query
  point $q_i$ contributed by the reference set $R \subseteq
  \mathcal{R}$, i.e. $\sum\limits_{r_{j_n} \in R} K_h(||q_i -
  r_{j_n}||)$.}
\item{$G^l(q_i, \mathcal{R})$: A lower bound on $G(q_i, \mathcal{R})$.}
\item{$G^u(q_i, \mathcal{R})$: An upper bound on $G(q_i, \mathcal{R})$.}
\item{$\widetilde{G}(q_i, R)$: An approximation to $G(q_i, R)$ for $R
  \subseteq \mathcal{R}$. This obeys the additive property for a
  family of pairwise disjoint sets $\{ R_i \}_{i=1}^{m}$:
  $\widetilde{G}\left( q_i, \bigcup\limits_{i=1}^m R_i \right) =
  \sum\limits_{i=1}^m \widetilde{G}(q_i, R_i)$.}
\item{$\widetilde{G}\left (q_i, \{ (R_j, A_j) \}_{j=1}^m \right )$: A
  refined notation of $\widetilde{G}\left (q_i, \bigcup\limits_{j=1}^m
  R_j \right)$ to specify the type of approximation used for each
  reference node $R_j$.}
\end{itemize}
Here we define some notations for representing postponed bound changes
to $G^l(q_{i_m}, \mathcal{R})$ and $G^u(q_{i_m}, \mathcal{R})$ for all
$q_{i_m} \in Q$.
\begin{itemize}
\item{$Q.\Delta^l$: Postponed lower bound changes on $G^l(q_{i_m},
  \mathcal{R})$ for a query node $Q$ in the query tree and $q_i \in
  Q$.}
\item{$Q.L$: Postponed changes to $\widetilde{G}(q_{i_m},
\mathcal{R_T}(q_{i_m}))$ for $q_{i_m} \in Q$.}
\item{$Q.\Delta^u$: Postponed upper bound changes on $G^l(q_{i_m},
  \mathcal{R})$ for a query node $Q$ in the query tree and $q_i \in
  Q$.}
\end{itemize}
These postponed changes to the upper and lower bounds must be
incorporated into each individual query $q_{i_m}$ belonging to the
sub-tree under $Q$.

Our series-expansion based algorithm uses four different approximation
methods, i.e. $A \in \{E, T(c, p), F(c, p), D(c, p) \}$. $E$ again
denotes the exhaustive computation of $\sum\limits_{r_{j_n} \in R}
K_h(||q_i - r_{j_n}||)$. $T(c, p)$ denotes the translation of the
order $p - 1$ far-field moments of $R$ to the local moments in the
query node $Q$ that owns $q_i$ about a representative centroid $c$
inside $Q$. $F(c, p)$ denotes the evaluation of the order $p - 1$
far-field expansion formed by the moments of $R$ expanded about a
representative point $c$ inside $R$. $D(c, p)$ denotes the $p - 1$th
order direct accumulation of the local moments due to $R$ about a
representative centroid $c$ inside $Q$ that owns $q_i$. We discuss
these approximation methods in
Section~\ref{sec:series_expansion_for_gaussian_kernel_sums}.

\subsection{Series Expansion for the Gaussian Kernel Sums}
\label{sec:series_expansion_for_gaussian_kernel_sums}
We would like to point out to our readers that we present the series
expansion in a way that sheds light to a working
implementation.~\cite{greengard1991fgt} chose a theorem-proof format
for explaining the essential operations. We present the series
expansion methods from the more informed computer science perspective
of divide-and-conquer and data structures, where the discrete aspects
of the methods are concerned.

One can derive the series expansion for the Gaussian kernel sums
(defined in Equation~\eqref{eq:gaussian_kernel_sums}) using
Equation~\eqref{eq:primal} and Equation~\eqref{eq:dual}. The basic
idea is to express the kernel sum contribution of a reference node as
a Taylor series of infinite terms and truncate it after some number of
terms, given that the truncation error meets the desired absolute
error tolerance.

The followings are two main types of Taylor series representations for
infinitely differentiable kernel functions
$K_h\left(\cdot\right)$'s. The key difference between two
representations is the location of the expansion center which is
either in a reference region or a query region. The center of the
expansion for both types of expansions is conveniently chosen to be
the geometric center of the region. For the node region $N$ bounded by
$[N.b[1].l, N.b[1].u] \times \cdots \times [N.b[D].l, N.b[D].u]$, the
center is $N.c = \left[ \frac{N.b[1].l + N.b[1].u}{2}, \cdots,
  \frac{N.b[D].l + N.b[D].u}{2} \right]^T$.
\begin{enumerate}
\item{{\bf Far-field expansion}: A {\it far-field expansion} (derived
  from Equation~\eqref{eq:primal}) expresses the kernel sum
  contribution from the reference points in the reference node $R$ for
  an arbitrary query point. It is expanded about $R.c$, a
  representative point of $R$. Equation~\eqref{eq:primal} is an
  infinite series, and thus we impose a truncation order $p$ in each
  dimension. Substituting $q_i$ for $t$, $r_j$ for $s$ and $R.c$ for
  $s_0$ into Equation~\eqref{eq:primal} yields: {\small
\begin{align*}
& G(q_i, R) = \sum\limits_{r_{j_n} \in R} e^{\frac{-||q_i -
r_{j_n}||^2}{2h^2}}\\ =& \sum\limits_{r_{j_n} \in R}
\prod\limits_{d=1}^D \left( \sum\limits_{\alpha[d] = 0}^{\infty}
\frac{1}{\alpha[d]!}  \left( \frac{r_{j_n}[d] - R.c[d]}{\sqrt{2h^2}}
\right)^{\alpha[d]} h_{\alpha[d]} \left( \frac{q_i[d] -
R.c[d]}{\sqrt{2h^2}} \right) \right)\\ =& \sum\limits_{r_{j_n} \in R}
\prod\limits_{d=1}^D \Bigg( \sum\limits_{\alpha[d] < p}
\frac{1}{\alpha[d]!}  \left( \frac{r_{j_n}[d] - R.c[d]}{\sqrt{2h^2}}
\right)^{\alpha[d]} h_{\alpha[d]} \left( \frac{q_i[d] -
R.c[d)]}{\sqrt{2h^2}} \right) +\\ & \ \ \ \ \ \ \ \ \ \ \ \ \
\sum\limits_{\alpha[d] \geq p} \frac{1}{\alpha[d]!}  \left(
\frac{r_{j_n}[d] - R.c[d]}{\sqrt{2h^2}} \right)^{\alpha[d]}
h_{\alpha[d]} \left( \frac{q_i[d] - R.c[d]}{\sqrt{2h^2}} \right)
\Bigg)
\end{align*}
}
Truncating after $p$ terms along each dimension yields:
{\small
\begin{align*}
& G(q_i, R) \approx \widetilde{G}(q_i, \{ (R, F(R.c, p)) \})\\ =
 &\sum\limits_{r_{j_n} \in R} \prod\limits_{d=1}^D \left(
 \sum\limits_{\alpha[d] < p} \frac{1}{\alpha[d]!}  \left(
 \frac{r_{j_n}[d] - R.c[d]}{\sqrt{2h^2}} \right)^{\alpha[d]}
 h_{\alpha[d]} \left( \frac{q_i[d] - R.c[d]}{\sqrt{2h^2}} \right)
 \right)\\ =& \sum\limits_{r_{j_n} \in R} \sum\limits_{\alpha < p}
 \frac{1}{\alpha!}  \left( \frac{r_{j_n} - R.c}{\sqrt{2h^2}}
 \right)^{\alpha} h_{\alpha} \left( \frac{q_i - R.c}{\sqrt{2h^2}}
 \right)\\ =& \sum\limits_{\alpha < p} \left[ \sum\limits_{r_{j_n} \in
 R} \frac{1}{\alpha!}  \left( \frac{r_{j_n} - R.c}{\sqrt{2h^2}}
 \right)^{\alpha} \right] h_{\alpha}\left( \frac{q_i -
 R.c}{\sqrt{2h^2}} \right)\\ =& \sum\limits_{\alpha < p} M_{\alpha}(R,
 R.c) h_{\alpha} \left( \frac{q_i - R.c}{\sqrt{2h^2}} \right)
\end{align*}
}
where we denote
{\small
\begin{equation}
M_{\alpha}(R, c) = \sum\limits_{r_{j_n} \in R} \frac{1}{\alpha!}
\left( \frac{r_{j_n} - c}{\sqrt{2h^2}} \right)^{\alpha}
\label{eq:far_field_moments}
\end{equation}
}which is a function of a reference node $R$ and an expansion center
$c$. We denote $\widetilde{G}(q_i, \{ R, F(c, p) \})$ as the {\bf
  far-field expansion of order $p - 1$ for the kernel sum contribution
  of $R$ expanded about $c$}. Ideally, we would like to choose the
smallest $p$ such that the truncation after the chosen order $p$
incurs tolerable error; this will be discussed in
Section~\ref{sec:error_bounds}.
\begin{figure}
\scalebox{0.62}{\includegraphics{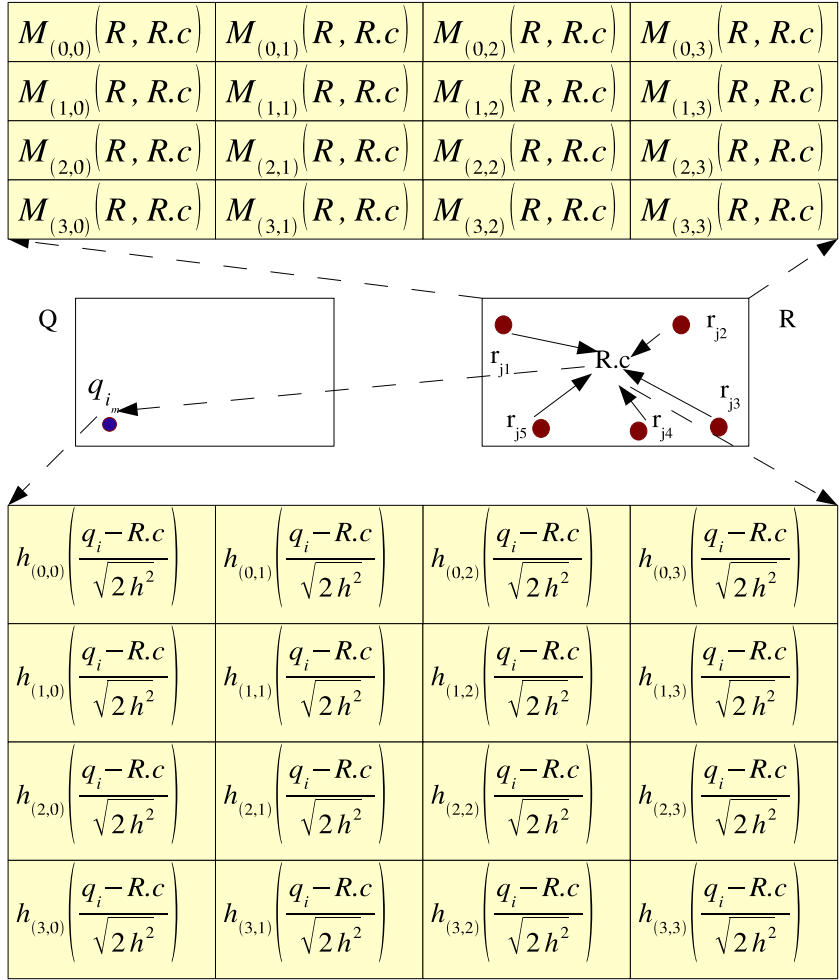}}
\caption{Given the query node $Q$ containing the query points
  $\{q_{i_m}\}_{m=1}^{|Q|}$ and the reference node $R$ containing the
  reference points $\{r_{j_n}\}_{n=1}^{|R|}$, evaluating the far-field
  expansion generated by the reference points at the given query point
  $q_{i_m}$ up to four terms in each dimension, $G(q_{i_m}, R) \approx
  \widetilde{G}(q_{i_m}, \{( R, F(R.c, 4) ) \} ) = \sum\limits_{\alpha
    < 4} \left[ \sum\limits_{r_{j_n} \in R} \frac{1}{\alpha!}  \left(
    \frac{r_{j_n} - R.c}{\sqrt{2h^2}} \right)^{\alpha} \right]
  h_{\alpha}\left( \frac{q_{i_m} - R.c}{\sqrt{2h^2}} \right)$,
  involves computing the sum of the element-wise product between the
  two-dimensional array of far-field coefficients with the
  query-dependent two-dimensional array.}
\label{fig:far_field_coeffs}
\end{figure}
Note that the far-field expansion for the Gaussian kernel separates
the interaction between a reference point and a query point (namely
$e^{-||q_i - r_{j_n}||^2 / (2h^2)}$) into a summation of two product
terms. For each multi-index $\alpha$, $M_{\alpha}(R, R.c)$, which
depends only on the {\it intrinsic} information for the reference node
(the reference points $r_{j_n} \in R$ and the reference centroid $R.c$
which is constant with respect to $R$), is called the {\it far-field
  moments/coefficients} of the reference region $R$. Because
$M_{\alpha}(R, R.c)$ part of the far-field expansion of the Gaussian
kernel sums is the same regardless of the query point $q_i$ used for
evaluation, they can be computed {\bf only once} and stored within $R$
for efficiently approximating the contribution of $R$ for different
query points (see Figure~\ref{fig:far_field_coeffs}). Precomputing the
far-field moments for a reference node $R$ up to $p^D$ terms
(i.e. computing $\sum\limits_{r_{j_n} \in R} \frac{1}{\alpha!}  \left(
\frac{r_{j_n} - R.c}{\sqrt{2h^2}} \right)^{\alpha}$ for each $\alpha <
p$) requires $O(|R| p^D)$ operations.

The far-field expansion of order $p - 1$ for the Gaussian kernel sums
is valid for any query locations $q_i$ given that the reference node
meets the certain size constraint (see
Section~\ref{sec:error_bounds}). However, for a fixed order $p$,
evaluating on query points that are far away from the reference
centroid in general incur smaller amount of error.
\begin{figure}[!t]
\scalebox{0.45}{\includegraphics{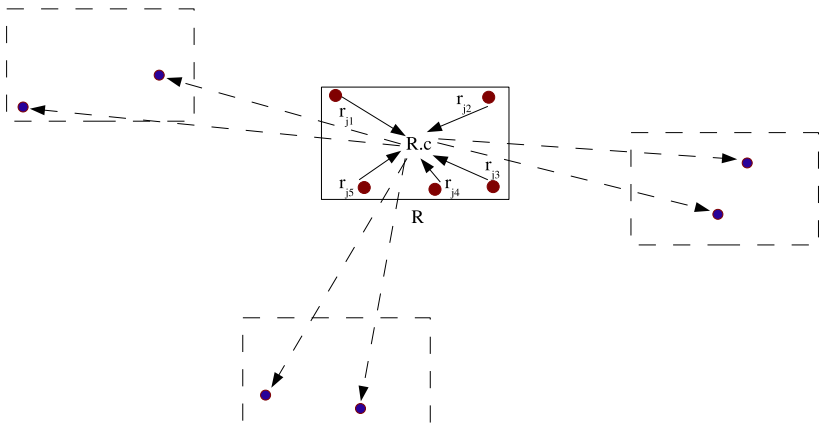}}
\caption{The Gaussian kernel sum series expansion represented by the
  far-field coefficients in $R$, $\sum\limits_{\alpha < p}
  M_{\alpha}(R, R.c) h_{\alpha} \left( \frac{r_{j_n} -
    R.c}{\sqrt{2h^2}} \right)$, is valid regardless of the location of
  the given query point, given the size constraint on the reference
  node (see Section~\ref{sec:error_bounds}). However, each query point
  location will incur different amount of error.}
\label{fig:far_field_validity}
\end{figure}
}
\item{{\bf Local expansion}: A {\it local expansion} (derived from
  Equation~\eqref{eq:dual}) is a Taylor expansion of the kernel sums
  about a representative point $Q.c$ in a query region $Q$. After
  substituting $q_i$ for $t$, $Q.c$ for $t_0$ and $r_{j_n}$ for $s$,
  the kernel sum contribution of all reference points in a reference
  region $R$ to a query point $q_i \in Q$ is given by: {\small
\begin{align*}
& G(q_i, R) = \sum\limits_{r_{j_n} \in R} e^{\frac{-||q_i -
r_{j_n}||^2}{2h^2}}\\ =& \sum\limits_{r_{j_n} \in R}
\prod\limits_{d=1}^D \left( \sum\limits_{n_d = 0}^{\infty}
\frac{(-1)^{n_d}}{n_d!} h_{n_d} \left( \frac{Q.c[d] -
r_{j_n}[d]}{\sqrt{2h^2}} \right) \left( \frac{q_i[d] -
Q.c[d]}{\sqrt{2h^2}}\right)^{\beta} \right)\\ =& \sum\limits_{r_{j_n}
\in R} \prod\limits_{d=1}^D \Biggl ( \sum\limits_{n_d < p }
\frac{(-1)^{n_d}}{n_d!} h_{n_d} \left( \frac{Q.c[d] -
r_{j_n}[d]}{\sqrt{2h^2}} \right) \left( \frac{q_i[d] -
Q.c[d]}{\sqrt{2h^2}}\right)^{\beta} +\\ & \ \ \ \ \ \ \ \ \ \ \ \ \
\sum\limits_{n_d \geq p} \frac{(-1)^{n_d}}{n_d!} h_{n_d} \left(
\frac{Q.c[d] - r_{j_n}[d]}{\sqrt{2h^2}} \right) \left( \frac{q_i[d] -
Q.c[d]}{\sqrt{2h^2}}\right)^{\beta} \Biggr )
\end{align*}
}
Again, truncating after $p$ terms along each dimension yields:
{\small
\begin{align*}
 & \widetilde{G}(q_i, \{ ( R, D(Q.c, p) ) \} )\\ =&
\sum\limits_{r_{j_n} \in R} \prod\limits_{d=1}^D \left(
\sum\limits_{n_d < p} \frac{(-1)^{n_d}}{n_d!} h_{n_d} \left(
\frac{Q.c[d] - r_{j_n}[d]}{\sqrt{2h^2}} \right) \left( \frac{q_i[d] -
Q.c[d]}{\sqrt{2h^2}}\right)^{\beta} \right) \\ =& \sum\limits_{r_{j_n}
\in R} \sum\limits_{\beta < p} \frac{(-1)^{\beta}}{\beta!}  h_{\beta}
\left( \frac{Q.c - r_{j_n}}{\sqrt{2h^2}} \right) \left( \frac{q_i -
Q.c}{\sqrt{2h^2}}\right)^{\beta}\\ =& \sum\limits_{\beta < p} \left[
\sum\limits_{r_{j_n} \in R}\frac{(-1)^{\beta}}{\beta!}  h_{\beta}
\left( \frac{Q.c - r_{j_n}}{\sqrt{2h^2}} \right) \right] \left(
\frac{q_i - Q.c}{\sqrt{2h^2}}\right)^{\beta}\\ =& \sum\limits_{\beta <
p} L_{\beta}(\{(R, D(Q.c, p)) \}) \left( \frac{q_i - Q.c}{\sqrt{2h^2}}
\right)^{\beta}
\end{align*}
}
where we denote:
{\small
\begin{equation}
L_{\beta}( \{ (R, D(c, p)) \} ) = \begin{cases} \sum\limits_{r_{j_n}
\in R}\frac{(-1)^{\beta}}{\beta!}  h_{\beta} \left( \frac{c -
r_{j_n}}{\sqrt{2h^2}} \right) &, \mbox{$\beta < p$} \\ 0 &,
\mbox{otherwise}
\end{cases}
\label{eq:local_moments}
\end{equation}
} $\{ L_{\beta}( \{ ( R, D(Q.c, p) ) \}) \}_{\beta}$ are the {\it
  direct local moments} of $R$ for $Q$. The error bound criterion will
be discussed in Section~\ref{sec:error_bounds}. Note that: {\small
\begin{align*}
& \widetilde{G}\left(q_i, \bigcup\limits_a \left \{ (R_a, D(p_a) )
\right \} \right ) = \sum\limits_{a} \widetilde{G}(q_i, \{ (R_a,
D(p_a)) \} )\\ =& \sum\limits_{a} \sum\limits_{\beta < p_a} \left[
\sum\limits_{r_{j_n} \in R_a}\frac{(-1)^{\beta}}{\beta!}  h_{\beta}
\left( \frac{Q.c - r_{j_n}}{\sqrt{2h^2}} \right) \right] \left(
\frac{q_i - Q.c}{\sqrt{2h^2}}\right)^{\beta}\\ =& \sum\limits_{\beta <
\max\limits_{a} p_a} \left[ \sum\limits_{a} L_{\beta}(\{ (R_a, D(Q.c,
p_a)) \} ) \right ] \left( \frac{q_i - Q.c}{\sqrt{2h^2}}
\right)^{\beta}\\ =& \sum\limits_{\beta < \max\limits_{a} p_a}
L_{\beta}\left( \bigcup\limits_a \left \{ ( R_a, D(Q.c, p_a) ) \right
\} \right) \left( \frac{q_i - Q.c}{\sqrt{2h^2}}\right)^{\beta}
\end{align*}
}In other words, the local moments for a fixed query node $Q$ are
additive (see Figure~\ref{fig:local_accumulation}) across a set of
disjoint portions of the reference dataset $\mathcal{R}$ since its
basis functions $\left \{ \left( \frac{q_i - Q.c}{\sqrt{2h^2}}
\right)^{\beta} \right \}_{\beta}$ remain the same for all reference
points regardless of their locations. For a given reference node $R$,
accumulating the local moments of $R$ up to $p^D$ terms (that is,
evaluating for each $\beta < p$) requires $O(|R| p^D)$
operations. These local coefficients are accumulated and stored within
the given query node. The local expansion represented by the local
coefficients is valid for all query points within the query node under
certain constraints.
\begin{figure}
\scalebox{0.59}{\includegraphics{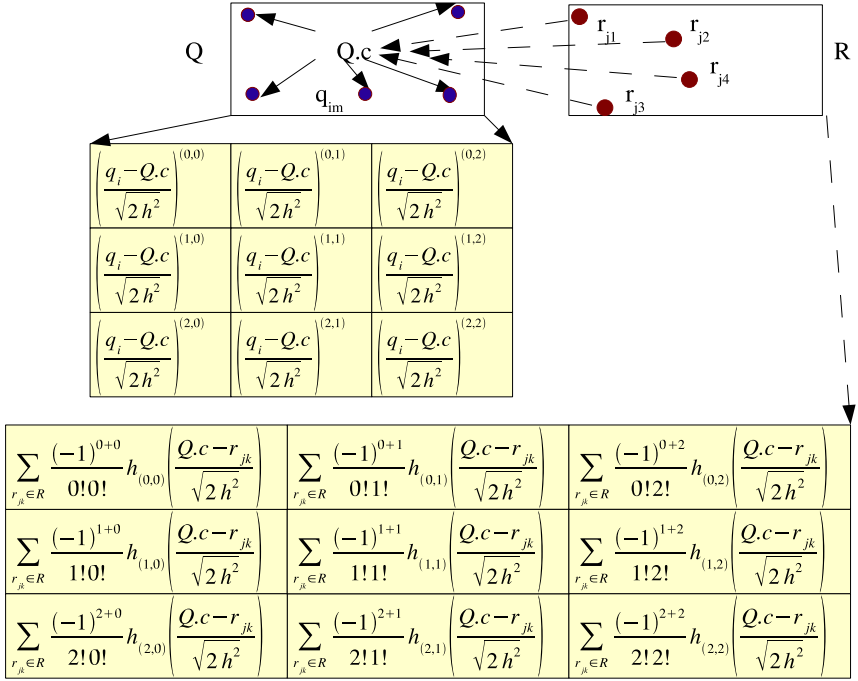}}
\caption{Given the query node $Q$ containing the query points
$\{q_{i_m}\}_{m=1}^{|Q|}$ and the reference node $R$ containing the
reference points $\{r_{j_n}\}_{n=1}^{|R|}$, evaluating the local
expansion generated by the reference points at the given query point
$q_{i_m}$ up to third terms in each dimension, $G(q_{i_m}, R) \approx
\widetilde{G}(q_{i_m}, \{ (R, D(Q.c, 3)) \}) = \sum\limits_{\beta < 3}
\left[ \sum\limits_{r_{j_n} \in R} \frac{(-1)^{\beta}}{\beta!}
h_{\beta}\left( \frac{Q.c - r_{j_n}}{\sqrt{2h^2}} \right) \right]
\left( \frac{q_{i_m} - Q.c}{\sqrt{2h^2}} \right)^{\beta}$, involves
taking the dot-product between the two-dimensional array of local
coefficients with the query-dependent two-dimensional array.}
\label{fig:local_coeffs}
\end{figure}

\begin{figure}[!t]
\scalebox{0.46}{\includegraphics{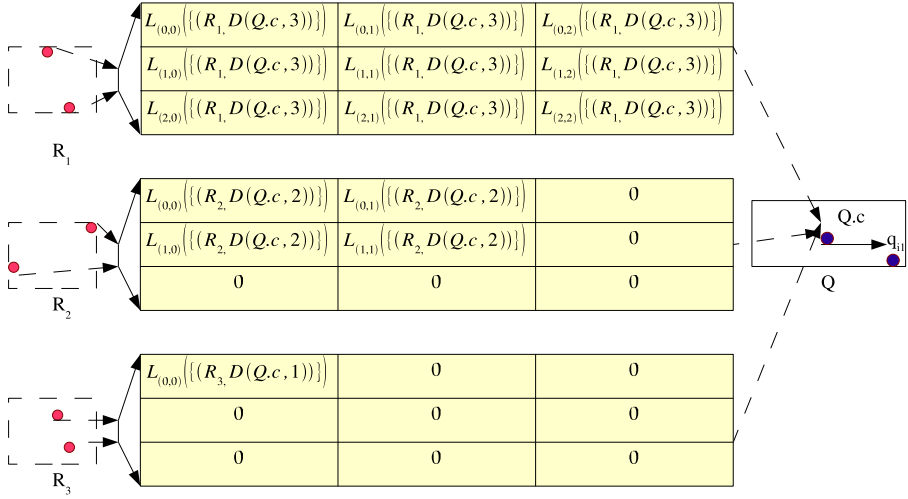}}
\caption{Accumulating direct local moments from three reference nodes
  with the nodes $R_1$, $R_2$, and $R_3$ contributing nine terms, four
  terms, and one term respectively to form the local moments
  containing the contribution from $R_1$, $R_2$, and $R_3$: $L(\{
  (R_1, D(Q.c, 3)), (R_2, D(Q.c, 2)), (R_3, D(Q.c, 1)) \})$. Zeros
  denote the positions that are not explicitly computed using
  Equation~\eqref{eq:local_moments}. $L(\{ (R_1, D(Q.c, 3)), (R_2,
  D(Q.c, 2)), (R_3, D(Q.c, 1)) \} ) = L(\{ (R_1, D(Q.c, 3)) \}) + L(\{
  (R_2, D(Q.c, 2)) \}) + L(\{ (R_3, D(Q.c, 1)) \})$ is added to the
  total local moments for $Q$.}
\label{fig:local_accumulation}
\end{figure}

\begin{figure}[t]
\scalebox{0.5}{\includegraphics{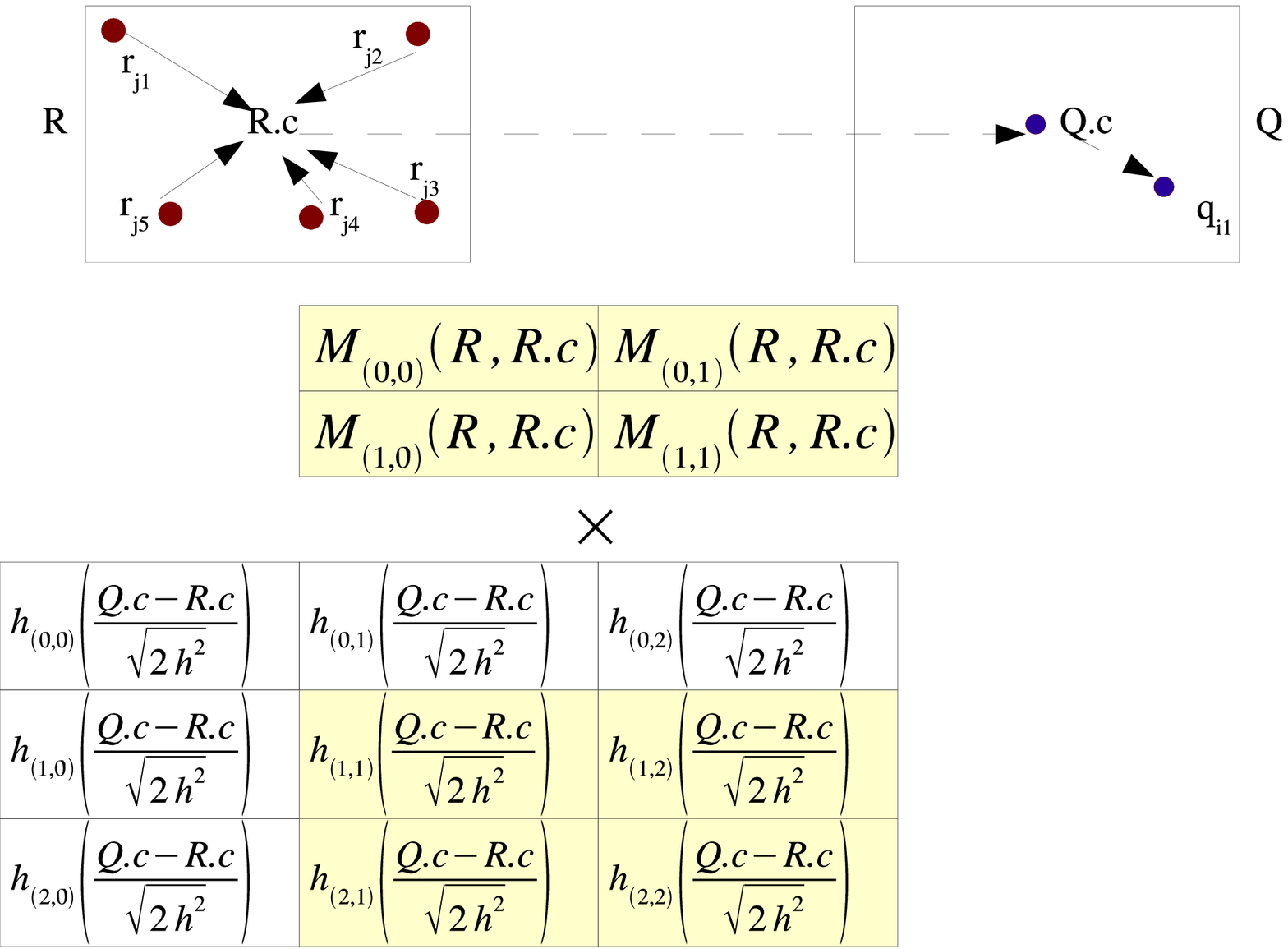}}
\caption{Two-dimensional far-field coefficients truncated after the
first two terms in each dimension can be converted into a set of local
moments using Equation~\eqref{eq:farfield_to_local_formula}. Computing
$L_{\beta}( \{(R, T(Q.c, 2))\})$ involves summing up the element-wise
product between the matrix (or tensor in higher dimensions) consisting
of the far-field moments and the two-by-two window over the Hermite
functions whose upper left multi-index is $\beta$. This figure shows
how to compute $L_{(1,1)}( \{(R, T(Q.c, 2))\})$.}
\label{fig:far_to_local}
\end{figure}
}
\end{enumerate}

\subsection{Gaussian Sum Approximation Using Series Expansion}
Now again assume we are given a query node $Q$ and a reference node
$R$. Here we describe three main methods that use the two expansion
types for approximating Gaussian summation, $\widetilde{G}(q, R)$, for
each $q \in Q$.
\begin{enumerate}
\item{{\bf Evaluating a far-field expansion of $R$: }Given the
pre-computed far-field moments $M_{\alpha}(R)$ up to $p^D$ terms, one
could evaluate the far-field expansion for a given query point $q$
(that is, approximate $\widetilde{G}(q, R)$) by forming a dot-product
between the query-dependent vector and the far-field moments, as shown
in Figure~\ref{fig:far_field_coeffs} and
Figure~\ref{fig:far_field_validity}. Approximating $\widetilde{G}(q,
R)$ for all $q \in Q$ requires $O( |Q| p^D )$ operations since
evaluating the far-field expansion each time requires $O(p^D)$
operations.}
\item{{\bf Computing and evaluating a local expansion inside $Q$ due
to the contribution of $R$: }one could iterate over each reference
point $r_{j_n} \in R$ and compute the local moments $L_{\beta}(\{(R,
D(Q.c, p)) \})$ due to $R$ up to $p^D$ terms, as shown in
Figure~\ref{fig:local_coeffs} and
Figure~\ref{fig:local_accumulation}. The {\it local accumulation} of
the contribution of the reference node $R$ requires $O(|R| p^D)$
operations, and evaluating the {\it local expansion} for each $q_{i_m}
\in Q$ requires a total of $O(|Q| p^D)$ operations.}
\item{{\bf Converting far-field moments of $R$ to a local expansion of
    $Q$: } Suppose $R$ has pre-computed far-field moments up to $p^D$
  terms. From the far-field moments, we can approximate the {\it local
    moments} of $R$ but with some amount of error. This can be seen as
  a generalization of centroid-based approximation. ~\cite{greengard1991fgt}
  describes this method as one of the {\it translation operators},
  called {\it far-field to local translation operator}, stated below:
\begin{lem}
\label{H2LTranslationOperator}
{\bf Far-field to local (F2L) translation operator for Gaussian
kernel} (as presented in Lemma 2.2 in \cite{greengard1991fgt}){\bf:} Given a
reference node $R$, a query node $Q$, and the truncated far-field
expansion centered at a centroid $R.c$ of $R$ up to $p^D$ terms:\\
$\widetilde{G}(q_{i_m}, \{ (R, F(R.c, p)) \}) = \sum\limits_{\alpha <
p} M_{\alpha}(R, R.c) h_{\alpha} \left( \frac{q_{i_m} -
R.c}{\sqrt{2h^2}} \right)$,\\ the Taylor expansion of the far-field
expansion at the centroid $Q.c$ in $Q$ is given by
$\widetilde{G}(q_{i_m}, \{ (R, F(R.c, p)) \}) = \sum\limits_{\beta
\geq 0} L_{\beta}(\{ (R, T(Q.c, p)) \}) \left( \frac{q_{i_m} - Q.c}
{\sqrt{2h^2}} \right) ^{\beta}$ where for $q_{i_m} \in Q$,
{\small
\begin{equation}
L_{\beta}(\{ (R, T(Q.c, p)) \}) = \frac{(-1)^{|\beta|}}{\beta!}
\sum\limits_{\alpha < p} M_{\alpha}(R, R.c) h_{\alpha + \beta} \left(
\frac{Q.c - R.c}{\sqrt{2h^2}} \right)
\label{eq:farfield_to_local_formula}
\end{equation}
}
\begin{proof}
The proof consists of replacing the Hermite function portion of the
expansion with its Taylor series:
{\small
\begin{align*}
& \widetilde{G}(q_{i_m}, \{ (R, F(R.c, p)) \}) =
\sum\limits_{\alpha < p} M_{\alpha}(R, R.c) h_{\alpha} \left(
\frac{q_{i_m} - R.c}{\sqrt{2h^2}} \right)\\ =& \sum\limits_{\alpha <
p} M_{\alpha}(R, R.c) \sum\limits_{\beta \geq 0} \frac{(-1)^{| \beta
|}}{\beta !}  h_{\alpha + \beta} \left( \frac{Q.c - R.c}{\sqrt{2h^2}}
\right) \left( \frac{q_{i_m} - Q.c}{\sqrt{2h^2}} \right)^{\beta}
\end{align*}
\begin{align*}
 =&
\sum\limits_{\beta \geq 0} \left [ \frac{(-1)^{| \beta| }}{\beta!}
\sum\limits_{\alpha < p} M_{\alpha}(R, R.c) h_{\alpha + \beta} \left(
\frac{Q.c - R.c}{\sqrt{2h^2}} \right) \right ] \left( \frac{q_{i_m} -
Q.c}{\sqrt{2h^2}} \right)^{\beta}
\end{align*}
}
\end{proof}
\end{lem}
However, note $\widetilde{G}(q_{i_m}, \{ (R, F(R.c, p)) \})$ has an
infinite number of terms, and must be truncated after $p^D$ terms. In
other words, the local moments accumulated for $Q$ are the
coefficients for $\widetilde{G}(q_{i_m}, \{ (R, T(Q.c, p)) \} ) =
\sum\limits_{\beta < p} L_{\beta}(\{ (R, T(Q.c, p)) \}) \left(
\frac{q_{i_m} - Q.c} {\sqrt{2h^2}} \right) ^{\beta}$, as shown in
Figure~\ref{fig:far_to_local}. To compute $\{ L_{\beta}(\{ (R, T(Q.c,
p)) \}) \}_{\beta < p}$, we need to iterate over all of $p^D$
far-field moments $\{ M_{\alpha}(R, R.c) \}_{\alpha < p}$ for each
$L_{\beta}(\{(R, T(Q.c, p)) \})$. This operation runs in $O(D p^{2D})$
operations.}
\end{enumerate}
In general, these approximations are valid only under certain
conditions which depend on how the error bounds associated with these
approximation methods are derived. Moreover, we have not discussed how
to choose the method of approximation given a query and reference node
pair, and how to determine the order of approximation, i.e. the number
of terms required to achieve a given level of error. We discuss the
details in Section~\ref{sec:error_bounds}.

\subsection{Truncation Error Bounds}
\label{sec:error_bounds}
Because the far-field and the local expansions are truncated after
taking $p^D$ terms, we incur an error in approximation. The original
error bounds for the Gaussian kernel in~\cite{greengard1991fgt} were
wrong and corrections were shown in~\cite{baxter2002new}. Here we
present the error bounds for (1) evaluating a truncated far-field
expansion of a reference node for any query point $q \in \mathbb{R}^D$
(2) evaluating a truncated local expansion of $Q$ due to the
contribution of a reference node $R$ for any query point $q_{i_m} \in
Q$ (3) evaluating a truncated local expansion formed from converting a
truncated far-field expansion of a reference node $R$ for any query
point $q_{i_m} \in Q$. Note that these error bounds place restrictions
on the size of the nodes in consideration: reference node, query node,
or both. First we start with the truncation error bound for evaluating
the far-field expansion formed for a given reference node.
\begin{lem}
\label{lem:truncating_far_field}
{\bf Error bound for evaluating a truncated far-field expansion} (as
presented in \cite{baxter2002new}){\bf:} Suppose we are given a far-field
expansion of a reference node $R$ about its centroid $R.c$:\\
$\widetilde{G}(q, \{(R, F(R.c, p)) \}) = \sum\limits_{\alpha < p}
M_{\alpha}(R, R.c) h_{\alpha} \left( \frac{q - R.c}{\sqrt{2h^2}}
\right)$ where\\$M_{\alpha}(R, R.c) = \sum\limits_{r_{j_n} \in
R}\frac{1}{\alpha!}  \left( \frac{r_{j_n} - R.c}{\sqrt{2h^2}}
\right)^{\alpha}$.
If $\forall r_{j_n} \in R$ satisfies $||r_{j_n} - R.c||_{\infty} < rh$
for $r < 1$, then for any $q \in \mathbb{R}^D$,
{\small \begin{equation}
\left| \widetilde{G}(q, \{(R, F(R.c, p)) \}) - G(q, R) \right| \leq
\frac{|R|}{(1 - r)^D} \sum\limits_{k = 0}^ {D - 1} \binom{D}{k} (1 -
r^p)^k \left( \frac{r^p}{\sqrt{p!}}  \right)^ {D - k}
\label{eq:error_bound_far_field}
\end{equation}
}
\begin{proof}
We expand the far-field expansion as a product of one-dimensional
Hermite functions, and utilize a bound on one-dimensional Hermite
functions due to \cite{szasz51}: $\frac{1}{n!}|h_n(x)| \leq
\frac{2^{\frac{n}{2}}}{\sqrt{n!}}  e^{\frac{-x^2}{2}}, n \geq 0, x \in
\mathbb{R}^1$.
{\small
\begin{align*}
u_{p_d}(q[d], r_{j_n}[d], R.c[d]) &= \sum\limits_{n_i = 0}^{p - 1}
\frac{1}{n_i!} \left( \frac{r_{j_n}[d] - R.c[d]}{\sqrt{2h^2}}
\right)^{n_i} h_{n_i} \left( \frac{q[d] - R.c[d]}{\sqrt{2h^2}}
\right)\\ v_{p_d}(q[d], r_{j_n}[d], R.c[d]) &= \sum\limits_{n_i =
p}^{\infty} \frac{1}{n_i!} \left( \frac{r_{j_n}[d] -
R.c[d]}{\sqrt{2h^2}} \right)^{n_i} h_{n_i} \left( \frac{q[d] -
R.c[d]}{\sqrt{2h^2}} \right)\\ e^{\frac{-||q - r_{j_n}||^2}{{2h^2}}}
&= \prod\limits_{d = 1}^D \left( u_{p_d}(q[d], r_{j_n}[d], R.c[d]) +
v_{p_d}(q[d], r_{j_n}[d], R.c[d]) \right)
\end{align*} }
We obtain for $1 \leq d \leq D$:
{\small
\begin{align*}
& u_{p_d}(q[d], r_{j_n}[d], R.c[d]) \leq \sum\limits_{n_i = 0}^{p - 1}
\frac{1}{n_i!} \left| \frac{r_{j_n}[d] - R.c[d]}{\sqrt{2h^2}}
\right|^{n_i} \left| h_{n_i} \left( \frac{q[d] - R.c[d]}{\sqrt{2h^2}}
\right) \right|\\ \leq & \sum\limits_{n_i = 0}^{p - 1} \left|
\frac{rh}{\sqrt{2h^2}} \right|^{n_i}
\frac{2^{\frac{n_i}{2}}}{\sqrt{n_i!}} \left(e^{-\frac{(q[d] -
R.c[d])^2}{4h^2}} \right)  \leq  \sum\limits_{n_i = 0}^{p - 1}
r^{n_i} \leq \frac{1 - r^p}{1 - r}\\ & v_{p_d}(q[d], r_{j_n}[d], R.c[d])
 \leq  \sum\limits_{n_i = p}^{\infty} \frac{1}{n_i!} \left|
\frac{r_{j_n}[d] - R.c[d]}{\sqrt{2h^2}} \right|^{n_i} \left| h_{n_i}
\left( \frac{q[d] - R.c[d]}{\sqrt{2h^2}} \right) \right|\\  \leq &
\sum\limits_{n_i = p}^{\infty} \left| \frac{rh}{\sqrt{2h^2}}
\right|^{n_i} \frac{2^{\frac{n_i}{2}}}{\sqrt{n_i!}}
\left(e^{-\frac{(q[d] - R.c[d])^2}{4h^2}} \right)  \leq 
\frac{1}{\sqrt{p!}}\sum\limits_{n_i = p}^{\infty} r^{n_i} \leq
\frac{1}{\sqrt{p!}}\frac{r^p}{1 - r}
\end{align*}
}
Therefore,
{\small
\begin{align*}
& \left|\prod\limits_{d = 1}^D u_{p_d}(q[d], r_{j_n}[d], R.c[d]) -
e^{\frac{-||q - r_{j_n}||^2}{2h^2}} \right| \\ \leq & (1-r)^{-D}
\sum\limits_{k = 0}^{D - 1}\binom{D}{k} (1 - r^p)^k \left(
\frac{r^p}{\sqrt{p!}}  \right)^{D - k}\\ & \left|\sum\limits_{\alpha <
p} M_{\alpha}(R, R.c) h_{\alpha} \left( \frac{q - R.c}{\sqrt{2h^2}}
\right) - \sum\limits_{r_{j_n} \in R} e^{\frac{-||q -
r_{j_n}||^2}{2h^2}} \right| \\ \leq & \frac{|R|}{(1-r)^{D}} \sum\limits_{k
= 0}^{D - 1}\binom{D}{k} (1 - r^p)^k \left( \frac{r^p}{\sqrt{p!}}
\right)^{D - k}
\end{align*}
}
\end{proof}
\end{lem}
Intuitively, this theorem implies that evaluating a truncated
far-field expansion for a query point (regardless of its location)
requires that the reference points used to form the expansion are
within the bandwidth $h$ in each dimension from the centroid $R.c$
(i.e. the reference node has a maximum side length of $2h$). 

The following gives the truncation bound for the local expansion
formed inside a query node whose bound is within a hypercube of some
side length.
\begin{lem}
\label{lem:truncating_local}
{\bf Error bound for evaluating a truncated local expansion:} Suppose
we are given the local expansion about the centroid $Q.c$ of the given
query node $Q$ accounting for the kernel sum contribution of the given
reference node $R$: $\widetilde{G}(q_{i_m}, \{ (R, D(Q.c, p)) \}) =
\sum\limits_{\beta < p} L_{\beta}(\{(R, D(Q.c, p)) \}) \left(
\frac{q_{i_m} - Q.c}{\sqrt{2h^2}} \right) ^{\beta}$ where $q_{i_m} \in
Q$ and $L_{\beta}(Q, \{(R, D(p)) \}) = \sum\limits_{r_{j_n} \in R}
\frac{(-1)^{|\beta|}}{\beta!}  h_{\beta} \left( \frac{Q.c -
r_{j_n}}{\sqrt{2h^2}} \right)$

If $\forall q_{i_m} \in Q$ satisfies $||q_{i_m} - Q.c||_{\infty} < rh$
for $r < 1$, then for any $q_{i_m} \in Q$:
{\small
\begin{equation}
\left| \widetilde{G}(q_{i_m}, \{(R, D(Q.c, p)) \}) - G(q_{i_m}, R)
\right| \leq \frac{|R|}{(1 - r)^D} \sum\limits_{k = 0}^ {D - 1}
\binom{D}{k} (1 - r^p)^k \left( \frac{r^p}{\sqrt{p!}}  \right)^{D - k}
\label{eq:error_bound_local}
\end{equation}
}
\begin{proof}
Taylor expansion of the Hermite function yields:
{\small
\begin{align*}
& e^{\frac{-||q_{i_m} - r_{j_n}||^2}{2h^2}} = \sum\limits_{\beta
\geq 0} \frac{(-1)^{|\beta|}}{\beta!} \sum\limits_{\alpha \geq 0}
\frac{1}{\alpha!} \left( \frac{r_{j_n} - R.c}{\sqrt{2h^2}}
\right)^{\alpha} h_{\alpha + \beta} \left( \frac{Q.c -
R.c}{\sqrt{2h^2}} \right)\left( \frac{q_{i_m} - Q.c}{\sqrt{2h^2}}
\right) ^{\beta}\\ =& \sum\limits_{\beta \geq 0}
\frac{(-1)^{|\beta|}}{\beta!}  \sum\limits_{\alpha \geq 0}
\frac{1}{\alpha!} \left( \frac{R.c - r_{j_n}}{\sqrt{2h^2}}
\right)^{\alpha}(-1)^{|\alpha|} h_{\alpha + \beta} \left( \frac{Q.c -
R.c}{\sqrt{2h^2}} \right)\left( \frac{q_{i_m} - Q.c}{\sqrt{2h^2}}
\right) ^{\beta}\\ =& \sum\limits_{\beta \geq 0}
\frac{(-1)^{|\beta|}}{\beta!}  h_{\beta} \left( \frac{Q.c -
r_{j_n}}{\sqrt{2h^2}} \right) \left( \frac{q_{i_m} - Q.c}{\sqrt{2h^2}}
\right) ^{\beta}
\end{align*}
}
Use $e^{\frac{-||q_{i_m} - r_{j_n}||^2}{2h^2}} = \prod\limits_{d =
1}^D \left( u_p(q_{i_m}[d], r_{j_n}[d], Q.c[d]) + v_p( q_{i_m}[d],
r_{j_n}[d], Q.c[d] ) \right)$ for $1 \leq d \leq D$, where
{\small
\begin{align*}
u_{p_d}(q_{i_m}[d], r_{j_n}[d], Q.c[d]) &= \sum\limits_{n_d = 0}^{p -
  1} \frac{(-1)^{n_d}}{n_d!} h_{n_d} \left( \frac{Q.c[d] -
  r_{j_n}[d]}{\sqrt{2h^2}} \right) \left( \frac{q_{m_i}[d] -
  Q.c[d]}{\sqrt{2h^2}} \right)^{n_d}\\ v_{p_d}( q_{i_m}[d],
  r_{j_n}[d], Q.c[d] ) &= \sum\limits_{n_i = p}^{\infty}
  \frac{(-1)^{n_d}}{n_d!}  h_{n_d} \left( \frac{Q.c[d] -
  r_{j_n}[d]}{\sqrt{2h^2}} \right) \left( \frac{q_{m_i}[d] -
  Q.c[d]}{\sqrt{2h^2}} \right)^{n_d}
\end{align*}
} These univariate functions respectively satisfy $
u_{p_d}(q_{i_m}[d], r_{j_n}[d], Q.c[d]) \leq \frac{1 - r^p}{1 - r}$
and $ v_{p_d}( q_{i_m}[d], r_{j_n}[d], Q.c[d] ) \leq
\frac{1}{\sqrt{p!}}  \frac{r^p}{1 - r}$, for $1 \leq d\leq D$,
achieving the multivariate bound. The proof is similar as in the one
given in Lemma~\ref{lem:truncating_far_field}.
\end{proof}
\end{lem}
Lastly, we present the error bound for evaluating a truncated local
expansion formed from a truncated far-field expansion, which requires
that both the query node and the reference node are ``small'':
\begin{lem}
\label{lem:far_field_to_local_error}
{\bf Error bound for evaluating a truncated local expansion converted
from an already truncated far-field expansion:} A truncated far-field
expansion centered about the centroid $R.c$ of a reference node $R$,
$$\widehat{G}(q, \{ (R, F(R.c, p)) \}) = \sum\limits_{\alpha < p}
M_{\alpha}(R, R.c) h_{\alpha} \left(\frac{q - R.c}{\sqrt{2h^2}}
\right)$$ has the following local expansion about the centroid $Q.c$
of a query node $Q$ for $q_{i_m} \in Q$: $\widehat{G}(q_{i_m}, \{ (R,
F(R.c, p)) \} ) = \sum\limits_{\beta \geq 0} L_{\beta}(\{(R, T(Q.c,
p)) \}) \left( \frac{q_{i_m} - Q.c}{\sqrt{2h^2}} \right)^{\beta}$
where:
$
L_{\beta}(\{(R, T(Q.c, p)) \}) = \frac{(-1)^{|\beta|}}{\beta!}
\sum\limits_{\alpha < p} M_{\alpha}(R, R.c) h_{\alpha + \beta} \left(
\frac{Q.c - R.c}{\sqrt{2h^2}} \right)
$

Let $\widetilde{G}(q_{i_m}, \{
( R, T(Q.c, p) ) \}) = \sum\limits_{\beta < p} L_{\beta}( \{ (R,
T(Q.c, p)) \}) \left( \frac{q_{i_m} - Q.c}{\sqrt{2h^2}}
\right)^{\beta}$, a\\truncation of the local expansion of
$\widehat{G}(q_{i_m}, \{ (R, F(R.c, p)) \})$ after $p^D$ terms.

If $\forall q_{i_m} \in Q$ satisfies $||q_{i_m} - Q.c||_{\infty} < rh$
and $\forall r_{j_n} \in R$ satisfies $||r_{j_n} - R.c||_{\infty} <
rh$ for $r < \frac{1}{2}$, then for any $q_{i_m} \in Q$:
{\footnotesize
\begin{align}
& \left | \widetilde{G}(q_{i_m}, \{ (R, T(Q.c, p)) \} ) - G(q_{i_m},
R) \right |  \notag \\ \leq & \frac{|R|}{(1-2r)^{2D}} \sum\limits_{k = 0}^{D -
1}\binom{D}{k} ((1 - (2r)^p)^2)^k \left( \frac{((2r)^p)(2 -
(2r)^p)}{\sqrt{p!}} \right)^{D - k}
\label{eq:error_bound_far_to_local}
\end{align}
}
\begin{proof}
We define for $1 \leq d \leq D$:
{\small
\begin{align*}
u_{p_d} &= u_p(q_{i_m}[d], r_{j_n}[d], Q.c[d], R.c[d])\\
v_{p_d} &= v_p(q_{i_m}[d], r_{j_n}[d], Q.c[d], R.c[d])\\
w_{p_d} &= w_p(q_{i_m}[d], r_{j_n}[d], Q.c[d], R.c[d])
\end{align*}}
{\scriptsize
\begin{align*}
u_{p_d} =& \sum\limits_{n_i = 0}^{p - 1} \frac{(-1)^{n_i}}{n_i!}
\sum\limits_{n_j = 0}^{p - 1} \frac{1}{n_j!} \left( \frac{R.c[d] -
r_{j_n}[d]}{\sqrt{2h^2}} \right)^{n_j} (-1)^{n_j} \\ & h_{n_i + n_j} \left(
\frac{Q.c[d] - R.c[d]}{\sqrt{2h^2}} \right) \left( \frac{q_{i_m}[d] -
Q.c[d]}{\sqrt{2h^2}} \right)^{n_i} \\
 v_{p_d} =& \sum\limits_{n_i =
0}^{p - 1} \frac{(-1)^{n_i}}{n_i!}  \sum\limits_{n_j = p}^{\infty}
\frac{1}{n_j!} \left( \frac{R.c[d] - r_{j_n}[d]}{\sqrt{2h^2}}
\right)^{n_j} (-1)^{n_j} \\ & h_{n_i + n_j} \left( \frac{Q.c[d] -
R.c[d]}{\sqrt{2h^2}} \right) \left( \frac{q_{i_m}[d] -
Q.c[d]}{\sqrt{2h^2}} \right)^{n_i} \\
 w_{p_d} =& \sum\limits_{n_i =
p}^{\infty} \frac{(-1)^{n_i}}{n_i!} \sum\limits_{n_j = 0}^{\infty}
\frac{1}{n_j!}  \left( \frac{R.c[d] - r_{j_n}[d]}{\sqrt{2h^2}}
\right)^{n_j} (-1)^{n_j} \\ & h_{n_i + n_j} \left( \frac{Q.c[d] -
R.c[d]}{\sqrt{2h^2}} \right) \left( \frac{q_{i_m}[d] -
Q.c[d]}{\sqrt{2h^2}} \right)^{n_i}
\end{align*}
} Note that $e^{\frac{-||q_{i_m} - r_{j_n}||^2}{2h^2}} =
\prod\limits_{d = 1}^D \left( u_{p_d} + v_{p_d} + w_{p_d} \right)$ for
$1 \leq d \leq D$.  Using the bound for Hermite functions and the
property of geometric series, we obtain the following upper bounds:
{\small
\begin{align*}
u_{p_d} & \leq \sum\limits_{n_i = 0}^{p - 1} \sum\limits_{n_j = 0}^{p
  - 1} (2r)^{n_i} (2r)^{n_j} = \left( \frac{1 - (2r)^p)}{1 - 2r}
\right)^2\\ v_{p_d} & \leq \frac{1}{\sqrt{p!}} \sum\limits_{n_i =
  0}^{p - 1} \sum\limits_{n_j = p}^{\infty} (2r)^{n_i} (2r)^{n_j} =
\frac{1}{\sqrt{p!}} \left( \frac{1 - (2r)^p}{1 - 2r} \right) \left(
\frac{(2r)^p}{1 - 2r} \right)\\ w_{p_d} &\leq \frac{1}{\sqrt{p!}}
\sum\limits_{n_i = p}^{\infty} \sum\limits_{n_j = 0}^{\infty}
(2r)^{n_i} (2r)^{n_j} = \frac{1}{\sqrt{p!}} \left( \frac{1}{1 - 2r}
\right) \left( \frac{(2r)^p}{1 - 2r} \right)
\end{align*}
}
Therefore,
{\scriptsize
\begin{align*}
& \left| \prod\limits_{d = 1}^D u_{p_d} - e^{\frac{-||q_{i_m} -
r_{j_n}||^2}{2h^2}} \right| \\ \leq & (1-2r)^{-2D} \sum\limits_{k = 0}^{D -
1}\binom{D}{k} ((1 - (2r)^p)^2)^k \left( \frac{((2r)^p)(2 -
(2r)^p)}{\sqrt{p!}} \right)^{D - k} \\
& \left|\widetilde{G}(q_{i_m}, \{( R, T(Q.c, p)) \}) - G(q_{i_m}, R)
\right |  \\ \leq & \frac{|R|}{(1-2r)^{2D}} \sum\limits_{k = 0}^{D -
1}\binom{D}{k} ((1 - (2r)^p)^2)^k \left( \frac{((2r)^p)(2 -
(2r)^p)}{\sqrt{p!}}  \right)^{D - k}
\end{align*}
}
\end{proof}
\end{lem}
~\cite{strain1991fast} proposes an interesting idea of using Stirling's
formula (for any non-negative integer $n$, $\left( \frac{n + 1}{e}
\right)^n \leq n!$) to lift the node size constraint. This could allow
approximation of larger regions that possibly contain more
points. Unfortunately, the error bounds derived
in~\cite{strain1991fast} were also incorrect. We have derived the
necessary corrected error bounds based on the techniques
in~\cite{baxter2002new}. However, we do not include the derivations
here since using these bounds actually degraded performance in our
algorithm.
\begin{algorithm}[t]
\caption{\mbox{\farfieldorder}$(Q, R, \tau)$: Determines the order of
  approximation needed for evaluating a far-field expansion of the
  reference node $R$.}
\begin{algorithmic}

\STATE{$r \leftarrow \max\limits_{1 \leq d \leq D} \frac{R.b[d].u -
R.b[d].l}{2h}$}

\IF{$r \geq 1$}

\RETURN{$\infty$}

\ELSE

\STATE{$p \leftarrow 0$}

\WHILE{$p < p_{\mathit{max}}$}

\STATE{$p \leftarrow p + 1$}

\IF{ $\frac{|R|}{(1 - r)^D} \sum\limits_{k = 0}^ {D - 1} \binom{D}{k}
(1 - r^p)^k \left( \frac{r^p}{\sqrt{p!}}  \right)^ {D - k} \leq \tau$}

\RETURN{$p$}

\ENDIF

\ENDWHILE

\RETURN{$\infty$}

\ENDIF

\end{algorithmic}
\label{alg:determine_far_field_order}
\end{algorithm}

\subsection{Determining the Approximation Order}
Note that Lemma~\ref{lem:truncating_far_field},
Lemma~\ref{lem:truncating_local}, and
Lemma~\ref{lem:far_field_to_local_error} answer the question of the
following form: given that we use $p^D$ terms in the appropriate
expansion type, what is the upper bound on the approximation error,
$\left| \widetilde{G}(q, R) - G(q, R) \right |$? Nevertheless, all
three lemmas can be re-phrased to answer the question in reverse:
given the maximum user-desired absolute error, what is the order of
approximation/number of terms required to achieve it? This question
rises naturally within our dual-tree based algorithm that bounds the
kernel sum approximation error on each part in a partition of the
reference dataset $\mathcal{R}$.

Algorithm~\ref{alg:determine_far_field_order} shows how to determine
the necessary order of the far-field expansion for the given reference
node $R$ such that $\left | \widetilde{G}(q, R) - G(q, R) \right |
\leq \tau$. That is, the approximation error due to the far-field
expansion of $R$ is bounded by the error allocated for approximating
the contribution of the reference node $R$. Using far-field expansion
based approximation requires a ``small'' reference node. Thus, the
algorithm first computes the ratio of the maximum side length of $R$
to twice the bandwidth $h$, and determines the least order required
for achieving the maximum absolute error $\tau$ by evaluating the
right-hand side of Equation~\eqref{eq:error_bound_far_field}
iteratively on different values of $p$.

Algorithm~\ref{alg:determine_local_order} shows how to determine the
necessary order of the local expansion formed by directly accumulating
the contribution of the given reference node $R$ onto the given query
node $Q$. This approximation method requires the query node $Q$ to
have the maximum side length within twice the bandwidth. The algorithm
determines the least order required for achieving the maximum absolute
error $\tau$ by evaluating the right-hand side of
Equation~\eqref{eq:error_bound_local} iteratively on different values
of $p$.
\begin{algorithm}[t]
\caption{\mbox{\localaccumulationorder}$(Q, R, \tau)$: Determining the
  order of approximation needed for forming a local expansion of the
  contribution from the given reference node $R$ for the query node
  $Q$.}
\begin{algorithmic}

\STATE{$r \leftarrow \max\limits_{1 \leq d \leq D} \frac{Q.b[d].u -
Q.b[d].l}{2h}$}

\IF{$r \geq 1$}

\RETURN{$\infty$}

\ELSE

\STATE{$p \leftarrow 0$}

\WHILE{$p < p_{\mathit{max}}$}

\STATE{$p \leftarrow p + 1$}

\IF{ $\frac{|R|}{(1 - r)^D} \sum\limits_{k = 0}^ {D - 1} \binom{D}{k}
  (1 - r^{p})^k \left( \frac{r^{p}}{\sqrt{p!}}  \right)^ {D - k} \leq
  \tau$}

\RETURN{$p$}

\ENDIF

\ENDWHILE

\RETURN{$\infty$}

\ENDIF

\end{algorithmic}
\label{alg:determine_local_order}
\end{algorithm}

Finally, Algorithm~\ref{alg:determine_f2l} determines the necessary
order of local expansion formed by converting a truncated far-field
expansion of the given reference node $R$. In contrast to the two
previous algorithms, this one requires both the query node $Q$ and the
reference node $R$ to have a maximum side length less than the
bandwidth $h$. After the node size requirements are satisfied, the
least order required for achieving the maximum absolute error $\tau$
is obtained by evaluating the right-hand side of
Equation~\eqref{eq:error_bound_far_to_local} iteratively on different
values of $p$.
\begin{algorithm}[t]
\caption{\mbox{\convertfarfieldtolocalorder}$(Q, R, \tau)$:
  Determining the order of approximation needed for evaluating a
  far-field expansion of the given reference node $R$.}
\begin{algorithmic}

\STATE{$r \leftarrow \max\limits_{1 \leq d \leq D} \max \{
\frac{Q.b[d].u - Q.b[d].l}{4h}, \frac{R.b[d].u - R.b[d].l}{4h} \}$}

\IF{$r \geq \frac{1}{2}$}

\RETURN{$\infty$}

\ELSE

\STATE{$p \leftarrow 0$}

\WHILE{$p < p_{\mathit{max}}$}

\STATE{$p \leftarrow p + 1$}

\IF{ $\frac{|R|}{(1-2r)^{2D}} \sum\limits_{k = 0}^{D - 1}\binom{D}{k}
((1 - (2r)^p)^2)^k \left( \frac{((2r)^p)(2 - (2r)^p)}{\sqrt{p!}}
\right)^{D - k} \leq \tau$}

\RETURN{$p$}

\ENDIF

\ENDWHILE

\RETURN{$\infty$}

\ENDIF

\end{algorithmic}
\label{alg:determine_f2l}
\end{algorithm}

\subsection{Deriving the Hierarchical FGT}
Until now, we have discussed the approximation methods developed for a
non-hierarchical version of fast Gauss transform described
in~\cite{greengard1991fgt}. In this section, we derive the two
additional translation operators that extend the original fast Gauss
transform to use a hierarchical data structure.
\begin{figure}[h]
\scalebox{0.35}{\includegraphics{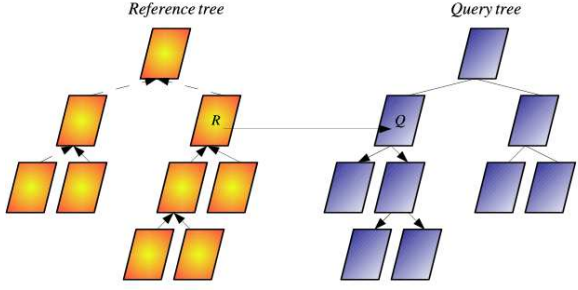}}
\caption{The solid arrows mark the flow of contribution from the
  reference tree to the query tree in case of a prune via a far-field
  to local translation between the reference node $R$ and query node
  $Q$. On the reference side, the far-field moments are formed in the
  bottom-up fashion; on the query side, the accumulated local moments
  will be propagated downwards during a post-processing step via
  local-to-local translations.}
\end{figure}
Here we consider the reference tree, which enables the consideration
of the different portions of the reference set $\mathcal{R}$ at a
different granularity. Given the computed far-field moments of $R^L$
and $R^R$, each centered at $R^L.c$ and $R^R.c$, how can we
efficiently compute the far-field moments of $R$ centered at $R.c$,
the parent of $R^L$ and $R^R$? The first operator allows the efficient
bottom-up pre-computation of the Hermite moments in the reference
tree.
\begin{lem}
\label{H2HTranslationOperator}
{\bf Shifting a far-field expansion of a reference node to a new
center (F2F translation operator for the Gaussian kernel):} Given the
far-field expansion centered at $R.c$ in a reference node $R$:
{\small
\begin{align*}
\widetilde{G}(q, \{ ( R, F(R.c, p)) \}) = \sum\limits_{\alpha < p}
M_{\alpha}(R, R.c) h_{\alpha} \left( \frac{q - R.c}{\sqrt{2h^2}}
\right)
\end{align*}}
this same far-field expansion shifted to a new location $c'$
is given by:
{\small
\begin{align*}
\widetilde{G}(q, \{ (R, F(R.c, p)) \} ) =
\widetilde{G}(q, \{ (R, F(c', p)) \}) = \sum\limits_{\gamma \geq 0}
M_{\gamma}(R, c') h_{\gamma} \left( \frac{q - c'}{\sqrt{2h^2}}
\right)
\end{align*}
}
where
{\small
\begin{equation}
M_{\gamma}(R, c') = \sum\limits_{0 \leq \alpha \leq \gamma}
\frac{1}{\left( \gamma - \alpha \right)!}  M_{\alpha}(R, R.c) \left(
\frac{R.c - c'}{\sqrt{2h^2}} \right)^{\gamma - \alpha}
\label{eq:far_to_far_translation}
\end{equation}
}
\begin{proof}
Replace the Hermite part of the expansion by a new Taylor
series: {\small
\begin{align*}
& \widetilde{G}(q, \{ (R, F(R.c, p)) \}) \\
=& \sum\limits_{\alpha < p}
M_{\alpha}(R, R.c) h_{\alpha} \left( \frac{q - R.c}{\sqrt{2h^2}}
\right) 
\end{align*}
\begin{align*}
 =& \sum\limits_{\alpha < p} M_{\alpha}(R, R.c)
\sum\limits_{\beta \geq 0} \frac{1}{\beta!} \left( \frac{c' -
R.c}{\sqrt{2h^2}} \right)^{\beta} (-1)^{|\beta|} h_{\alpha + \beta}
\left( \frac{q - c'}{\sqrt{2h^2}} \right) \\
 =& \sum\limits_{\alpha <
p} \sum\limits_{\beta \geq 0} M_{\alpha}(R, R.c) \frac{1}{\beta!}
\left( \frac{c' - R.c}{\sqrt{2h^2}} \right)^{\beta} (-1)^{|\beta|}
h_{\alpha + \beta} \left( \frac{q - c'}{\sqrt{2h^2}} \right)\\
=&
\sum\limits_{\alpha < p} \sum\limits_{\beta \geq 0} M_{\alpha}(R, R.c)
\frac{1}{\beta!}  \left( \frac{R.c - c'}{\sqrt{2h^2}} \right)^{\beta}
h_{\alpha + \beta} \left( \frac{q - c'}{\sqrt{2h^2}} \right) \\
=&
\sum\limits_{\gamma < p} \left[ \sum\limits_{0 \leq \alpha \leq
\gamma} \frac{1}{\left(\gamma - \alpha \right)!}  M_{\alpha}(R, R.c)
\left( \frac{R.c - c'}{\sqrt{2h^2}} \right)^{\gamma - \alpha} \right]
h_{\gamma} \left( \frac{q - c'}{\sqrt{2h^2}} \right)
\end{align*}
}
\noindent where $\gamma = \alpha + \beta$.
\end{proof}
\end{lem}
\begin{figure}[t]
\scalebox{0.5}{\includegraphics{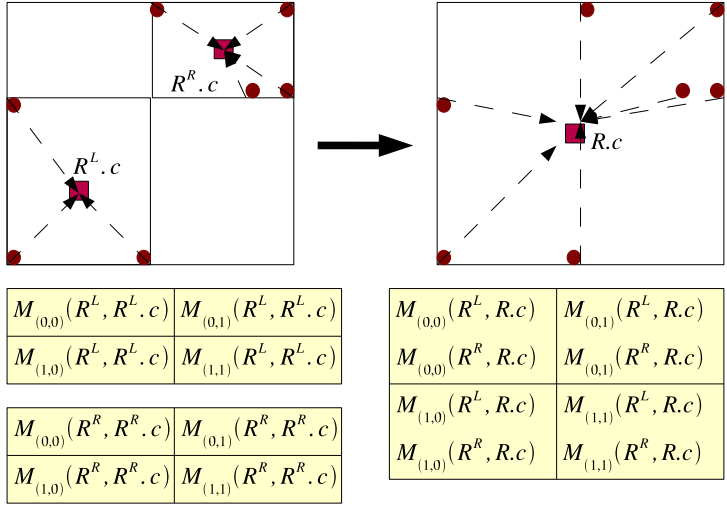}}
\caption{Given the far-field moments of $R^L$ and $R^R$ illustrated in
the first two tables, Theorem~\ref{H2HTranslationOperator} can
re-center each set of far-field moments of $R^L$ and $R^R$ at centroid
$R.c$. The re-centered far-field moments are shown in the third table
with two numbers, each contributed by $R^L$ and $R^R$. The far-field
moments of $R$ are then computed by adding up the two re-centered
moments entry-wise.}
\end{figure}
Using Lemma~\ref{H2HTranslationOperator}, we can compute the far-field
moments of $Q$ centered at $Q.c$ by translating the moments $\{
M_{\gamma}(R^L, R^L.c) \}_{\gamma < p}$ and $\{ M_{\gamma}(R^R, R^R.c)
\}_{\gamma < p}$ to form the moments $\{ M_{\gamma}(R^L, R.c)
\}_{\gamma < p}$ and $\{ M_{\gamma}(R^R, R.c) \}_{\gamma < p}$. Then,
the far-field moments of $R = R_1 \cup R_2$ are $\{ M_{\gamma}(R^L,
R.c) + M_{\gamma}(R^R, R.c) \}$ and
{\small
\begin{align*}\widetilde{G}(q, \{ (R, F(R.c, p)) \} ) = \sum\limits_{\gamma < p} (
M_{\gamma} (R^L, R.c) + M_{\gamma} (R^R, R.c) ) h_{\gamma} \left(
\frac{q - R.c}{\sqrt{2h^2}} \right)
\end{align*}} Computing each $M_{\gamma}(R^L,
R.c)$ from $M_{\gamma}(R^L, R^L.c)$ (and each $M_{\gamma}(R^R, R.c)$
from $M_{\gamma}(R^R, R^R.c)$) requires iterating over at most $p^D$
terms. This operation runs in $O(D p^{2D})$, which can be more
efficient than computing the far-field moments of $R$ centered at
$R.c$ from scratch (which is $O(|R| D p^D)$).

The next translation operator acts as a ``clean-up'' routine in a
hierarchical algorithm.  Since we can approximate at different scales
in the query tree, we must somehow combine all the approximations at
the end of the computation.  By performing a breadth-first traversal
of the query tree, the L2L operator shifts a node's local expansion to
the centroid of each child.
\begin{lem}
\label{L2LTranslationOperator}
{\bf Shifting a combined local expansion of a query node to a new
  center (L2L translation operator for Gaussian kernel):} Given a
combined local expansion centered at $Q.c$ of the given query node
$Q$: 
{\small
\begin{align*}
\widetilde{G}(q, \mathcal{R_D}(Q) \cup \mathcal{R_T}(Q)) =
\sum\limits_{\beta < p} L_{\beta}(Q.c, \mathcal{R_D}(Q) \cup
\mathcal{R_T}(Q)) \left( \frac{q - Q.c}{\sqrt{2h^2}} \right)^{\beta}
\end{align*}
}
\noindent Shifting this local expansion to the new center $c' \in Q$ yields:{\small
\begin{align*}
& \widetilde{G}(q, \mathcal{R_D}(Q) \cup
\mathcal{R_T}(Q)) \\
= &\sum\limits_{\alpha < p} \left[ \sum\limits_{\beta
    \geq \alpha} \frac{\beta!}{\alpha! (\beta - \alpha)!}
  L_{\beta}(Q.c, \mathcal{R_D}(Q) \cup \mathcal{R_T}(Q)) \left(
  \frac{c' - Q.c}{\sqrt{2h^2}} \right) ^{\beta - \alpha} \right]
\left( \frac{q - c'}{\sqrt{2h^2}} \right)^{\alpha}
\end{align*}}
where we denote
{\small
\begin{equation}
L_{\beta}(c', \mathcal{R_D}(Q) \cup \mathcal{R_T}(Q)) =
\sum\limits_{\beta \geq \alpha} \frac{\beta!}{\alpha! (\beta -
\alpha)!}  L_{\beta}(Q.c, \mathcal{R_D}(Q) \cup \mathcal{R_T}(Q))
\left( \frac{c' - Q.c}{\sqrt{2h^2}} \right) ^{\beta - \alpha}
\label{eq:local_to_local_translation}
\end{equation}
}
\begin{proof}
Use the multinomial theorem to expand about the new center $c'$:
{\small
\begin{align*}
& \widetilde{G}(q, \mathcal{R_D}(Q) \cup \mathcal{R_T}(Q)) =
\sum\limits_{\beta < p} L_{\beta}(Q.c, \mathcal{R_D}(Q) \cup
\mathcal{R_T}(Q)) \left( \frac{q - Q.c}{\sqrt{2h^2}} \right)^{\beta}\\
=& \sum\limits_{\beta < p} \sum\limits_{\alpha \leq \beta}
L_{\beta}(Q.c, \mathcal{R_D}(Q) \cup \mathcal{R_T}(Q))
\frac{\beta!}{\alpha!(\beta - \alpha)!} \left( \frac{c' -
Q.c}{\sqrt{2h^2}} \right)^{\beta - \alpha} \left( \frac{q -
c'}{\sqrt{2h^2}} \right)^{\alpha}
\end{align*}
}
\noindent whose summation order can be interchanged to achieve the result. 
\end{proof}
\end{lem}
\begin{figure}[!t]
\scalebox{0.52}{\includegraphics{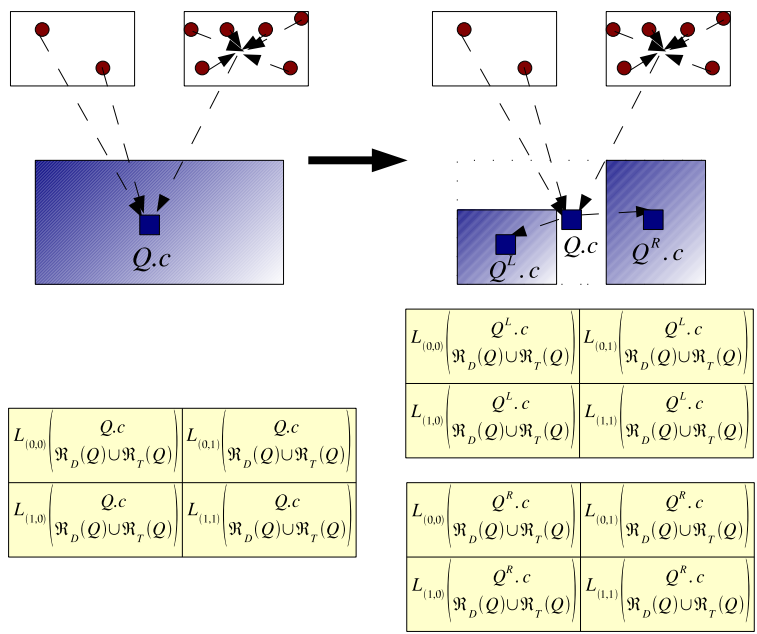}}
\caption{Given the local moments centered at $Q.c$,
  Theorem~\ref{L2LTranslationOperator} can re-center them at the two
  centroids $Q^L.c$ and $Q^R.c$.}
\end{figure}
Using Lemma~\ref{L2LTranslationOperator}, we can shift the local
moments of $Q$ centered at $Q.c$ to a different expansion center, such
as an expansion center of one of the child nodes of $Q$. Let $p$ be
the maximum approximation order used among the reference nodes pruned
via far-to-local translation ($\mathcal{R_T}(Q)$) and direct local
accumulation ($\mathcal{R_D}(Q)$). The local moment propagation to
both child nodes of $Q$ is achieved by the following operations:
{\small
\begin{align*}
\{ L_{\beta}(Q^L.c, \mathcal{R_D}(Q^L) \cup \mathcal{R_T}(Q^L))
\}_{\beta < p} \leftarrow & \{ L_{\beta}(Q^L.c, \mathcal{R_D}(Q^L)
\cup \mathcal{R_T}(Q^L)) \}_{\beta < p} +\\ & \{ L_{\beta}(Q^L.c,
\mathcal{R_D}(Q) \cup \mathcal{R_T}(Q)) \}_{\beta < p}\\ \{
L_{\beta}(Q^R.c, \mathcal{R_D}(Q^R) \cup \mathcal{R_T}(Q^R)) \}_{\beta
< p} \leftarrow & \{ L_{\beta}(Q^R.c, \mathcal{R_D}(Q^R) \cup
\mathcal{R_T}(Q^R)) \}_{\beta < p} +\\ & \{ L_{\beta}(Q^L.c,
\mathcal{R_D}(Q) \cup \mathcal{R_T}(Q)) \}_{\beta < p}
\end{align*}
}
\noindent where the addition operation is an element-wise operation that
combines the two scalars with the same multi-index position.

\subsection{Choosing the Best Approximation Method}
Suppose we are given a query node $Q$ and a reference node $R$ pair
during the invocation of
Algorithm~\ref{alg:dtkde}. $\mbox{\cansummarize}$ function for the
higher-order DFGT algorithm has four approximation methods available:
$A \in \{E, T(c, p), F(c, p),$ $D(c, p) \}$ (see
Section~\ref{sec:algorithmic_notations}). Because we would like to
avoid exhaustive computations, the higher-order DFGT algorithm uses
only three of the approximation methods and defers exhaustive
computations until query/reference leaf pairs are
encountered. Algorithm~\ref{alg:choose_best_method} tests whether the
given query node and reference node pair can be approximated by
evaluating the far-field moments of $R$, computing direct local
accumulation due to $R$, and translating some of the terms that
constitute the far-field moments of $R$ (far-field-to-local
translation operator) and evaluates the asymptotic cost of each
approximation. Algorithm~\ref{alg:choose_best_method} then determines
the approximation method with the lowest asymptotic cost. This idea
was originally introduced in~\cite{greengard1991fgt} in the
description of the original fast Gauss transform algorithm. The key
difference is that even if Algorithm~\ref{alg:choose_best_method}
returns $E$ (when none of the other approximation methods can beat the
cost of the exhaustive method), our hierarchical algorithm will not
default to exhaustive evaluations and will consider the query points
and reference points at a finer granularity, as shown in
Algorithm~\ref{alg:dtkde}.
\begin{algorithm}[t]
\caption{$\mbox{\choosebestmethod}(Q, R, \tau)$: Chooses the FMM-type
  approximation with the least cost for a query and reference node
  pair.}

\begin{algorithmic}

\STATE{$p_F \leftarrow \farfieldorder(Q, R, \tau)$}

\STATE{$p_D \leftarrow \localaccumulationorder(Q, R,
\tau)$}

\STATE{$p_T \leftarrow \convertfarfieldtolocalorder(Q, R, \tau)$}

\STATE{$c_F \leftarrow N_Q D^{p_F + 1}$, $c_D \leftarrow N_R D^{p_D +
1}$, $c_T \leftarrow D^{2 p_T + 1}$, $c_E \leftarrow D N_Q N_R$}

\IF{$c_F = \min\{ c_F, c_D, c_T, c_E\} $}

\RETURN{$F(R.c, p_F)$}

\ELSIF{$c_D = \min\{ c_F, c_D, c_T, c_E\}$}

\RETURN{$D(Q.c, p_D)$}

\ELSIF{$c_T = \min\{ c_F, c_D, c_T, c_E\}$}

\RETURN{$T(Q.c, p_T)$}

\ELSE

\RETURN{$E$}

\ENDIF

\end{algorithmic}
\label{alg:choose_best_method}
\end{algorithm}
\begin{figure}[!h]
\scalebox{0.5}{\includegraphics{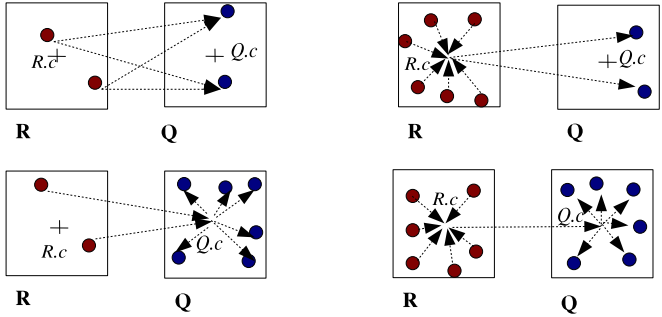}}
\caption{Four ways of approximating the contribution of a reference
  node to a query node. {\bf Top left:} exhaustive computations (few
  reference points/few query points); {\bf Top right: }far-field
  moment evaluating (many reference points/few query points); {\bf
    Bottom left:} direct local moment accumulation (few reference
  points/many query points); {\bf Bottom right: }far-field-to-local
  translation (many reference points/many query points).}
\end{figure}

\subsection{Hierarchical FGT}
Given the analytical machinery developed in the previous section, we
now describe how to extend the centroid-based
dual-tree~\cite{gray2003vfm,gray2003nde} to do higher-order
approximations. The main structure of the algorithm is shown in
Algorithm~\ref{alg:dtkdemain}. We provide only a high-level overview
of our algorithm and defer the discussion on the implementation
details to Appendix.
\begin{algorithm}[t]
\caption{$\mbox{\dtkdemain}(\mathcal{Q}, \mathcal{R})$: The main KDE routine.}

\begin{algorithmic}

\STATE{$Q^{\mathit{root}} \leftarrow \kdtree(\mathcal{Q})$, \ \ \ \ \ \
$R^{\mathit{root}} \leftarrow \kdtree(\mathcal{R})$}

\STATE{$\mbox{\dtkdeinitquerytree}(Q^{\mathit{root}})$, \ \ \ \ \ \
$\mbox{\dtkdeinitreferencetree}(R^{\mathit{root}})$}

\STATE{
$\mbox{\dtkde}(Q^{\mathit{root}},R^{\mathit{root}})$, \ \ \ \ \ \
$\mbox{\dtkdepost}(Q^{\mathit{root}})$}

\end{algorithmic}
\label{alg:dtkdemain}
\end{algorithm}
\ \\{\bf Initialization of the query tree. }Each query node maintains
a vector storing $\left( p_{\mathit{max}} \right)^D$ terms, where
$p_{\mathit{max}}$ is a pre-determined limit on the approximation
order\footnote{We impose this limit because the number of terms scales
  exponentially with the dimensionality $D$, $O(p^D)$.} depending on
the dimensionality of the query set $\mathcal{Q}$ and the reference
set $\mathcal{R}$. For the experimental results, we have fixed
$p_{\mathit{max}} = 6$ for $D = 2$, $p_{\mathit{max}} = 4$ for $D =
3$, $p_{\mathit{max}} = 2$ for $D = 4$ and $D = 5$, $p_{\mathit{max}}
= 1$ for $D \geq 6$.
\begin{algorithm}[h]
\caption{$\mbox{\dtkdeinitquerytree}(Q)$: Initializes query bound
  summary statistics.}

\begin{algorithmic}

\STATE{\COMMENT{Initialize the node bound summary statistics.}}

\STATE {$G^{l}(Q, \mathcal{R}) \leftarrow 0$, \ \ \ $G^{u}(Q, \mathcal{R})
  \leftarrow |\mathcal{R}|$, \ \ \ $Q.\Delta^{l} \leftarrow 0$, \ \ \
  $Q.\Delta^{u} \leftarrow 0$}

\STATE{\COMMENT{Initialize translated local moments to be a vector of length
$\left( p_{\mathit{max}}^D \right)$.}}

\STATE{$Q.L_{0 \leq i < \left( p_{\mathit{max}}\right)^D} \leftarrow
0$}

\IF {$Q$ is a leaf node}

\STATE{\COMMENT{Initialize for each query point.}}

\FOR {each $q_{i_m} \in Q$}

\STATE {$G^{l}(q_{i_m}, \mathcal{R}) \leftarrow 0$, \ \ \ \ \ \ $G^{u}(q_{i_m},
\mathcal{R}) \leftarrow | \mathcal{R} |$}

\STATE{ $\widetilde{G}(q_{i_m},
\mathcal{R}) \leftarrow 0$, \ \ \ \ \ \ $\widetilde{G}(q_{i_m},
\mathcal{R_E}(q_{i_m})) \leftarrow 0$}

\STATE{$\widetilde{G}(q_{i_m}, \mathcal{R_F}(q_{i_m})) \leftarrow 0$,
  \ \ \ \ \ \ $\widetilde{G}(q_{i_m}, \mathcal{R_D}(q_{i_m}) \cup
  \mathcal{R_T}(q_{i_m})) \leftarrow 0$}

\ENDFOR

\ELSE 

\STATE{ $\mbox{\dtkdeinitquerytree}(Q^L)$, \ \ \ \ \ \ 
$\mbox{\dtkdeinitquerytree}(Q^R)$}

 \ENDIF
\end{algorithmic}
\label{alg:higher_order_dtkde_init_query_tree}
\end{algorithm}

\ \\{\bf Pre-computation of far-field moments. }Before the main KDE
computation can begin, we pre-compute the far-field moments of each
reference node in the reference tree up to $\left( p_{\mathit{max}}
\right)^D$ terms. We show how to efficiently pre-compute the far-field
moments of each reference node in the reference tree in
Algorithm~\ref{alg:dtkdeinitreferencetree}. The algorithm uses
Equation~\eqref{eq:far_field_moments} for the leaf node and
Equation~\eqref{eq:far_to_far_translation} for translating the moments
of the child nodes for the internal node case. We describe the
implementation details in Appendix.
\begin{algorithm}[t]
\caption{$\mbox{\dtkdeinitreferencetree}(R)$: Pre-computes far-field
  moments.}
\begin{algorithmic}

\STATE{\COMMENT{Initialize the far-field moments of $R$ to be empty.}}

\FOR{$i = 0$ to $i < (p_{\mathit{max}})^D$}

\STATE{$M_{\mbox{\positiontomultiindex}(i, p_{\mathit{max}})}(R, R.c)
\leftarrow 0$}

\ENDFOR

\IF {$R$ is a leaf node}

\STATE{\COMMENT{Accumulate far-field moment from each point (Equation~\eqref{eq:far_field_moments}).}}

\STATE{$\mbox{\accumulatefarfieldmoment}(R)$}

\ELSE 

\STATE{\COMMENT{Recursively compute the moments of the child nodes and
    combine them.}}

\STATE{$\mbox{\dtkdeinitreferencetree}(R^L)$, \ \ \ \ \ \
$\mbox{\dtkdeinitreferencetree}(R^R)$}

\STATE{$\mbox{\translatefartofarfield}(R^L, R)$, \ \ \ \ \ \
$\mbox{\translatefartofarfield}(R^R, R)$}

\ENDIF
\end{algorithmic}
\label{alg:dtkdeinitreferencetree}
\end{algorithm}

\ \\{\bf Determining the prunability of the given query and reference
  pair} (shown in
Algorithm~\ref{alg:higher_order_dfgt_summarize_functions}). Note that
the function $\mbox{\summarize}$ includes calls to the following
functions (see Appendix):
\begin{enumerate}
\item{$\mbox{\evaluatefarfieldexpansion}$: evaluates the far-field
  moments stored in $R$ at each query point in $Q$ up to $(p_F)^D$
  terms. See Algorithm~\ref{alg:evaluate_far_field_expansion}.}

\item{$\mbox{\accumulatedirectlocalmoment}$: computes direct
local moment contribution of $R$ centered at $Q.c$ in $Q$. See
Algorithm~\ref{alg:accumulate_direct_local_moment}.}

\item{$\mbox{\translatefartolocal}$: translates the far-field moments
of $R$ up to $(p_T)^D$ terms to the local moment centered $Q.c$ in
$Q$. See Algorithm~\ref{alg:translate_far_to_local}.}
\end{enumerate}
\begin{algorithm}
\caption{$\mbox{\cansummarize}(Q, R, \epsilon)$: Determines the
  prunability of the given query node $Q$ and reference node $R$}
\begin{algorithmic}

\RETURN{$\mbox{\choosebestmethod}\left(Q, R,
\frac{\epsilon |R| G^{l, \mathit{new}}(Q, \mathcal{R})}{| \mathcal{R}
|} \right) \not = E$}

\end{algorithmic}
\end{algorithm}
\begin{algorithm}
\caption{$\mbox{\summarize}(Q, R)$: Summarizes the contribution of $R$.}
\begin{algorithmic}

\STATE{\COMMENT{Add bound changes.}}

\STATE{$Q.\Delta^l \leftarrow Q.\Delta^l + \delta^l(Q, R)$,
$Q.\Delta^u \leftarrow Q.\Delta^u + \delta^u(Q, R)$}

\IF{$A$ is of the form $F(R.c, p_F)$}

\STATE{$\mbox{\evaluatefarfieldexpansion}(R, Q, p_F)$}

\ELSIF{$A$ is of the form $D(Q.c, p_D)$}

\STATE{$\mbox{\accumulatedirectlocalmoment}(R, Q, p_D)$}

\ELSE

\STATE{$\mbox{\translatefartolocal}(R, Q, p_T)$}

\ENDIF

\end{algorithmic}
\label{alg:higher_order_dfgt_summarize_functions}
\end{algorithm}

\begin{algorithm}[t]
\caption{$\mbox{\dtkde}(Q, R)$: The core dual-tree routine for computing KDE.}

\begin{algorithmic}

\STATE{$\delta^l(Q, R) = |R| K_h(d^u(Q, R))$, $\delta^u(Q, R) = |R|
  (K_h(d^l(Q, R)) - 1)$}

\STATE{\COMMENT{Add postponed contributions/bound changes from the
    current pair.}}

\STATE{$G^{l, \mathit{new}}(Q, \mathcal{R}) \leftarrow G^l(Q,
\mathcal{R}) + Q.\Delta^l + \delta^l(Q, R)$}

\STATE{$G^{u, \mathit{new}}(Q, \mathcal{R}) \leftarrow G^u(Q,
\mathcal{R}) + Q.\Delta^u + \delta^u(Q, R)$}

\IF {$\mbox{\cansummarize}(Q, R, \epsilon)$}

\STATE{$\mbox{\summarize}(Q, R)$}

\ELSE

\IF{$Q$ is a leaf node}

\IF{$R$ is a leaf node}

\STATE $\mbox{\dtkdebase}(Q, R)$

\ELSE

\STATE{$\mbox{\dtkde}(Q, R^L)$,$\mbox{\dtkde}(Q, R^R)$}

\ENDIF

\ELSE

\STATE{\COMMENT{Push down postponed bound changes owned by $Q$ to the
    children.}}

\STATE{$Q^L.\Delta^l \leftarrow Q^L.\Delta^l +
Q.\Delta^l$, \ \ \ \ \ $Q^R.\Delta^l \leftarrow Q^R.\Delta^l + Q.\Delta^l$}

\STATE{$Q^L.\Delta^u \leftarrow Q^L.\Delta^u + Q.\Delta^u$,
  \ \ \ \ \ $Q^R.\Delta^u \leftarrow Q^R.\Delta^u + Q.\Delta^u$}

\STATE{$Q.\Delta^l \leftarrow 0$, \ \ \ \ \ $Q.\Delta^u \leftarrow 0$}

\IF{$R$ is a leaf node}

\STATE{$\mbox{\dtkde}(Q^L, R)$, $\mbox{\dtkde}(Q^R, R)$}

\ELSE

\STATE{$\mbox{\dtkde}(Q^L, R^L)$,$\mbox{\dtkde}(Q^L,
R^R)$,$\mbox{\dtkde}(Q^R, R^L)$,$\mbox{\dtkde}(Q^R, R^R)$}

\ENDIF

\STATE{\COMMENT{Refine the bounds based on the recursion results.}}

\STATE{$G^l(Q, \mathcal{R}) \leftarrow \min \{ G^l(Q^L, \mathcal{R}) +
Q^L.\Delta^l, G^l(Q^R, \mathcal{R}) + Q^R.\Delta^l
\}$\label{alg:dtkde:lower_bound_refinement}}

\STATE{$G^u(Q, \mathcal{R}) \leftarrow \max \{ G^u(Q^L, \mathcal{R}) +
Q^L.\Delta^u, G^u(Q^R, \mathcal{R}) + Q^R.\Delta^u
\}$\label{alg:dtkde:upper_bound_refinement}}

\ENDIF

\ENDIF

\end{algorithmic}
\label{alg:dtkde}
\end{algorithm}
\ \\ {\noindent \bf Dual-tree Recursion. }Algorithm~\ref{alg:dtkde}
shows the basic structure of the dual-tree based KDE computation (see
Figure~\ref{fig:dualtree}). This procedure is first called with $Q$
and $R$ as the root nodes of the query and the reference tree
respectively. $\mbox{\cansummarize}$ takes three parameters: the
current query node $Q$, the current reference node $R$, and the global
relative error tolerance $\epsilon$. This function tests whether the
the contribution of the given reference node for each query point in
the given query node can be approximated within the error tolerance.
If the approximation is not possible, then the algorithm continues to
consider the query and the reference data at a finer granularity. The
basic idea is to terminate the recursion as soon as possible by
considering large ``chunks'' of the query data and the reference data
and avoiding the number of exhaustive leaf-leaf computations. We can
achieve this if we utilize approximation schemes that yield high
accuracy and have cheap computational costs.

Each prune made for a pair of a query and a reference node is
summarized in the given query node by incorporating the lower and the
upper bound changes $\delta^l(Q, R)$ and $\delta^u(Q, R)$ contributed
by the reference node into $Q.\Delta^{l}$ and $Q.\Delta^{u}$. These
two bound updates due to a prune can be regarded as a new piece of
information which is known only locally to the given query node
$Q$. All of the bounds in the entire subtree of $Q$ should reflect
this information. One way to achieve this effect is to pass the lower
bound and the upper bound changes owned by $Q$ (i.e., $Q.\Delta^l$ and
$Q.\Delta^u$) to $Q$'s immediate children, whenever the algorithm
needs to consider the query dataset at a finer granularity by
recursing to the left and the right child of $Q$.

\begin{algorithm}[t]
\caption{$\mbox{\dtkdebase}(Q, R)$: Computes exact contribution of $R$
  to $Q$.}

\begin{algorithmic}

\STATE{$G^l(Q, \mathcal{R}) \leftarrow \infty$, \ \ \ \ \ \ $G^u(Q,
  \mathcal{R}) \leftarrow -\infty$}

\FOR {each $q_{i_m} \in Q$}

\STATE{\COMMENT{Add postponed changes passed down from the
    ancestor node of $Q$.}}

\STATE{$G^l(q_{i_m}, \mathcal{R}) \leftarrow G^l(q_{i_m}, \mathcal{R})
  + Q.\Delta^l$, \ \ \ \ \ \ $G^u(q_{i_m}, \mathcal{R}) \leftarrow
  G^u(q_{i_m}, \mathcal{R}) + Q.\Delta^u$}

\FOR {each $r_{j_n} \in R$}

\STATE{$v \leftarrow K_h(\|q_{i_m} - r_{j_n}\|)$, \ \ \ \ \ \ \ \
$G^{l}(q_{i_m}, \mathcal{R}) \leftarrow G^{l}(q_{i_m},
\mathcal{R}) + v$}

\STATE{$\widetilde{G}(q_{i_m}, \mathcal{R_E}(q_{i_m})) \leftarrow
\widetilde{G}(q_{i_m}, \mathcal{R_E}(q_{i_m})) + v$}

\STATE{$G^{u}(q_{i_m}, \mathcal{R}) \leftarrow G^{u}(q_{i_m},
\mathcal{R}) + (v - 1)$}

\ENDFOR

\STATE{\COMMENT{Refine the bound summary statistics owned by $Q$.}}

\STATE{$G^l(Q, \mathcal{R}) \leftarrow \min\{ G^l(Q, \mathcal{R}),
G^l(q_{i_m}, \mathcal{R}) \}$}

\STATE{$G^u(Q, \mathcal{R}) \leftarrow \max\{ G^u(Q, \mathcal{R}),
G^u(q_{i_m}, \mathcal{R}) \}$}

\ENDFOR

\STATE{$Q.\Delta^l \leftarrow 0$, \ \ \ \ \ \ $Q.\Delta^u \leftarrow 0$}

\end{algorithmic}
\label{alg:dtkdebase}
\end{algorithm}

\ \\ {\bf Base-case Computation. }If a given leaf query and leaf
reference node pair could not be pruned, then $\mbox{\dtkdebase}$
(shown in Algorithm~\ref{alg:dtkdebase}) is called. Because all kernel
evaluations are computed exactly, we can refine the bound summary
statistics of the given query node $Q$ (that is, $G^l(Q, \mathcal{R})$
and $G^u(Q, \mathcal{R})$) further and hence we reset them to $\infty$
and $-\infty$ respectively. For each query point $q_{i_m} \in Q$, we
first incorporate the postponed bound changes passed down from the
ancestor node of $Q$. We loop over each reference point $r_{j_n} \in
R$ and compute the kernel value between $q_{i_m}$ and $r_{j_n}$ and
accumulate the lower bound $G^l(q_{i_m}, \mathcal{R})$, the kernel sum
computed exhaustively $\widetilde{G}(q_{i_m},
\mathcal{R_E}(q_{i_m}))$, and the upper bound $G^u(q_{i_m},
\mathcal{R})$

Note that we subtract one for updating $G^u(q_{i_m},
\mathcal{R})$ for correcting the prior assumption that $K_h(||q_{i_m}
- r_{j_n}||) = 1$, while the lower bound $G^l(q_{i_m}, \mathcal{R})$
and $\widetilde{G}(q_{i_m}, \mathcal{R_E}(q_{i_m}))$ are incremented
by $K_h(||q_{i_m} - r_{j_n}||)$. As the contribution of the reference
node $R$ is added onto the query point $q_{i_m}$'s sum, we can refine
the bound summary statistics owned by $Q$ such that $G^l(Q,
\mathcal{R}) = \min\limits_{q_{i_m} \in Q} G^l(q_{i_m}, \mathcal{R})$
and $G^u(Q, \mathcal{R}) = \max\limits_{q_{i_m} \in Q} G^u(q_{i_m},
\mathcal{R})$. Finally, we reset the postponed bound changes stored in
$Q$ to zero.

\ \\ {\bf Post-processing} (shown in
Algorithm~\ref{alg:dtkdepost}). For the non-leaf case, the
local-to-local translation operator ($\mbox{\translatelocaltolocal}$)
is called to re-center the local moments at the current level and
passes them down to the child nodes. For the leaf-case,
$\mbox{\evaluatelocalexpansion}$ is called to convert local moments to
a single scalar that represents the contribution to a given query
point.
\begin{algorithm}[t]
\caption{$\mbox{\dtkdepost}(Q)$: The post-processing routine.}

\begin{algorithmic}

\IF {$Q$ is a leaf node}

\STATE{$G^l(Q, \mathcal{R}) \leftarrow \infty$, \ \ \ \ \ \ $G^u(Q, \mathcal{R})
\leftarrow -\infty$}

\FOR {each $q_{i_m} \in Q$}

\STATE{\COMMENT{Add bound changes for the query node at the given
    query point $q_{i_m}$.}}

\STATE{$G^l(q_{i_m}, \mathcal{R}) \leftarrow G^l(q_{i_m}, \mathcal{R})
+ Q.\Delta^l$, \ \ \ \ $G^u(q_{i_m}, \mathcal{R}) \leftarrow G^u(q_{i_m},
\mathcal{R}) + Q.\Delta^u$}

\STATE{\COMMENT{Refine summary statistics for lower and upper bounds.}}

\STATE{$G^l(Q, \mathcal{R}) \leftarrow \min \{G^l(Q, \mathcal{R}),
G^l(q_{i_m}, \mathcal{R}) \}$}

\STATE{$G^u(Q, \mathcal{R}) \leftarrow \max
\{G^u(Q, \mathcal{R}), G^u(q_{i_m}, \mathcal{R}) \}$}

\STATE{\COMMENT{Compute the contributions from the accumulated local moments.}}

\STATE{$\widetilde{G}(q_{i_m}, \mathcal{R_T}(q_{i_m})) \leftarrow
\mbox{\evaluatelocalexpansion}(Q)$}

\STATE{\COMMENT{Sum the contribution from the local moments (direct or
    translated), the far-field evaluations, and exhaustive
    evaluations.}}

\STATE{{\small$\widetilde{G}(q_{i_m}, \mathcal{R}) \leftarrow
\widetilde{G}(q_{i_m}, \mathcal{R_D}(q_{i_m}) \cup
\mathcal{R_T}(q_{i_m})) + \widetilde{G}(q_{i_m},
\mathcal{R_F}(q_{i_m})) + \widetilde{G}(q_{i_m}, \mathcal{R_E})$}}

\ENDFOR

\STATE{$\Delta^l(Q) \leftarrow 0$, \ \ \ \ $\Delta^u(Q)\leftarrow 0$,
  \ \ \ \ $Q.L \leftarrow 0$}

\ELSE

\STATE{$\mbox{\translatelocaltolocal}(Q, Q^L)$, \ \ \ \ \ \ 
$\mbox{\translatelocaltolocal}(Q, Q^R)$}

\STATE{$Q^L.\Delta^l \leftarrow Q^L.\Delta^l + Q.\Delta^l$, \ \ \ \
$Q^R.\Delta^l \leftarrow Q^R.\Delta^l + Q.\Delta^l$}

\STATE{$Q^L.\Delta^u \leftarrow Q^L.\Delta^u + Q.\Delta^u$, \ \ \ \
$Q^R.\Delta^u \leftarrow Q^R.\Delta^u + Q.\Delta^u$}

\STATE{$Q.L \leftarrow 0$, \ \ \ \ $Q.\Delta^l \leftarrow 0$, \ \ \ \ $Q.\Delta^u
\leftarrow 0$}

\STATE {$\mbox{\dtkdepost}(Q^L)$, \ \ \ \ $\mbox{\dtkdepost}(Q^R)$}

\STATE{\COMMENT{Refine the bounds based on the results of the recursion.}}

\STATE{$G^l(Q, \mathcal{R}) \leftarrow \min \{ G^l(Q^L, \mathcal{R}),
G^l(Q^R, \mathcal{R}) \}$}

\STATE{$G^u(Q, \mathcal{R}) \leftarrow \max \{
G^u(Q^L, \mathcal{R}), G^u(Q^R, \mathcal{R}) \}$}

\ENDIF
\end{algorithmic}
\label{alg:dtkdepost}
\end{algorithm}
\subsection{Basic Properties of DFGT Algorithms} 
\begin{thm}
Lower/upper bounds are maintained properly at all times for each $q
\in \mathcal{Q}$ and each query node $Q$ during the function call
$\mbox{\dtkdemain}$.
\begin{proof}
We show that the bounds are maintained properly for three main parts
in the function $\mbox{\dtkdemain}$: $\mbox{\dtkdeinitquerytree}$,
$\mbox{\dtkde}$, and $\mbox{\dtkdepost}$.

\noindent The function call $\mbox{\dtkdeinitquerytree}$: It is clear that for
all $q_i \in \mathcal{Q}$, $0 = G^l(q_i, \mathcal{R}) \leq G(q_i,
\mathcal{R}) \leq G^u(q_i, \mathcal{R}) = | \mathcal{R}
|$. Furthermore, for each query node $Q$, $0 = G^l(Q, \mathcal{R})
\leq G(q_{i_m}, \mathcal{R}) \leq G^u(Q, \mathcal{R}) = | \mathcal{R}
|$ for each $q_{i_m} \in Q$.\\

\noindent The function call $\mbox{\dtkdebase}$: Let $Q$ and $R$ be the query
node and the reference node respectively. For each query point
$q_{i_m} \in Q$, $G^l(q_{i_m}, \mathcal{R})$ is incremented by
$Q.\Delta^l + \sum\limits_{r_{j_n} \in R} K_h(||q_{i_m} - r_{j_n}||)$,
and $G^u(q_{i_m}, \mathcal{R})$ by $Q.\Delta^u + $

\noindent
$\sum\limits_{r_{j_n} \in R} \left( K_h(||q_{i_m} - r_{j_n}||) - 1
\right)$; this operation incorporates the passed-down contribution for
$q_{i_m} \in Q$, and un-does the assumption made during the
initialization phase of $\mbox{\dtkdeinitquerytree}$. $G^l(Q,
\mathcal{R})$ and $G^u(Q, \mathcal{R})$ are updated to be the minimum
among $G^l(q_{i_m}, \mathcal{R})$ and the maximum among $G^u(q_{i_m},
\mathcal{R})$ respectively. The postponed bound changes $Q.\Delta^l$
and $Q.\Delta^u$ are cleared to avoid double-counting when $Q$ may be
visited later.\\

\noindent The function call $\mbox{\dtkde}$: We induct on the number of points
owned by the query node $Q$ and the reference node $R$ in
consideration (i.e. $|Q| + |R|$). The only possible places that change
$G^l(q_{i_m}, \mathcal{R})$, $G^u(q_{i_m}, \mathcal{R})$, $G^l(Q,
\mathcal{R})$ and $G^u(Q, \mathcal{R})$ are the call to the base case
function $\mbox{\dtkdebase}$ and the last two lines of the function
$\mbox{\dtkde}$. The correctness of $\mbox{\dtkdebase}$ function is
proven already, so we consider the second case. The two function calls
$\mbox{\dtkde}(Q^L, R)$ and $\mbox{\dtkde}(Q^R, R)$ (in case $R$ is a
leaf node) and the four function calls\\
\noindent $\mbox{\dtkde}(Q^L, R^L)$, $\mbox{\dtkde}(Q^L, R^R)$,
$\mbox{\dtkde}(Q^R, R^L)$, and $\mbox{\dtkde}(Q^R, R^R)$ (in case $R$
is an internal node) are smaller subproblems than $(Q, R)$ pair. By
the induction hypothesis, these calls maintain the lower and the upper
bounds properly. The lower bound is set to the minimum of the ``best''
lower bound owned by the children of $Q$: $\min\{ G^l(Q^L,
\mathcal{R}) + Q^L.\Delta^l, G^l(Q^R, \mathcal{R}) + Q^R.\Delta^l
\}$. Similarly, the upper bound is set to the maximum of the ``best''
upper bound owned by the children of $Q$: $\max\{ G^u(Q^L,
\mathcal{R}) + Q^L.\Delta^u, G^u(Q^R, \mathcal{R}) + Q^R.\Delta^u
\}$.

\ \\
\noindent The function call $\mbox{\dtkdepost}$: We again induct on
the number of points owned by the query node $Q$ passed in as the
argument to this function. If the query node $Q$ is a leaf node, each
query point $q_{i_m} \in Q$ incorporates the passed-down bound changes
$Q.\Delta^l$ and $Q.\Delta^u$. The bounds $G^l(Q, \mathcal{R})$ and
$G^u(Q, \mathcal{R})$ are (correctly) set to the minimum among
$G^l(q_{i_m}, \mathcal{R})$ and the maximum among $G^u(q_{i_m},
\mathcal{R})$. If $Q$ is not a leaf node: we know the sub-calls
$\mbox{\dtkdepost}(Q^L)$ and $\mbox{\dtkdepost}(Q^R)$ maintains
correct lower and upper bounds by the induction hypothesis since $Q^L$
and $Q^R$ contain a smaller number of points. Setting the lower and
upper bounds for $Q$ by the operations: $G^l(Q, \mathcal{R})
\leftarrow \min \{ G^l(Q^L, \mathcal{R}), G^l(Q^R, \mathcal{R}) \}$,
$G^u(Q, \mathcal{R}) \leftarrow \max \{ G^u(Q^L, \mathcal{R}),
G^u(Q^R, \mathcal{R}) \}$ is valid.
\end{proof}
\end{thm}
\begin{thm}
After calling $\mbox{\dtkdepost}$ (Algorithm~\ref{alg:dtkdepost}) in
$\mbox{\dtkdemain}$ (Algorithm~\ref{alg:dtkdemain}), each {\it query
  point} $q_i \in \mathcal{Q}$ accounts for every {\it reference
  point} $r_j \in \mathcal{R}$ in its Gaussian kernel sum
approximation $\widetilde{G}(q_i, \mathcal{R})$.
\begin{proof}
In Algorithm~\ref{alg:dtkde}, for each $q_i \in \mathcal{Q}$, each
$r_j \in \mathcal{R}$ is either accounted by an exhaustive computation
in $\mbox{\dtkdebase}$ or a prune in $\mbox{\summarize}$. All
exhaustive computations for $q_i \in Q$ directly update
$\widetilde{G}(q_i, \mathcal{R_E}(q_i))$, while any pruned
contributions will be incorporated into each $\widetilde{G}(q_i,
\mathcal{R_T}(q_i))$ (hence into $\widetilde{G}(q_i,
\mathcal{R}(q_i))$) and when they are pushed down (to the leaf node to
which $q_i$ belongs) during the $\mbox{\dtkde}$ recursion or $\mbox{\dtkdepost}$.
\end{proof}
\label{thm:partition}
\end{thm}
\begin{thm}
For each query point $q_i \in \mathcal{Q}$, the approximated kernel
sum $\widetilde{G}(q_i, \mathcal{R})$ satisfies the global relative
error tolerance $\epsilon$.
\begin{proof}
For simplicity, let us limit the available approximation methods to $A
\in \{ E, T(c, 1) \}$ where $E$ denotes the exhaustive computation and
$T(c, 1)$ denotes the centroid-based approximation about $c$.

Given $q_i \in \mathcal{Q}$, let $Q'$ be the (unique) leaf node that
owns $q_i$. Let $\{ R_{T_a} \}_{a=1}^{N_a}$ denote the set of
reference nodes whose kernel sum contribution were accounted via
centroid approximation and $\{R_{E_b}\}_{b=1}^{N_b}$ the set of
reference nodes whose kernel sum contribution were computed
exhaustively. Then it is clear that $\mathcal{R} = \left(
\bigcup\limits_{a=1}^{N_a} R_{T_a} \right) \cup \left(
\bigcup\limits_{b=1}^{N_b} R_{E_b} \right)$ with $R_{T_{a'}} \cap
R_{T_{a''}} = \emptyset$, $R_{E_{b'}} \cap R_{E_{b''}} = \emptyset$,
$R_{T_{a'}} \cap R_{E_{b'}} = \emptyset$ for $1 \leq a', a'' \leq N_a$
and $1 \leq b', b'' \leq N_b$. Let $Q_{T_a}$ be the query node that
owns $q_i$ and is considered with the reference node $R_{T_a}$ and
pruned. Let $G^{l(a)}(Q_{T_a}, \mathcal{R})$ be a ``snapshot'' of the
running lower bound on the kernel sum for query points owned by
$Q_{T_a}$ at the time the query node $Q_{T_a}$ and the reference node
$R_{T_a}$ were considered (and subsequently pruned). By the triangle
inequality:
{\footnotesize
\begin{align*}
& \left | \widetilde{G}(q_i, \mathcal{R}) - G(q_i, \mathcal{R}) \right
| \\ =& \Biggl | \widetilde{G}\left (q_i, \left(
\bigcup\limits_{a=1}^{N_a} \left \{ ( R_{T_a}, T(Q.c, 1) ) \right \}
\right) \cup \left( \bigcup\limits_{b=1}^{N_b} \left \{ ( R_{E_b}, E )
\right \} \right) \right ) - \\
& G \left (q_i, \left(
\bigcup\limits_{a=1}^{N_a} R_{T_a} \right) \cup \left(
\bigcup\limits_{b=1}^{N_b} R_{E_b} \right) \right) \Biggr | \\
 \leq&
\Biggl | \left( \sum\limits_{a=1}^{N_a} \widetilde{G}\left( q_i, \{ (
R_{T_a}, T(Q.c, 1) )\} \right ) - G(q_i, R_{T_a}) \right) + \\
& \left(
\sum\limits_{b=1}^{N_b} \widetilde{G}\left ( q_i, \{ (R_{E_b}, E ) \}
\right) - G(q_i, R_{E_b}) \right) \Biggr | \\
\leq & \sum\limits_{a=1}^{N_a} \left| \widetilde{G}\left(q_i, \{
(R_{T_a}, T(Q.c, 1)) \} \right) - G(q_i, R_{T_a}) \right| +
\sum\limits_{b=1}^{N_b} \left | \widetilde{G}\left( q_i, \{ (R_{E_b},
E) \} \right ) - G(q_i, R_{E_b}) \right|\\ \leq &
\sum\limits_{a=1}^{N_a} | R_{T_a} | \max \left \{
\begin{array}{c} \left | K_h(d^u(Q_{T_a}, R_{T_a})) - K_h(||Q_{T_a}.c
  - R_{T_a}.c||) \right | ,\\ \left | K_h(d^l(Q_{T_a}, R_{T_a})) -
  K_h(||Q_{T_a}.c - R_{T_a}.c||) \right |
\end{array} \right \} + \sum\limits_{b=1}^{N_b} | R_{E_b} | \cdot 0\\
\leq & \sum\limits_{a=1}^{N_a} \frac{| R_{T_a} |
\epsilon}{|\mathcal{R}|} G^{l(a)}(Q_{T_a}, \mathcal{R}) +
\sum\limits_{b=1}^{N_b} \frac{| R_{E_b}| \epsilon}{| \mathcal{R} |}
G^{l(b)}(Q', \mathcal{R})\\ \leq & \sum\limits_{a=1}^{N_a} \frac{|
R_{T_a} | \epsilon}{|\mathcal{R}|} G(q_i, \mathcal{R}) +
\sum\limits_{b=1}^{N_b} \frac{| R_{E_b}| \epsilon}{| \mathcal{R} |}
G(q_i, \mathcal{R}) \leq \epsilon G(q_i, \mathcal{R})
\end{align*}
} The proof can be easily extended to the case with four available
approximation methods $A \in \{ E, T(c, p), F(c, p), D(c, p) \}$. 
\end{proof}
\label{thm:global_guarantee}
\end{thm}

\begin{figure}[t]
\small\noindent
\begin{tabular}{|p{0.48in}|p{0.35in}|p{0.35in}|p{0.35in}|p{0.35in}|p{0.35in}|p{0.35in}|p{0.35in}|p{0.35in}|p{0.35in}}
\hline
Alg$\backslash$Scale & {\it 0.001} & {\it 0.01} & {\it 0.1} & {\it 1} & {\it 10} & {\it 100} & {\it 1000} & {\it $\Sigma$}\\
\hline
\multicolumn{9}{|c|}{sj2-50000-2, $D = 2, N = 50000, h_{CV_{LS}}^* = 0.00139506$}\\
\hline
{\it Naive} & 241 & 241 & 241 & 241 & 241 & 241 & 241 & 1687\\
\hline
{\it FFT} & $\infty$ & $\infty$ & $\infty$ & $\infty$ & $\infty$ & 1.02 & 0.03 & $\infty$\\
\hline
{\it FGT} & X & X & X & 2.63 & 1.48 & 0.33 & 0.18 & X \\
\hline
{\it IFGT} & $\infty$ & $\infty$ & $\infty$ & 155 & 7.26 & 0.40 & 0.03 & $\infty$\\
\hline
{\it DFD} & 1.58 & 1.63 & 2.14 & 4.33 & 39.7 & 29.5 & 1.51 & 80.39\\
\hline
{\it DFGT} & 0.43 & 0.47 & 1.00 & 3.48 & 21 & 2.48 & 0.96 & 29.8\\
\hline
\multicolumn{9}{|c|}{colors50k, $D = 2, N = 50000, h_{CV_{LS}}^* = 0.0016911$}\\
\hline
{\it Naive} & 241 & 241 & 241 & 241 & 241 & 241 & 241 & 1687\\
\hline
{\it FFT} & $\infty$ & $\infty$ & $\infty$ & $\infty$ & $\infty$ & $\infty$ & 0.16 & $\infty$\\
\hline
    {\it FGT} & X & X & X & 120 & 10 & 4 & 0.22 & X\\
\hline
{\it IFGT} & $\infty$ & $\infty$ & $\infty$ & $\infty$ & $\infty$ & 0.54 & 0.07 & $\infty$\\
\hline
{\it DFD} & 1.62 & 1.76 & 2.36 & 12.5  & 102 & 17.0 & 2.41 & 139.65\\
\hline
{\it DFGT} & 0.44 & 0.60 & 1.21 & 15.6 & 20 & 4.20 & 0.67 & 42.7\\
\hline
\multicolumn{9}{|c|}{bio5, $D = 5, N = 103010, h_{CV_{LS}}^* = 0.000308646$}\\
\hline
{\it Naive} & 1310 & 1310 & 1310 & 1310 & 1310 & 1310 & 1310 & 9170\\
\hline
{\it FFT} & X & X & X & X & X & X & X & X\\
\hline
{\it FGT} & X & X & X & X & X & X & X & X\\
\hline
{\it IFGT} & $\infty$ & $\infty$ & $\infty$ & $\infty$ & $\infty$ & $\infty$ & 1.04 & $\infty$\\
\hline
{\it DFD} & 0.34 & 0.36 & 0.92 & 6.31 & 113 & 643 & 125  & 888.93\\
\hline
{\it DFGT} & 0.35 & 0.37 & 0.94 & 6.51 & 102 & 304 & 121 &  535.17\\
\hline
\end{tabular}
\caption{Empirical comparison of six different algorithms on different
  magnitudes of bandwidths on three different datasets. Each entry in
  the table has a timing number (if finite), $\infty$ symbol (if no
  parameter tweaking could achieve the error tolerance), $X$ symbol
  (if the algorithm segfaulted).}
\end{figure}
\section{Experimental Results}
\label{sec:experimental_results}
We evaluated empirical performance of six algorithms:
\begin{itemize}
\item{Naive: the brute-force algorithm (Algorithm~\ref{alg:bruteforce_alg}).}
\item{FFT: Fast fourier transform based kernel density estimate~\cite{wand94}.}
\item{FGT: Fast Gauss transform~\cite{greengard1991fgt}.}
\item{IFGT: improved fast Gauss transform~\cite{yang2003improved,raykar2005fast}.}
\item{DFD: the dual-tree centroid-based approximation method~\cite{gray2003vfm,gray2003nde}.}
\item{DFGT: our new algorithm (Algorithm~\ref{alg:dtkdemain}).}
\end{itemize}
\begin{figure}[t]
\small\noindent
\begin{tabular}{|p{0.48in}|p{0.35in}|p{0.35in}|p{0.35in}|p{0.35in}|p{0.35in}|p{0.35in}|p{0.35in}|p{0.35in}|p{0.35in}}
\hline
Alg$\backslash$Scale & {\it 0.001} & {\it 0.01} & {\it 0.1} & {\it 1} & {\it 10} & {\it 100} & {\it 1000} & {\it $\Sigma$}\\
\hline
\multicolumn{9}{|c|}{edsgc-radec, $D = 2, N = 1495877, h_{CV_{LS}}^* = 0.000473061$}\\
\hline
{\it Naive} & 2.2e5 & 2.2e5 & 2.2e5 & 2.2e5 & 2.2e5 & 2.2e5 & 2.2e5 & 1.5e6 \\
\hline
{\it DFD} & 4.9e1 & 4.9e1 & 6.3e1 & 1e2 & 1.5e3 & 2e4 & 1.3e3 & 2.3e4\\
\hline
{\it DFGT} & 6.8e0 & 7.4e0 & 2.1e1 & 5.9e1 & 1.7e3 & 3.5e3 & 1.4e2 & 5.4e3\\
\hline
\multicolumn{9}{|c|}{mockgalaxy-D-1M, $D = 3, N = 1000000, h_{CV_{LS}}^* = 0.00010681$}\\
\hline
{\it Naive} & 9.6e4 & 9.6e4 & 9.6e4 & 9.6e4 & 9.6e4 & 9.6e4 & 9.6e4 & 6.7e5\\
\hline
{\it DFD} & 2.4e0 & 2.4e0 & 2.6e0 & 1.5e1 & 9.7e1 & 1.7e2 & 4.4e3 & 4.7e3\\
\hline
{\it DFGT} & 2.4e0 & 2.4e0 & 2.6e0 & 1.5e1 & 1.1e2 & 2.1e2 & 4e3 & 4.3e3\\
\hline
\multicolumn{9}{|c|}{psf1-psf4-stargal-2d-only, $D = 2, N = 3056092, h_{CV_{LS}}^* = 0.00489463$}\\
\hline
{\it Naive} & 9e5 & 9e5 & 9e5 & 9e5 & 9e5 & 9e5 & 9e5 & 6.3e6\\
\hline
{\it DFD} & 1.1e2 & 1.5e2 & 1.2e3 & 2.2e4 & 3.9e4 & 2.9e3 & 1.1e2 & 6.5e4\\
\hline
{\it DFGT} & 3.9e1 & 8.1e1 & 1.4e3 & 1.6e4 & 2.3e3 & 1.9e2 & 4.2e1 & 1.9e4\\
\hline
\end{tabular}
\caption{Empirical comparison of three algorithms on different
  magnitudes of bandwidths on three larger datasets. All timings are
  reported in seconds.}
\end{figure}
We used the following six real-world datasets:
\begin{itemize}
\item{{\it sj2-50000-2}: two-dimensional astronomy position dataset.}
\item{{\it colors50k}: two-dimensional astronomy color dataset.}
\item{{\it bio5}: five-dimensional pharmaceutical dataset.}
\item{{\it edgsc-radec}: two-dimensional astronomy angle dataset.}
\item{{\it mockgalaxy-D-1M}: three-dimensional astronomy position
  dataset.}
\item{{\it psf1-psf4-stargal-2d-only}: two-dimensional astronomy
  dataset.}
\end{itemize}
Note that the last three datasets contain over 1 million points and
demonstrate the scalability of our fast algorithm.  For each dataset,
we evaluated the empirical performance on computing kernel density
estimates at seven different bandwidths ranging from $10^{-3}$ to
$10^3$ times the optimal bandwidths according to the standard
least-squares cross-validation score~\cite{silverman1986des}. We
measured the time required for computing KDE estimates that guarantee
the global relative error criterion: $\left | \widetilde{G}(q_i,
\mathcal{R}) - G(q_i, \mathcal{R}) \right | \leq \epsilon G(q_i,
\mathcal{R})$. We used $\epsilon = 0.01$. Each entry in the table has
a timing number (if finite), $\infty$ symbol (if no parameter tweaking
could achieve the error tolerance), $X$ symbol (if the algorithm
segfaulted; this is common in grid-based algorithms in higher
dimension). The entries under $\Sigma$ symbol denote the total time
for least-squares cross-validation.
\begin{figure}[t]
\scalebox{0.635}{\includegraphics{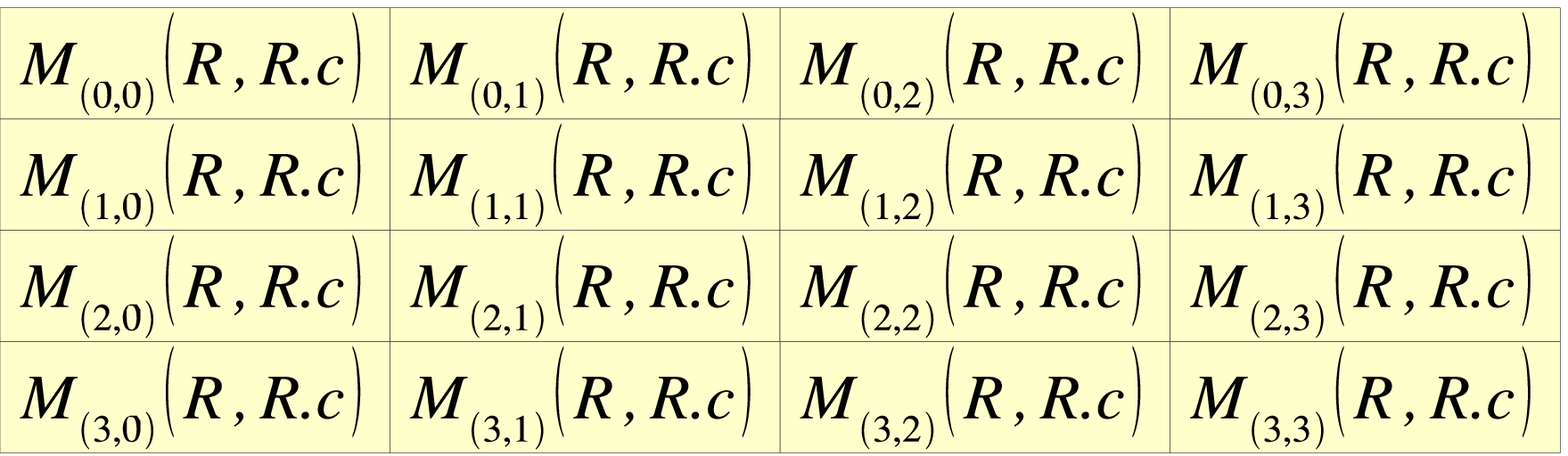}}
\scalebox{0.5}{\includegraphics[angle=-270]{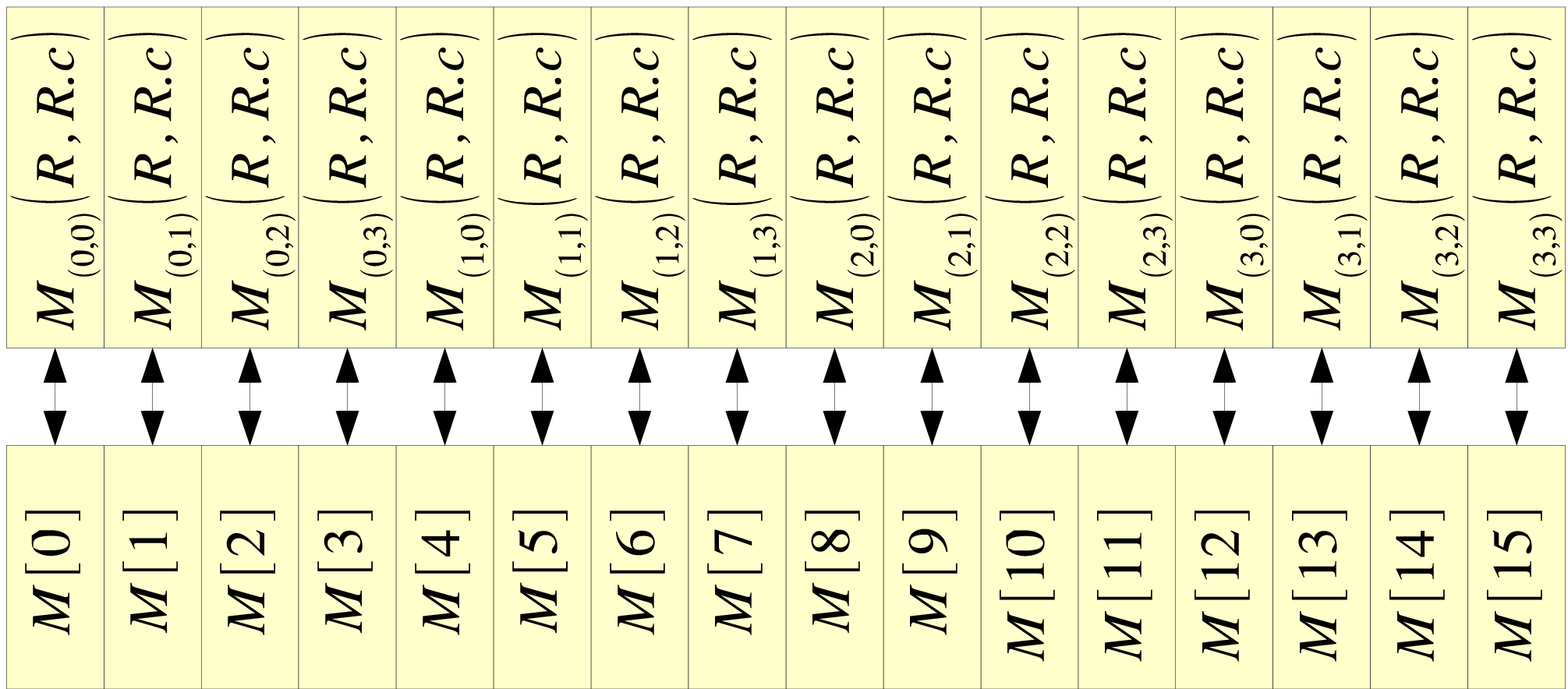}}
\caption{{\bf Top:} It is conceptually easy to visualize the moments
  to be stored in a multi-dimensional array conceptually. Each
  dimension iterates over $p_{\mathit{max}}$ scalars, giving a total
  count of $\left( p_{\mathit{max}} \right)^D$ scalars. {\bf Bottom:
  }The linear layout for the storing the coefficients.}
\label{fig:linear_array_layout}
\end{figure}
Note that the FGT ensures: $\left |
\widetilde{G}(q_i, \mathcal{R}) - G(q_i, \mathcal{R}) \right | \leq
\tau$. Therefore, we first set $\tau = \epsilon$, halving $\tau$ until
the error tolerance $\epsilon$ was met; the time for verifying the
global error guarantee (which includes comparison against the naively
computed results) was not included in the timing. For the FFT, we
started with 16 grid points along each dimension, and doubled the
number of grid points until the error guarantee was met. For the IFGT,
we took the most recent version of the algorithm that does automatic
parameter tuning described in~\cite{raykar2005fast}. Our algorithms
based on dual-tree methods guarantees the error bound automatically
via a direct parameter $\epsilon$.

The naive timings for the last datasets have been extrapolated from
the performances on the smaller datasets. Our results demonstrate that
our new algorithm can be as 15 times as fast as the original dual-tree
algorithm. As expected, the grid-based original fast Gauss transform
and the fast Fourier transformed based method fails in dimensions
above two.

\section{Conclusion}
In this paper, we combined the two methods: the dual-tree
KDE~\cite{gray2001nbp} and the original fast Gauss
transform~\cite{greengard1991fgt} to form the hierarchical form of
the fast Gauss transform, the Dual-tree Fast Gauss Transform. Our
results demonstrate that the $O(p^D)$ expansion helps reduce the
computational time on datasets of dimensionality up to 5.

\section*{Appendix: Implementing the Gaussian Series-expansion}
\begin{figure}[h]
\scalebox{0.5}{\includegraphics{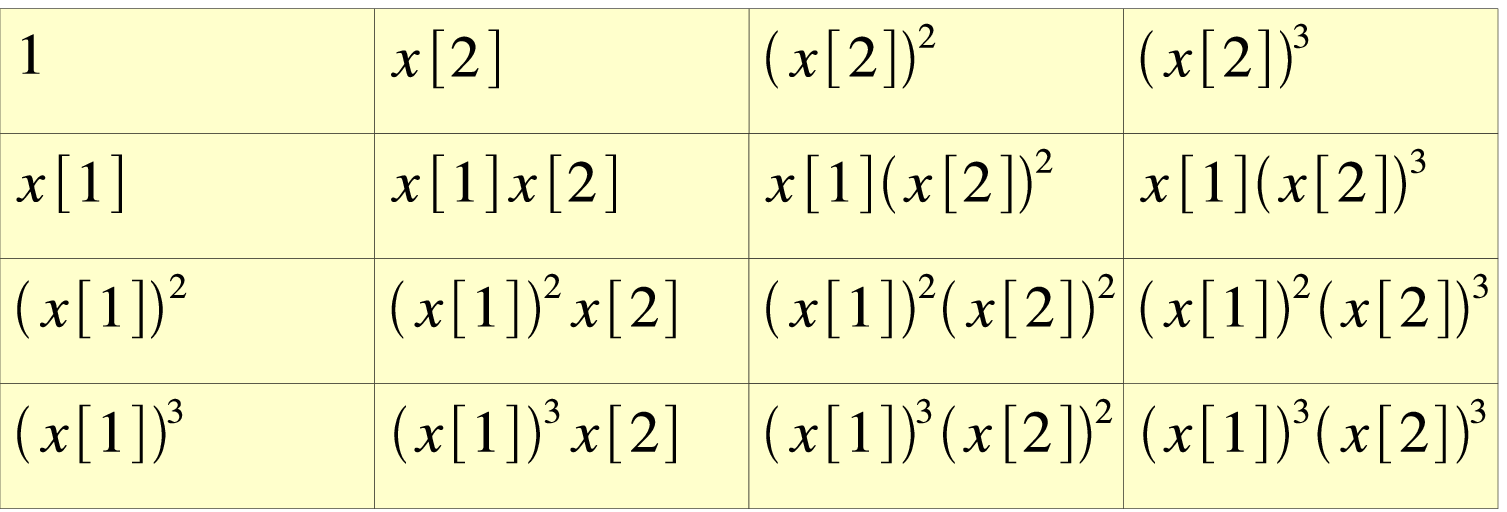}}
\caption{The multi-index expansion of a 2-D vector $x = [ x[1], x[2]
  ]^T$ up to 16 terms.}
\label{fig:multi_index_expansion}
\end{figure}

This section explains how to implement the series-expansion mechanisms
in computer languages such as C/C++.
\ \\{\bf Storing the far-field/local moments as a linear
array. }Although the moments are inherently multi-dimensional, we store
all coefficients in a C-style one-dimensional array.
Each query node stores $\left( p_{\mathit{max}} \right)^D$ local
moment terms. Similarly, each reference node stores $\left(
p_{\mathit{max}} \right)^D$ far-field moment terms. These are
allocated as a linear array during the construction of the two trees,
as shown in Figure~\ref{fig:linear_array_layout} which implies a
bijective mapping between $D$-digit radix-$p_{\mathit{max}}$ numbers
and decimal numbers between 0 and $p_{\mathit{max}}^D$ - 1 inclusive.

\ \\{\bf Converting between a position and a multi-index in the linear
array. }Algorithm~\ref{alg:position_to_multiindex} shows the mapping
from a position in the linear array of $\left( p_{\mathit{max}}
\right)^D$ terms to its corresponding multi-index. The algorithm
converts the given position (given in base 10) to a number in base
$p$.
\begin{algorithm}[h]
\caption{$\mbox{\positiontomultiindex}(i, p)$: Converts the position
  of a linear array of length $p^D$ to its multi-index.}

\begin{algorithmic}

\STATE{\COMMENT{$i$-th position maps to the multi-index $\alpha$.}}

\STATE{$\alpha_{i = 1, \cdots, D} \leftarrow 0$}

\FOR{$d = D$ to $d = 1$}

\STATE{$\alpha[d - (D - 1)] \leftarrow \left \lfloor \frac{i}{p}
\right \rfloor$}

\STATE{$i \leftarrow i \mod p$}

\ENDFOR

\RETURN{$\alpha$}

\end{algorithmic}
\label{alg:position_to_multiindex}
\end{algorithm}
Algorithm~\ref{alg:multiindex_to_position} converts the given
multi-index to its corresponding position in the linear array of
length $\left( p_{\mathit{max}} \right)^D$. It is basically an
algorithm to convert a radix-$p_{\mathit{max}}$ number to its decimal
representation.
\begin{algorithm}[h]
\caption{$\mbox{\multiindextoposition}(\alpha)$: Converts the given
  multi-index to its corresponding position in the linear array of
  length $\left( p_{\mathit{max}}\right)^D$.}

\begin{algorithmic}

\STATE{\COMMENT{Converted position from the multi-index.}}

\STATE{$x \leftarrow 0$, \ \ \ \ \ \ \ \ $f \leftarrow 1$}

\FOR{$d = D$ to $d = 1$}

\STATE{$x \leftarrow x + f \cdot \alpha[d]$}

\STATE{$f \leftarrow f \cdot p_{\mathit{max}}$}

\ENDFOR

\RETURN{$x$}

\end{algorithmic}
\label{alg:multiindex_to_position}
\end{algorithm}

\ \\ {\bf Computing a multi-index expansion of a vector. }A
multi-index expansion of a vector $x \in \mathbb{R}^D$ up to $p^D$
terms is basically the set of coefficients $\{ x^{\alpha} \}_{\alpha <
  p}$. See Figure~\ref{fig:multi_index_expansion}. This is used in the
process of forming a far-field moment contribution of a single
reference point in $\mbox{\accumulatefarfieldmoment}$ and evaluating a
local expansion in $\mbox{\evaluatelocalexpansion}$.
\begin{algorithm}[h]
\caption{$\mbox{\computemultiindexexpansion}(x, p, M')$: Computes $M'
  = \left \{ x ^{\alpha} \right \}_{\alpha < p}$.}

\begin{algorithmic}

\STATE{$M'[0] \leftarrow 1$}

\FOR{each $i = 0$ to $i = p^D - 1$}

\STATE{\COMMENT{Retrieve the multi-index mapping of the current
    position.}}

\STATE{$\alpha \leftarrow \mbox{\positiontomultiindex}(i, p)$}

\STATE{$j \leftarrow$ the first index of $\alpha$ such that $\alpha[j]
\geq 1$.}

\STATE{\COMMENT{Found a direct ancestor of the multiindex map
    $\alpha$.}}

\STATE{$\alpha' \leftarrow \alpha$, \ \ \ \ $\alpha'[j] \leftarrow \alpha'[j] - 1$}

\STATE{\COMMENT{Recursively compute the $\alpha$-th multi-index
    component based on $\alpha'$-th.}}

\STATE{$M'[i] \leftarrow M'[\mbox{\multiindextoposition}(\alpha')]
\cdot x[j]$}

\ENDFOR

\end{algorithmic}
\label{alg:compute_multi_index_expansion}
\end{algorithm}

\ \\{\bf Implementing the far-field moment accumulation
  (Equation~\eqref{eq:far_field_moments}). }This is straightforward
given the implementation of the function\\
$\mbox{\computemultiindexexpansion}$. Basically, it computes the
multi-index of each reference point in the given reference node and
accumulates each contribution and normalizes the sum. See
Algorithm~\ref{alg:accumulatefarfieldmoment}.
\begin{algorithm}[t]

\caption{$\mbox{\accumulatefarfieldmoment}(R)$: Implements
  Equation~\eqref{eq:far_field_moments}.}

\begin{algorithmic}

\STATE{\COMMENT{Temporary space that is equal in size to $\{
    M_{\alpha}(R, R.c) \}_{\alpha < p_{\mathit{max}}}$.}}

\STATE{$M'_{i = 0, \cdots, \left( p_{\mathit{max}} \right)^D - 1}
\leftarrow 0$}

\FOR{each $r_{j_n} \in R$}

\STATE{\COMMENT{Add $M' = \left \{ \left( \frac{r_{j_n} -
      R.c}{\sqrt{2h^2}} \right)^{\alpha} \right \}_{\alpha <
      p_{\mathit{max}}}$ onto $\{ M_{\alpha} (R, R.c) \}_{\alpha <
      p_{\mathit{max}}}$.}}

\STATE{$\mbox{\computemultiindexexpansion}\left (\frac{r_{j_n} -
R.c}{\sqrt{2h^2}}, p_{\mathit{max}}, M' \right )$}

\STATE{$\{ M_{\alpha} (R, R.c) \}_{\alpha < p_{\mathit{max}}}
\leftarrow \{ M_{\alpha} (R, R.c) \}_{\alpha < p_{\mathit{max}}} +
M'$}

\ENDFOR

\FOR{$i = 0$ to $i = \left( p_{\mathit{max}} \right)^D - 1$}

\STATE{$M_{\alpha}(R, R.c) \leftarrow M_{\alpha}(R, R.c) \cdot
 \frac{1}{\alpha!}$}

\ENDFOR

\end{algorithmic}
\label{alg:accumulatefarfieldmoment}
\end{algorithm}

\ \\ {\bf Implementing the far-to-far translation operator} (shown in
Algorithm~\ref{alg:translatefartofarfield}). This consists of a
doubly-nested for-loop over accumulated far-field moments.
\begin{algorithm}[h]
\caption{$\mbox{\translatefartofarfield}(R', R)$: Implements
  Equation~\eqref{eq:far_to_far_translation}.}

\begin{algorithmic}

\STATE{\COMMENT{Allocate space for and compute $\left \{ \left(
    \frac{R'.c - R.c}{\sqrt{2h^2}} \right)^{\alpha} \right \}_{\alpha
      < p_{\mathit{max}}}$.}}

\STATE{$C_{i = 0, \cdots, \left( p_{\mathit{max}} \right)^D - 1}
\leftarrow 0$}

\STATE{$\mbox{\computemultiindexexpansion}\left( \frac{R'.c -
R.c}{\sqrt{2h^2}}, p_{\mathit{max}}, C \right)$}

\STATE{}

\FOR{$i = 0$ to $i < (p_{\mathit{max}})^D$}

\STATE{$\gamma \leftarrow \mbox{\positiontomultiindex}(i,
p_{\mathit{max}})$}

\FOR{$j = 0$ to $j < (p_{\mathit{max}})^D$}

\STATE{$\alpha \leftarrow \mbox{\positiontomultiindex}(j,
p_{\mathit{max}})$}

\IF{$\alpha \leq \gamma$}

\STATE{$M_{\gamma}(R, R.c) \leftarrow  M_{\gamma}(R, R.c) +$}

\STATE{\ \ \ \ \ \ \ \ \ $\frac{1}{(\gamma - \alpha)!} M_{\alpha}(R', R'.c) \ \ 
  \cdot \ \ C[\mbox{\multiindextoposition}(\gamma - \alpha)]$}

\ENDIF

\ENDFOR

\ENDFOR

\end{algorithmic}
\label{alg:translatefartofarfield}
\end{algorithm}

\ \\ {\bf Computing the multivariate Hermite functions. }We exploit
the fact that the multivariate Hermite functions is a product of $D$
univariate Hermite
functions. Algorithm~\ref{alg:compute_partial_derivatives} computes
partial derivatives of the Gaussian kernel evaluated at the given
point $x$ along each dimension up to $p$-th order. $h_{\alpha}(x) =
\prod\limits_{d=1}^D h_{\alpha[d]}(x)$ is a simple product of the
univariate functions (see
Algorithm~\ref{alg:compute_hermite_function}).
\begin{algorithm}[t]
\caption{$\mbox{\computepartialderivatives}(a, p, H)$: Evaluates the
  partial derivatives of $e^{-x^2 / (2h^2)}$ up to $(p - 1)$-th order
  at each coordinate of $a$.}

\begin{algorithmic}

\FOR{$d = 1$ to $D$}

\STATE{$H[d][0] \leftarrow e^{-(a[d])^2}$}

\IF{$p > 1$}

\STATE{$H[d][1] \leftarrow 2 \cdot a[d] \cdot e^{-(a[d])^2}$}

\IF{$p > 2$}

\FOR{$k = 1$ to $k = p - 2$}

\STATE{$H[d][k + 1] \leftarrow 2 \cdot a[d] \cdot H[d][k] - 2 \cdot
k \cdot H[d][k - 1]$}

\ENDFOR

\ENDIF

\ENDIF

\ENDFOR

\end{algorithmic}
\label{alg:compute_partial_derivatives}
\end{algorithm}

\begin{algorithm}[t]
\caption{$\mbox{\computehermitefunction}(H, \alpha)$: Computes the
  Hermite function $h_{\alpha}( \cdot)$ using the pre-computed partial
  derivatives $H$.}

\begin{algorithmic}

\STATE{$f \leftarrow 1$}

\FOR{$d = 1$ to $D$}

\STATE{$f \leftarrow f \cdot H[d][\alpha[d]]$}

\ENDFOR

\RETURN{$f$}

\end{algorithmic}
\label{alg:compute_hermite_function}
\end{algorithm}

\ \\ {\bf Evaluating a far-field expansion. }Once the functions for
computing the Hermite functions
(Algorithm~\ref{alg:compute_partial_derivatives} and
Algorithm~\ref{alg:compute_hermite_function}), we can implement the
function for evaluating a far-field expansion up to $p^D$ terms, as
shown in Algorithm~\ref{alg:evaluate_far_field_expansion}. The basic
structure is one outer-loop over each query point and the inner loop
iterating over each far-field moment. The contribution to each query
point is computed as a dot-product between the far-field moment and
the computed Hermite functions (see
Figure~\ref{fig:far_field_coeffs}).
\begin{algorithm}[t]
\caption{$\mbox{\evaluatefarfieldexpansion}(R, Q, p)$: Evaluates the
  far-field expansion of the given reference node $R$ up to $p^D$
  terms.}

\begin{algorithmic}

\STATE{\COMMENT{Allocate space for holding the partial derivatives.}}

\STATE{$H_{\substack{d = 1, \cdots, D\\ k = 0, \cdots, p - 1}}
\leftarrow 0$}

\STATE{}

\FOR{each $q_{i_m} \in Q$}

\STATE{\COMMENT{Compute partial derivatives up to $(p - 1)$-th
    order along each dimension.}}

\STATE{$\mbox{\computepartialderivatives}\left ( \frac{q_{i_m} -
R.c}{\sqrt{2h^2}}, p, H \right)$}

\STATE{$w \leftarrow 0$}

\FOR{$i = 0$ to $i = p^D -1$}

\STATE{$\alpha \leftarrow \mbox{\positiontomultiindex}(i, p)$}

\STATE{$f \leftarrow \mbox{\computehermitefunction}(H, \alpha)$}

\STATE{$w \leftarrow w + M_{\alpha}(R, R.c) \cdot f$}

\ENDFOR

\STATE{$\widetilde{G}(q_{i_m}, \mathcal{R_F}(q_{i_m})) \leftarrow
\widetilde{G}(q_{i_m}, \mathcal{R_F}(q_{i_m})) + w$}

\ENDFOR

\end{algorithmic}
\label{alg:evaluate_far_field_expansion}
\end{algorithm}

\ \\ {\bf Implementing the far-to-local translation operator. }The
basic structure of the algorithm is a doubly nested for-loop, each
over the coefficients. The doubly-nested for-loop first translate a
portion of the accumulated far-field moments of $R$ up to $p^D$ terms
into the local moments. The final step of the algorithm is to add the
translated moments $\{ L_{\beta} ( \{ (R, T(Q.c, p)) \} ) \}$ to the
local moments stored in $Q$, $L_{\beta} (Q.c, \mathcal{R_D}(Q) \cup
\mathcal{R_T}(Q) )$. See Algorithm~\ref{alg:translate_far_to_local}.
\begin{algorithm}[t]
\caption{$\mbox{\translatefartolocal}(R, Q, p)$: Implements
  Equation~\eqref{eq:farfield_to_local_formula}.}

\begin{algorithmic}

\STATE{$H_{\substack{d=1, \cdots, D\\k = 0, \cdots, 2 (p - 1)}}
\leftarrow 0$}

\STATE{$\mbox{\computepartialderivatives}\left(\frac{Q.c -
R.c}{\sqrt{2h^2}}, 2p - 1, H \right)$}

\STATE{}

\FOR{$i = 0$ to $i = p^D - 1$}

\STATE{$\beta \leftarrow \mbox{\positiontomultiindex}(i, p)$}

\FOR{$j = 0$ to $j = p^D - 1$}

\STATE{$\alpha \leftarrow \mbox{\positiontomultiindex}(j, p)$}

\STATE{$f \leftarrow \mbox{\computehermitefunction}(H, \alpha +
\beta)$}

\STATE{$L_{\beta}(\{( R, T(Q.c, p)) \}) \leftarrow L_{\beta}(\{ (R,
T(Q.c, p)) \}) + M_{\alpha} (R, R.c) \cdot f$}

\ENDFOR

\STATE{$L_{\beta}(\{( R, T(Q.c, p)) \}) \leftarrow
\frac{(-1)^{|\beta|}}{\beta!} L_{\beta}(\{ (R, T(Q.c, p)) \})$}

\ENDFOR

\STATE{$\{ L_{\beta}(Q.c, \mathcal{R_D}(Q) \cup \mathcal{R_T}(Q))
\}_{\beta < p} \leftarrow \{ L_{\beta}(Q.c, \mathcal{R_D}(Q) \cup
\mathcal{R_T}(Q)) \}_{\beta < p} + \left \{ L_{\beta}(\{(R, T(Q.c, p))
\}) \right \}_{\beta < p}$}

\end{algorithmic}
\label{alg:translate_far_to_local}
\end{algorithm}

\ \\ {\bf Implementing the direct local accumulation operation. }The
basic structure is a doubly-nested for-loop, the outer-loop over the
reference points whose moments are to be accumulated as local moments
and the inner loop over the coefficient positions. See
Algorithm~\ref{alg:accumulate_direct_local_moment}.
\begin{algorithm}[t]

\caption{$\mbox{\accumulatedirectlocalmoment}(R, Q, p)$: Implements
  Equation~\eqref{eq:local_moments}.}

\begin{algorithmic}

\STATE{$H_{\substack{d = 1, \cdots, D\\ k = 0, \cdots, p - 1}}
\leftarrow 0$, \ \ \ \ \ \ \ \ \ \
$\{ L_{\beta}(\{ (R, D(Q.c, p)) \})\}_{\beta < p} \leftarrow
0$}

\FOR{each $r_{j_n} \in R$}

\STATE{$\mbox{\computepartialderivatives}\left( \frac{Q.c -
r_{j_n}}{\sqrt{2h^2}}, p, H \right)$}

\FOR{$i = 0$ to $p^{D} - 1$}

\STATE{$\alpha \leftarrow \mbox{\positiontomultiindex}(i, p)$}

\STATE{$f \leftarrow \computehermitefunction(H, \beta)$}

\STATE{$L_{\beta}(\{ (R, D(Q.c, p)) \}) \leftarrow L_{\beta}(\{ (R,
D(Q.c, p)) \}) + f$}

\ENDFOR

\ENDFOR

\STATE{$\{ L_{\beta}(\{ (R, D(Q.c, p)) \})\}_{\beta < p} \leftarrow
\{L_{\beta}(\{ (R, D(Q.c, p)) \}) \}_{\beta < p} * \frac{(-1)^{| \beta
|}}{\beta!}$}

\STATE{$\{ L_{\beta}(Q.c, \mathcal{R_D}(Q) \cup \mathcal{R_T}(Q))
\}_{\beta < p} \leftarrow \{ L_{\beta}(Q.c, \mathcal{R_D}(Q) \cup
\mathcal{R_T}(Q)) \}_{\beta < p} + \left \{ L_{\beta}(\{(R, D(Q.c, p))
\}) \right \}_{\beta < p}$}

\end{algorithmic}
\label{alg:accumulate_direct_local_moment}
\end{algorithm}

\ \\ {\bf Implementing the local-to-local translation operator. }We
direct readers' attention to the first step of the algorithm, which
retrieves the maximum order among used in local moment
accumulation/translation. Then the algorithm proceeds with a
doubly-nested for-loop over the local moments applies
Equation~\eqref{eq:local_to_local_translation}. See
Algorithm~\ref{alg:translate_local_to_local}.
\begin{algorithm}[t]
\caption{$\mbox{\translatelocaltolocal}(Q', Q)$: Implements
  Equation~\eqref{eq:local_to_local_translation}.}

\begin{algorithmic}

\STATE{\COMMENT{$p$ is the maximum approximation order used among the
    reference nodes pruned via far-to-local and direct local
    accumulations for $Q'$.}}

\STATE{$p \leftarrow \max \left \{ \max\limits_{R \in
\mathcal{R_D}(Q')} p_D, \max\limits_{R \in \mathcal{R_T}(Q')} p_T \right
\}$}

\STATE{\COMMENT{Temporary space that is equal in size to $\{ L_{\beta} \}$.}}

\STATE{$X \leftarrow 0$}

\STATE{$\mbox{\computemultiindexexpansion}\left( \frac{Q.c -
Q'.c}{\sqrt{2h^2}}, p, X \right)$}

\FOR{$j = 0$ to $p^D - 1$}

\STATE{$\alpha \leftarrow \mbox{\positiontomultiindex}(j, p)$}

\FOR{$k = 0$ to $p^D - 1$}

\STATE{$\beta \leftarrow \mbox{\positiontomultiindex}(k, p)$}

\IF{$\beta \geq \alpha$}

\STATE{$L_{\beta}(Q.c, \mathcal{R_D}(Q') \cup \mathcal{R_T}(Q'))
\leftarrow L_{\beta}(Q.c, \mathcal{R_D}(Q') \cup \mathcal{R_T}(Q'))
+$}

\STATE{\ \ \ \ \ \ $\frac{\beta!}{\alpha! (\beta - \alpha)!}
L_{\beta}(Q'.c, \mathcal{R_D}(Q') \cup \mathcal{R_T}(Q')) X_{\beta -
\alpha}$}

\ENDIF

\ENDFOR

\ENDFOR

\STATE{$\{ L_{\beta}(Q.c, \mathcal{R_D}(Q) \cup \mathcal{R_T}(Q))
\}_{\beta < p} \leftarrow \{ L_{\beta}(Q.c, \mathcal{R_D}(Q) \cup
\mathcal{R_T}(Q)) \}_{\beta < p} + \{ L_{\beta}(Q.c, \mathcal{R_D}(Q')
\cup \mathcal{R_T}(Q')) \}_{\beta < p}$}

\end{algorithmic}
\label{alg:translate_local_to_local}
\end{algorithm}

\ \\ {\bf Evaluating the local expansion of the given query
  node. }This function (see
Algorithm~\ref{alg:evaluate_local_expansion}) is consisted of one
outer-loop over reference points and the inner-loop over the local
moments up to $p^D$ terms, where $p$ is the maximum approximation
order used among the reference nodes pruned via far-to-local and
direct local accumulations for $Q$.
\begin{algorithm}[t]
\caption{$\mbox{\evaluatelocalexpansion}(Q)$: Evaluates the
  accumulated local expansion of the given query node $Q$.}
\begin{algorithmic}

\STATE{\COMMENT{$p$ is the maximum approximation order used among the
    reference nodes pruned via far-to-local and direct local
    accumulations for $Q$.}}

\STATE{$p \leftarrow \max \left \{ \max\limits_{R \in
\mathcal{R_D}(Q)} p_D, \max\limits_{R \in \mathcal{R_T}(Q)} p_T \right
\}$}

\STATE{\COMMENT{Temporary space to hold the multi-index expansion of each
$\left( \frac{q_{i_m} - Q.c}{\sqrt{2h^2}} \right )^{\alpha}$.}}

\STATE{$X_{i = 0, \cdots, p^D - 1} \leftarrow 0$}

\STATE{}

\FOR{each $q_{i_m} \in Q$}

\STATE{$z \leftarrow 0$}

\STATE{\COMMENT{Compute the multi-index expansion of $\frac{q_{i_m} -
Q.c}{\sqrt{2h^2}}$ up to $p^D$ terms.}}

\STATE{$\mbox{\computemultiindexexpansion}\left( \frac{q_{i_m} -
Q.c}{\sqrt{2h^2}}, p, X \right)$}

\FOR{$i = 0$ to $i = p^D - 1$}

\STATE{$\beta \leftarrow \mbox{\positiontomultiindex}(i, p)$}

\STATE{$z \leftarrow z + L_{\beta}(Q.c, \mathcal{R_D}(Q) \cup
\mathcal{R_T}(Q)) \cdot z$}

\ENDFOR

\STATE{$\widetilde{G}(q_{i_m}, \mathcal{R_D}(q_{i_m}) \cup
\mathcal{R_T}(q_{i_m})) \leftarrow \widetilde{G}(q_{i_m},
\mathcal{R_D}(q_{i_m}) \cup \mathcal{R_T}(q_{i_m})) + z$}

\ENDFOR

\end{algorithmic}
\label{alg:evaluate_local_expansion}
\end{algorithm}

\bibliographystyle{abbrv}
\bibliography{../bibtexfile/global.bib}

\begin{thebibliography}{10}

\bibitem{appel1985efficient}
A.~Appel.
\newblock {An efficient program for many-body simulation}.
\newblock {\em SIAM Journal on Scientific and Statistical Computing}, 6:85,
  1985.

\bibitem{barnes1986hof}
J.~Barnes and P.~Hut.
\newblock {A Hierarchical $O(N log N)$ Force-Calculation Algorithm}.
\newblock {\em Nature}, 324, 1986.

\bibitem{baxter2002new}
B.~Baxter and G.~Roussos.
\newblock {A new error estimate of the fast Gauss transform}.
\newblock {\em SIAM Journal on Scientific Computing}, 24:257, 2002.

\bibitem{bentley1975multidimensional}
J.~L. Bentley.
\newblock {Multidimensional Binary Search Trees used for Associative
  Searching}.
\newblock {\em {Communications of the ACM}}, 18:509--517, 1975.

\bibitem{callahan1995dwh}
P.~B. Callahan.
\newblock {\em Dealing with Higher Dimensions: The Well-Separated Pair
  Decomposition and its Applications}.
\newblock PhD thesis, Johns Hopkins University, Baltimore, Maryland, 1995.

\bibitem{gray2003rem}
A.~Gray and A.~Moore.
\newblock {Rapid evaluation of multiple density models}.
\newblock {\em Artificial Intelligence and Statistics}, 2003.

\bibitem{gray2003vfm}
A.~Gray and A.~Moore.
\newblock {Very fast multivariate kernel density estimation via computational
  geometry}.
\newblock In {\em Joint Stat. Meeting}, 2003.

\bibitem{gray2001nbp}
A.~Gray and A.~W. Moore.
\newblock {N-Body Problems in Statistical Learning}.
\newblock In T.~K. Leen, T.~G. Dietterich, and V.~Tresp, editors, {\em
  {Advances in Neural Information Processing Systems 13 (December 2000)}}. MIT
  Press, 2001.

\bibitem{gray2003nde}
A.~G. Gray and A.~W. Moore.
\newblock {Nonparametric Density Estimation: Toward Computational
  Tractability}.
\newblock In {\em {SIAM International Conference on Data Mining 2003}}, 2003.

\bibitem{greengard1987fap}
L.~Greengard and V.~Rokhlin.
\newblock {A Fast Algorithm for Particle Simulations}.
\newblock {\em Journal of Computational Physics}, 73, 1987.

\bibitem{greengard1991fgt}
L.~Greengard and J.~Strain.
\newblock {The Fast Gauss Transform}.
\newblock {\em SIAM Journal of Scientific and Statistical Computing},
  12(1):79--94, 1991.

\bibitem{lee2006dtf}
D.~Lee, A.~Gray, and A.~Moore.
\newblock Dual-tree fast gauss transforms.
\newblock In Y.~Weiss, B.~Sch\"{o}lkopf, and J.~Platt, editors, {\em Advances
  in Neural Information Processing Systems 18}, pages 747--754. MIT Press,
  Cambridge, MA, 2006.

\bibitem{raykar2005fast}
V.~C. Raykar, C.~Yang, R.~Duraiswami, and N.~Gumerov.
\newblock Fast computation of sums of gaussians in high dimensions.
\newblock Technical Report CS-TR-4767, Department of Computer Science,
  University of Maryland, CollegePark, 2005.

\bibitem{silverman1982kernel}
B.~Silverman.
\newblock {Kernel Density Estimation using the Fast Fourier Transform}.
\newblock {\em Journal of the Royal Statistical Society Series C: Applied
  Statistics}, 33, 1982.

\bibitem{silverman1986des}
B.~W. Silverman.
\newblock {\em {Density Estimation for Statistics and Data Analysis}}.
\newblock {Chapman and Hall/CRC}, 1986.

\bibitem{strain1991fast}
J.~Strain.
\newblock The fast {Gauss} transform with variable scales.
\newblock {\em SIAM Journal on Scientific and Statistical Computing},
  12(5):1131--1139, 1991.

\bibitem{szasz51}
O.~Sz\'asz.
\newblock On the relative extrema of the hermite orthogonal functions.
\newblock {\em J. Indian Math. Soc.}, 15:129--134, 1951.

\bibitem{wand94}
M.~P. Wand.
\newblock {Fast Computation of Multivariate Kernel Estimators}.
\newblock {\em Journal of Computational and Graphical Statistics}, 1994.

\bibitem{yang2003improved}
C.~Yang, R.~Duraiswami, N.~A. Gumerov, and L.~Davis.
\newblock Improved fast gauss transform and efficient kernel density
  estimation.
\newblock {\em International Conference on Computer Vision}, 2003.

\end{thebibliography}

\end{document}